\documentclass[%
 reprint,
superscriptaddress,
twocolumn,
f
 amsmath,amssymb,
 aps,
prb,
]{revtex4-2}
\usepackage[normalem]{ulem}
\usepackage{braket}
\usepackage{graphicx}  
\usepackage{epstopdf}
\usepackage{dcolumn}   
\usepackage{bm}        
\usepackage{amsmath}
\usepackage{amssymb}   
\usepackage{wrapfig}
\usepackage{array}
\usepackage{comment}
\usepackage{bbding}
\usepackage[bookmarks=false,linkcolor=blue,urlcolor=blue,colorlinks,citecolor=blue]{hyperref}
\usepackage[caption=false]{subfig}
\usepackage{color, colortbl}
\newcommand{\di}[1]{d_{#1\uparrow}^\dagger}
\newcommand{\dai}[1]{d_{#1,\sigma}}
\newcommand{\dci}[1]{d_{#1,\sigma}^\dagger}
\newcommand{\dddi}[3]{d_{#1 \uparrow}^\dagger d_{#2 \uparrow}^\dagger d_{#3 \uparrow}^\dagger}
\renewcommand{\hat}[1]{{\bf {\widehat #1}}}

\begin{document}
\title{Giant effective magnetic moments of chiral phonons from orbit-lattice coupling}
\author{Swati Chaudhary}
\email{swati.chaudhary@austin.utexas.edu}
\affiliation{Department of Physics, The University of Texas at Austin, Austin, Texas 78712, USA}
\affiliation{Department of Physics, Northeastern University, Boston, Massachusetts 02115, USA}
\affiliation{Department of Physics, Massachusetts Institute of Technology, Cambridge, Massachusetts 02139, USA}

\author{Dominik M. Juraschek}
\affiliation{School of Physics and Astronomy, Tel Aviv University, Tel Aviv 6997801, Israel}

\author{Martin Rodriguez-Vega}
\affiliation{Department of Physics, The University of Texas at Austin, Austin, Texas 78712, USA}
\affiliation{Department of Physics, Northeastern University, Boston, Massachusetts 02115, USA}
\author{Gregory A.\ Fiete}
\affiliation{Department of Physics, Northeastern University, Boston, Massachusetts 02115, USA}
\affiliation{Department of Physics, Massachusetts Institute of Technology, Cambridge, Massachusetts 02139, USA}

\begin{abstract}
Circularly polarized lattice vibrations carry angular momentum and lead to magnetic responses in applied magnetic fields or when resonantly driven with ultrashort laser pulses. Recent measurements have found responses that are orders of magnitude larger than those calculated in prior theoretical studies. 
Here, we present a microscopic model for the effective magnetic moments of chiral phonons in magnetic materials that is able to reproduce the experimentally measured magnitudes and that allows us to make quantitative predictions for materials with giant magnetic responses using microscopic parameters. Our model is based on orbit-lattice couplings that hybridize optical phonons with orbital electronic transitions. 
We apply our model to two types of materials: $4f$ rare-earth halide paramagnets and $3d$ transition-metal oxide magnets. In both cases, we find that chiral phonons can carry giant effective magnetic moments of the order of a Bohr magneton, orders of magnitude larger than previous predictions.

\end{abstract}

\maketitle


\section{Introduction}

In circularly polarized lattice vibrations (chiral phonons), the atoms move on closed orbits and can therefore carry angular momentum \cite{Zhang2014,Zhu2018,Streib2021,ishito2023truly,ueda2023chiral}. In an ionic lattice, the orbital motions of the ions create atomistic circular currents and therefore produce a collective phonon magnetic moment \cite{Rebane1983,juraschek2:2017,Juraschek2019,Xiong2022}. These phonon magnetic moments lead to different magnetic responses when an external magnetic field is applied or when they are resonantly excited with an ultrashort terahertz pulse. In an applied static magnetic field, the frequencies between right- and left-handed circular polarizations split up in a phonon Zeeman effect \cite{Anastassakis1971,Holz1972,juraschek2:2017,Juraschek2019} and phonons with opposite chirality are deflected in different directions when propagating in a phonon Hall effect \cite{strohm:2005,sheng:2006,Zhang2010,Grissonnanche2019,Grissonnanche2020,Park2020_phononhall,Saito2019,Flebus2021,Flebus2022}. In turn, when the chiral phonons are infrared-active, they can be resonantly excited with an ultrashort terahertz pulse, which generates a macroscopic phonon magnetic moment and therefore effective magnetic field in a phonon inverse Faraday or phonon Barnett effect \cite{nova:2017,juraschek2:2017,Juraschek2019,Juraschek2020,Juraschek2021,Juraschek2022_giantphonomag,Geilhufe2021,Geilhufe2022,Geilhufe2022electronmagmom,Basini2022,Davies2022,luo2023large}.

The phonon magnetic moment produced by an ionic charge current scales with the gyromagnetic ratio of the ions, $\gamma=Z^*/(2M)$, which depend on the effective charge, $Z^*$, and the ionic mass, $M$. Previous studies based on density functional theory have computed phonon magnetic moments in various nonmagnetic materials to yield up to a nuclear magneton \cite{ceresoli:2002,juraschek2:2017,Hamada2018,Juraschek2019,Geilhufe2021,Zabalo2022}. Intriguingly, a number of early and recent experiments have measured magnitudes of phonon Zeeman effects in paramagnets \cite{schaack:1976,schaack:1977} and in materials with non-trivial quantum geometry in electronic bands \cite{Baydin2022,Cheng2020,hernandez2022chiral} that suggest the presence of phonon magnetic moments ranging from fractions to a few Bohr magnetons, orders of magnitude larger than the nuclear magneton. Furthermore, very recent pump-probe experiments have shown that coherently driven chiral phonons produce effective magnetizations in nonmagnetic materials \cite{Basini2022} and paramagnets \cite{Davies2022,luo2023large} that are compatible with phonon magnetic moments on the Bohr magneton scale. These findings indicate an effective contribution to the phonon magnetic moment arising from electron-phonon or spin-phonon coupling.

Theories based on electron-phonon coupling have so far involved two types of explanations for materials with non-trivial electronic band topology: First, an adiabatic evolution of the electronic states alongside the circularly polarized phonon modes that induces an adiabatic electronic orbital magnetization \cite{Sengupta2020,Xiao2021,Ren2021,Xiao-Wei2023}, and second, a coupling of the cyclotron motion of electrons close to a Dirac point to the chiral phonon mode \cite{Cheng2020}. Theories based on spin-phonon coupling have focused both on nonmagnetic and paramagnetic materials: In paramagnets, phonon modes can couple to the electron spin through modifications of the crystal electric field (CEF) and subsequently through the spin-orbit coupling \cite{Thalmeier1977,thalmeier:1978,Juraschek2022_giantphonomag}. For nonmagnetic materials, a recent proposal suggests that the spin channels of doping-induced conduction electrons couple to the phonon magnetic moment, resulting in phonon-induced electronic polarization \cite{Geilhufe2022electronmagmom}. Despite these developments, a microscopic theory that can quantitatively predict the experimentally found giant effective magnetic moments of phonons is still missing.

In this study, we develop a microscopic model for effective phonon magnetic moments in paramagnetic and magnetic materials. Our model is based on orbit-lattice coupling, where chiral phonon modes induce transitions between different orbital states, similar to the Raman mechanism of spin relaxation \cite{Orbach1961}. We perform a comprehensive group-theoretical analysis to identify the possible couplings between chiral phonon modes and orbital transitions and apply it to two distinct cases: First, we apply our model to the well-known case of $4f$ paramagnetic rare-earth trihalides, in which the spin-orbit coupling is much larger than the CEF splitting. In these materials, a CEF excitation hybridizes with circularly polarized phonons, which allows them to obtain a large phonon magnetic moment. Here, our model is, for the first time, able to quantitatively predict the giant phonon Zeeman splittings that were measured already half a century ago \cite{schaack:1976,schaack:1977}, using only microscopic parameters in combination with results from first-principles calculations. Second, our model predicts that a similar phonon Zeeman effect can also occur in $d$-orbital magnets, where the CEF splitting of the $e_g$ and $t_{2g}$ orbits is much larger than the spin-orbit coupling. In $d$-orbital systems, the hybridization occurs between orbital excitations connecting multiplets split by spin-orbit coupling or lattice distortions and circularly polarized phonons. We predict that this mechanism can lead to large effective phonon magnetic moments and, therefore, phonon Zeeman splittings when the energies of the orbital transitions and the phonons become comparable.

The manuscript is organized as follows. In Sec.~\ref{sec:II}, we develop a microscopic model for the coupling of doubly degenerate modes to orbital transitions and show that it leads to a splitting into circularly polarized phonons with opposite chirality in an applied magnetic field. We furthermore derive an expression for the effective magnetic moment that the chiral phonons obtain through this coupling and show that it naturally leads to the phenomenological expression for the phonon Zeeman splitting in the limit of small magnetic fields. In Sec.~\ref{sec:III}, we apply our model to the $4f$ paramagnet CeCl$_3$ and predict the effective phonon magnetic moment and the phonon Zeeman splitting for its doubly degenerate chiral phonon modes. In Sec.~\ref{sec:IV}, we apply our model to $3d$ magnets and predict the effective phonon magnetic moment and phonon Zeeman splitting for the example of the transition-metal oxide CoTiO$_3$ in both its paramagnetic and antiferromagnetic phases. In Sec.~\ref{sec:V}, we conclude with a discussion of the results.


\section{Phonon Zeeman splitting and effective magnetic moments\label{sec:II}}

In this section, we discuss a detailed microscopic model for the Zeeman splitting of doubly degenerate zone-center phonon modes. 
This splitting arises from the hybridization of doubly degenerate phonon modes with orbital excitations on a magnetic ion when the degeneracy of Kramer pairs is lifted. The new phonon modes have finite and opposite chirality decided by the sign of the time-reversal symmetry breaking term. This model is based on the early work of Ref.~\cite{Thalmeier1977} and can be described by Hamiltonian:
\begin{equation}
H=H_{el}+H_{ph}+H_{el-ph}.
\end{equation}
Here, $H_{el}$ is the electronic Hamiltonian for the magnetic ion, $H_{ph}$ is the phonon Hamiltonian for a degenerate phonon mode, and $H_{el-ph}$ is the electron-phonon coupling term. We consider a magnetic ion with doubly degenerate electronic levels (Kramers doublets), which are eigenstates of the total electronic angular momentum $J$, and which are split in energy due to CEF or spin-orbit coupling. The electronic Hamiltonian can therefore be written as
\begin{equation}
H_{el}=\sum_i\varepsilon_i \ket{\psi_i}\bra{\psi_i},
\label{elham}
\end{equation}
where $\varepsilon_i$ is the energy of state $i$. These states are given by the ground-state and excited-state Kramers doublets, and we will look at two Kramers doublets at a time. We denote the states of the ground-state doublet by
\begin{align}
\left|\psi_1\right>&=\ket{J=J_\alpha,m_j=m_j^\alpha},\\
\left|\psi_2\right>&=\ket{J=J_\alpha,m_j=-m_j^\alpha},
\end{align}
and those of the excited-state doublet by
\begin{align}
\left|\psi_3\right>&=\ket{J=J_\beta,m_j=m_j^\beta},\\
\left|\psi_4\right>&=\ket{J=J_\beta,m_j=-m_j^\beta}.
\end{align}
In the absence of a magnetic field, $\varepsilon_1=\varepsilon_2$ and $\varepsilon_3=\varepsilon_4$.

\begin{figure}[t]
    \centering
    \includegraphics[scale=0.35]{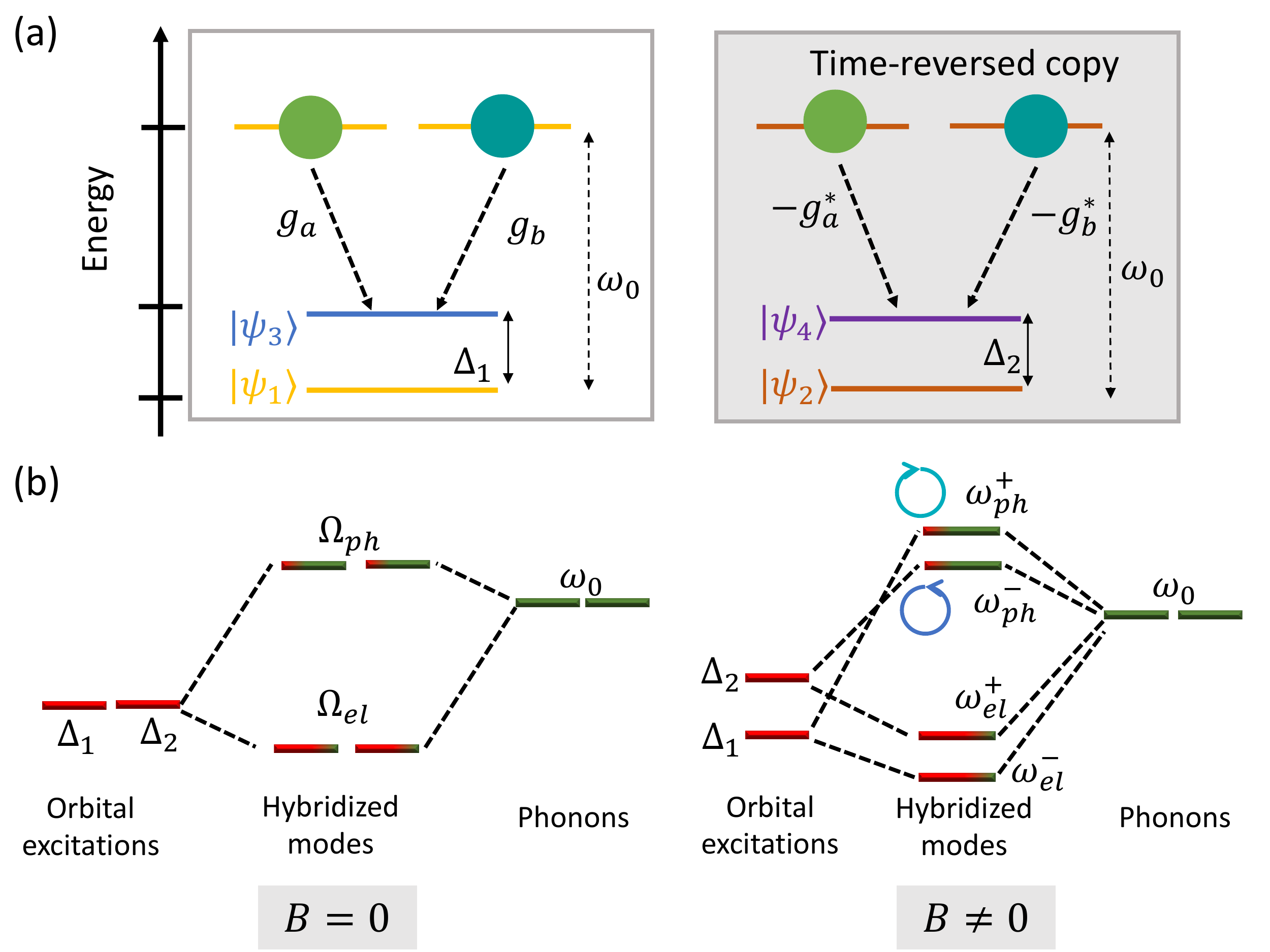}
    \caption{\textbf{Orbit-lattice coupling based mechanism for phonon chirality.}
    (a) Left: The two components, $a$ and $b$ (green spheres), of a doubly degenerate phonon with frequency $\omega_0$ couple to the orbital transition between the ground and excited CEF states, $\psi_1$ and $\psi_3$, with frequency $\Delta_1$. The coupling strengths are given by $g_a$ and $g_b$. Right: The time-reversal partners of the states on the left also couple to the two components of this phonon. When time-reversal symmetry is preserved, states $\psi_1$ and $\psi_2$ are degenerate, as well as $\psi_3$ and $\psi_4$, and therefore $\Delta_1=\Delta_2$. (b) Left: Energy level diagram of the coupling between the orbital transitions, $\Delta_1$ and $\Delta_2$, and a doubly degenerate phonon, $\omega_0$, forming hybrid orbit-lattice modes, $\Omega_{ph}$ and $\Omega_{el}$, with mainly phononic and electronic admixtures respectively. Although the frequencies are normalized, the two phonons remain degenerate Right: A magnetic field lifts the degeneracy of the orbital transitions (Eq.~(\ref{eq:delta_split})) and therefore the degeneracy of the hybrid modes. The hybrid states with mainly phonon admixture can be seen as circularly polarized phonon states with frequencies $\omega_{ph}^\pm$.}
    \label{mainfig}
\end{figure}

For the phonon part, we focus on one doubly degenerate phonon mode at a time, where the Hamiltonian containing both orthogonal components, $a$ and $b$, is given by
\begin{equation}
H_{ph}=\omega_0 (a^\dagger a+ b^\dagger b).
\label{phn}
\end{equation}
where $\omega_0$ is the frequency of the two components of the doubly degenerate phonon mode and we set $\hbar=1$. Here, $a^\dagger, b^\dagger$ ($a, b$) are the bosonic creation (annihilation) operators of the two orthogonal components. These phonons can interact with the electronic states of the magnetic ion and mix different Kramers doublets due to orbit-lattice coupling. On a microscopic level, this orbit-lattice coupling originates from the modification of the crystal electric field by the lattice vibrations.  It can be obtained by expanding the crystal field to first order in the atomic displacements along the eigenvectors of a phonon mode 
and can be written as
\begin{equation}
H_{el-ph}=\sum_{\Gamma_\alpha} Q_{\Gamma_\alpha}\hat{O}_{\Gamma_\alpha}
\end{equation}
where $\hat{O}_{\Gamma_\alpha}$ is an operator acting on electronic states and $Q_{\Gamma_\alpha}$ is the displacement associated with phonon mode $\alpha$ with irreducible representation $\Gamma$. For CEF excitations, where both doublets belong to the same multiplet, these operators take the form of Steven's operators~\cite{stevens1952matrix} but here we keep it  more general to include the possibility of spin-orbit excitations. In our microscopic model, these operators are computed from the changes of the Coulomb potential around the magnetic ion, where we treat all ions as point charges. We ignore higher-order corrections in the lattice displacement, which would lead to higher-order scattering processes that are not considered here.
Within this expansion, the electron-phonon coupling term can be written as
\begin{equation}
H_{el-ph}=(a^\dagger+a)\hat{O}_a+(b^\dagger+b)\hat{O}_b,
\label{elph1}
\end{equation}
where  $\hat{O}_{a/b}$ can couple different electronic states. The form of these operators is determined by time-reversal symmetry and can be expressed as 
\begin{align}
\hat{O}_a & =g_a\ket{\psi_1}\bra{\psi_3}-g_a^*\ket{\psi_2}\bra{\psi_4}+\mathrm{h.c.},\label{coupling1}\\
\hat{O}_b & =g_b\ket{\psi_1}\bra{\psi_3}-g_b^*\ket{\psi_2}\bra{\psi_4}+\mathrm{h.c.}
\label{coupling2}
\end{align}
where the value of $g_{a/b}$ depends on the strength of the orbit-lattice coupling. These couplings are illustrated in Fig.~\ref{mainfig}(a).

There are further coupling terms that mix the states $\psi_1 (\psi_2)$ with $\psi_4 (\psi_3)$, which in most cases turn out to be zero, however. Heuristically, this can be understood on the basis of angular momentum transfer. The chiral superposition of two components of a doubly degenerate mode possesses angular momentum $\pm l\hbar$ and thus only mixes electronic states for which the change in angular momentum is given by $|\Delta m_j|=l$. This restricts the number of terms in Eq.~(\ref{coupling2}), and it thus suffices to take into consideration only the transitions between $\ket{\psi_1}(\ket{\psi_2})$ and $\ket{\psi_3}(\ket{\psi_4})$. We will discuss the particular form of this mixing for specific examples in Secs.~\ref{sec:III} and \ref{sec:IV}.

The electron-phonon interaction, therefore, manifests as orbit-lattice coupling and hybridizes phonons and electronic excitations, which modifies the phonon frequencies. The modification of the phonon spectrum can be obtained using a Green's-function formalism. For the non-interacting case, described by $H_{ph}$, the bare phonon Green's function is given by
\begin{equation}
\mathbf{D}_0(\omega)=\begin{pmatrix}
D_0^{aa}(\omega)&0\\0&D_0^{bb}(\omega)
\end{pmatrix},
\end{equation}
where 
$D_0^{aa}(\omega)=D_0^{bb}(\omega)=\frac{2\omega_0}{\omega^2-\omega_0^2}$,
and the phonon energies are trivially retrieved by solving $\mathrm{Det}\left(\mathbf{D}_0^{-1}(\omega)\right)=0$. Including interactions, the full phonon Green's function is given by
\begin{equation}\label{eq:fullphonon}
    \mathbf{D}^{-1}(\omega) =\mathbf{D}_0^{-1}(\omega)-\mathbf{\Pi}(\omega),
\end{equation}
where the phonon self-energy matrix $\mathbf{\Pi}(\omega)$ contains corrections from the orbit-lattice coupling that are given by
\begin{align}
\Pi^{aa}&=4\pi |g_a|^2\left(\frac{ f_{13}\varepsilon_{13}}{\omega^2-\varepsilon_{31}^2}+\frac{ f_{24}\varepsilon_{24}}{\omega^2-\varepsilon_{42}^2}\right),\\
\Pi^{bb}&=4\pi |g_b|^2\left(\frac{ f_{13}\varepsilon_{13}}{\omega^2-\varepsilon_{31}^2}+\frac{ f_{24}\varepsilon_{24}}{\omega^2-\varepsilon_{42}^2}\right),\\
\Pi^{ab}&=(\Pi^{ba})^*=\Pi^{ab}_\mathrm{Re}+i \Pi^{ab}_\mathrm{Im}.
\end{align}
The real and imaginary parts of the mixed term, $\Pi^{ab}$, are given by
\begin{align}
\Pi^{ab}_\mathrm{Re}&=4\pi\mathrm{Re}(g_ag_b^*)\left(\frac{ f_{13}\varepsilon_{13}}{\omega^2-\varepsilon_{31}^2}+\frac{ f_{24}\varepsilon_{24}}{\omega^2-\varepsilon_{42}^2}\right),\\
\Pi^{ab}_\mathrm{Im}&=4\pi\mathrm{Im}(g_ag_b^*)\left(\frac{ f_{13}\omega}{\omega^2-\varepsilon_{31}^2}-\frac{ f_{24}\omega}{\omega^2-\varepsilon_{42}^2}\right),\\
\Pi^{ab}&=2\pi \left(\frac{g_ag_b^* f_{13}}{\omega-\varepsilon_{13}}-\frac{g_a^*g_b f_{13}}{\omega-\varepsilon_{31}}+\frac{g_a^*g_b f_{24}}{\omega-\varepsilon_{24}}-\frac{g_ag_b^* f_{24}}{\omega-\varepsilon_{42}}\right).
\end{align}
Here, $\varepsilon_{ij}=\varepsilon_i-\varepsilon_j$ is the energy difference between the states $i$ and $j$ and $f_{ij}=f_i-f_j$ is the difference in their occupancies, which are determined by thermal populations of these levels.

\begin{figure}[t]
    \centering
    \includegraphics[scale=0.3]{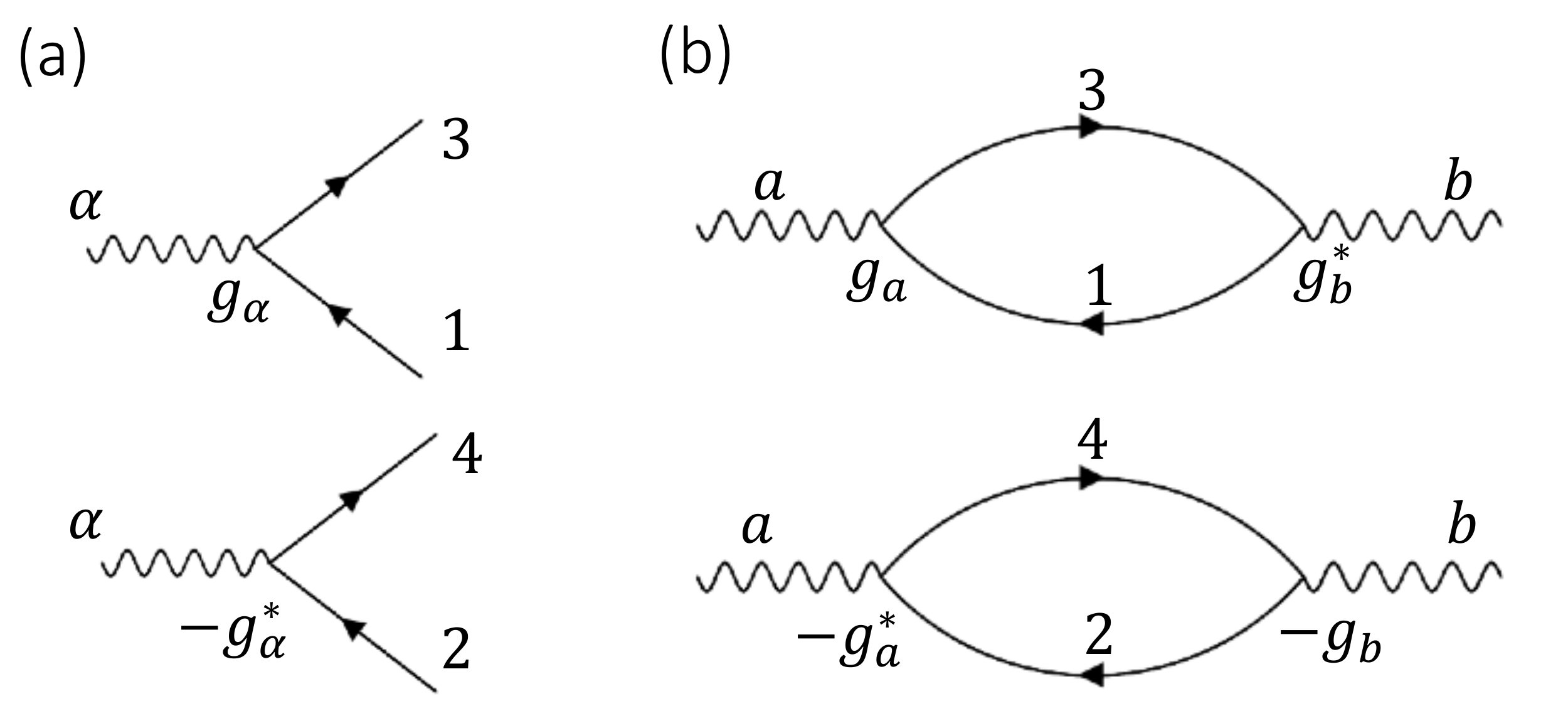}
    \caption{Scattering mechanism of electron-phonon interactions. (a) Electron-phonon interaction vertices for a doubly degenerate phonon mode $\alpha=(a,b)$ as defined by Eq.~\eqref{coupling1} and \eqref{coupling2}. (b) Feynman diagrams representing the off-diagonal contributions to the phonon self-energy matrix, mixing the two orthogonal components, $a$ and $b$, of the doubly degenerate phonon mode. }
    \label{Feynman}
\end{figure}

All of the terms in the self-energy matrix  introduce corrections to phonon energies. The phonon degeneracy can be lifted if the two diagonal terms are unequal. Eq.~(\ref{eq:fullphonon}) therefore becomes 
\begin{widetext}
\begin{equation}
    \mathbf{D}^{-1}(\omega)=\begin{pmatrix}
    \frac{\omega^2-\omega_0^2}{2\omega_0}-\tilde{g}^2\left(\frac{f_1\Delta_1}{\omega^2-\Delta_1^2}+\frac{f_2\Delta_2}{\omega^2-\Delta_2^2}\right)&i\tilde{g}^2\left(-\frac{f_1\omega}{\omega^2-\Delta_1^2}+\frac{f_2\omega}{\omega^2-\Delta_2^2}\right)\\
    -i\tilde{g}^2\left(-\frac{f_1\omega}{\omega^2-\Delta_1^2}+\frac{f_2\omega}{\omega^2-\Delta_2^2}\right)&\frac{\omega^2-\omega_0^2}{2\omega_0}-\tilde{g}^2\left(\frac{f_1\Delta_1}{\omega^2-\Delta_1^2}+\frac{f_2\Delta_2}{\omega^2-\Delta_2^2}\right)
    \end{pmatrix},
    \label{newg}
\end{equation}
\end{widetext}
where we redefine $\tilde{g}^2\equiv4\pi g^2$, $\Delta_1\equiv\varepsilon_{31}$, $\Delta_2\equiv\varepsilon_{42}$, and we assume the excited-state Kramers doublet to be unoccupied, $f_3=f_4=0$. Please see the Appendix for a detailed derivation of Eq.~(\ref{newg}). The off-diagonal elements arise from the orbital transitions shown in Fig.~\ref{mainfig}(a) and can be understood in terms of the Feynman diagrams shown in Fig.~\ref{Feynman}. 
With no external magnetic field applied, the occupancies of states 1 and 2 are equal, $f_1=f_2\equiv f_0/2$, ( $f_0$ is the occupancy of the ground-state manifold) as well as the energies of both transitions, $1\rightarrow 3$ and $2\rightarrow 4$, $\Delta_1=\Delta_2\equiv \Delta$.
Under these conditions, the two contributions shown in the two panels of Fig.~\ref{mainfig}(a) are equal in magnitude and opposite in sign, and thus the off-diagonal terms of the self-energy matrix vanish. As a result, phonons and electronic excitations hybridize to form doubly degenerate states with primarily electronic character and lower energies and states with primarily phononic character and higher energies, as illustrated in Fig.~\ref{mainfig}(b) for $B=0$. We solve $\mathrm{Det}(\mathbf{D}^{-1}(B=0))=0$ to obtain the frequencies of the hybridized states with primarily phononic and electronic character, $\Omega_{ph}$ and $\Omega_{el}$,
\begin{align}
    \Omega_{ph} & \equiv \omega_{ph}(B=0) \nonumber\\
    & =\left(\frac{\omega_0^2+\Delta^2}{2}+\sqrt{\left(\frac{\omega_0^2-\Delta^2}{2}\right)^2+2\tilde{g}^2f_0\omega_0\Delta}\right)^{1/2} \\
    \Omega_{el} & \equiv \omega_{el}(B=0) \nonumber\\
    & = \left(\frac{\omega_0^2+\Delta^2}{2}-\sqrt{\left(\frac{\omega_0^2-\Delta^2}{2}\right)^2+2\tilde{g}^2f_0\omega_0\Delta}\right)^{1/2}.
\end{align}
The energy levels are depicted schematically in Fig.~\ref{mainfig}.

We next apply an external magnetic field, $\mathbf{B}=B\,\hat{z}$, which lifts the degeneracies of the Kramers doublets, $\varepsilon_{12}\neq 0$ and $\varepsilon_{34}\neq 0$. This subsequently modifies the electronic transition energies as
\begin{equation}
    \Delta_1=\Delta-\gamma B,~~\Delta_2=\Delta+\gamma B,
    \label{eq:delta_split}
\end{equation}
where $\gamma=g^{el}_{ex}-g^{el}_{gs}$ contains the $g$-factors of the ground- and excited-state doublets. Lifting the degeneracy of the ground-state doublet leads to an asymmetric population of the ground-state energy levels, $f_{12}\neq 0$. This population difference is an odd function of the magnetic field $B$, and we will show in the following that it is directly proportional to the magnetization of the system. 
The energies of phonon and electronic excitation branch can be obtained by solving $\mathrm{Det}(\mathbf{D}^{-1}(B\neq 0))=0$, which yields 
 \begin{multline}
    (\omega^2-\Omega_{ph}^2)(\omega^2-\Omega_{el}^2) 
    \pm 2 \omega \left(\gamma B(\omega^2-\omega^2_0)+\tilde{g}^2\omega_0 f_{21}\right) \\
     +\gamma B\left(\gamma B (\omega^2-\omega^2_0)+2\tilde{g}^2\omega_0f_{21}\right)=0.
    \label{sc2}
 \end{multline}
For small magnetic fields, we can assume a solution of the form
 \begin{align}
     w_{ph}^{\pm}=\Omega_{ph}\left(1\mp \eta \right)
\end{align}
and by substituting it in Eq.~\eqref{sc2}, we get
\begin{equation}
    \Omega_{ph}\eta=\frac{\gamma B(\Omega_{ph}^2-\omega^2_0)+\tilde{g}^2\omega_0 f_{21}}{\Omega_{ph}^2-\Omega_{{el}}^2+\gamma^2B^2}
    \label{splt1}
\end{equation}
Please see the Appendix for a detailed derivation. Consequentially, we obtain an expression for the splitting of the phonon frequencies,
\begin{equation}
   \frac{\omega_{ph}^+-\omega_{ph}^-}{\Omega_{ph}}=2\frac{\gamma B(\Omega_{ph}^2-\omega^2_0)/\omega_0+\tilde{g}^2 f_{21}}{\sqrt{\left(\omega_0^2-\Delta^2\right)^2+8\tilde{g}^2f_0\omega_0\Delta}+\gamma^2 B^2}.
   \label{split1a}
\end{equation}

The complex hybridization of energy levels leading to this splitting is depicted in Fig.~\ref{mainfig}(b), and it arises from a combination of two factors: 1) a Zeeman shift of the electronic energy levels that is determined by the $g$-factor of the Kramers states in each manifold, and 2) a population imbalance between the ground-state energy levels that is directly related to a change in spin polarization (and subsequently magnetization) of the ion.

The net spin polarization of the ground state of the system depends on magnetic field $B$, temperature $T$, and also on the exchange interactions in the system. We will derive an explicit form for the population asymmetry in Secs.~\ref{sec:III} and \ref{sec:IV}, when considering the examples of paramagnets and magnets. For example, the population difference for the paramagnetic case is simply given by $f_{21}=\tanh(g^{el}_{gs}B/(k_B T))$. On the other hand, for ferromagnetic or antiferromagnetic cases, its form is more complicated and can be derived by adding the exchange mean field. In all cases, in the limit $B\rightarrow 0$, we can write the population difference as linear in the magnetic field,
\begin{equation}
    f_{21}\approx \chi B,
\end{equation}
where $\chi$ is directly related to the magnetic susceptibility of the system. 
As a result, for a small magnetic field, both the terms in Eq.~\eqref{splt1} lead to Zeeman splitting of previously doubly degenerate phonon mode. We notice that the splitting becomes more pronounced as the non-interacting phonon energy, $\omega_0$ comes closer to electronic excitation energy $\Delta$. Here, we can consider two different scenarios:
\begin{enumerate}
    \item Resonant case: $\Delta\approx \omega_0$ such that $|\Delta-\omega_0|\ll \tilde{g}$. In this case, the relative splitting\begin{equation}
       \frac{\omega_{ph}^+-\omega_{ph}^-}{\Omega_{ph}} \approx \frac{\gamma+\frac{\tilde{g}}{\sqrt{2}}\chi}{\omega_0} B
       \end{equation}
       which depends linearly on the orbit-lattice coupling strength~$\tilde{g}$.
       \item Off-resonant case:~$|\Delta-\omega_0|\gg \tilde{g}$. In this case, $\Omega_{ph}^2-\omega_0^2\approx \tilde{g}^2\omega_0/|\Delta-\omega_0|$, and thus
       the relative splitting is
       \begin{equation}
       \frac{\omega_{ph}^+-\omega_{ph}^-}{\Omega_{ph}(B=0)}\approx 
       \tilde{g}^2\frac{\frac{\gamma}{|\Delta-\omega_0|}+\chi}{(\omega_0-\Delta)(\omega_0+\Delta_0)}  B
       \label{off_resonant}
    \end{equation} 
    \end{enumerate}
    where the splitting depends quadratically on orbit-lattice strength $\tilde{g}$.
Due to the general mismatch of CEF transition frequencies and phonon frequencies, the off-resonant scenario can be more commonly found in materials. In the off-resonant case, the splitting diminishes as the energy difference between phonon and electronic excitations increases, which requires $\omega_0$ and $\Delta$ to be at least of similar order of magnitude to yield a significant effect.

So far, we discussed how the phonon energies are shifted but we have not investigated the consequences of the orbit-lattice coupling on the displacements associated with the eigenmodes. Assuming the off-resonant scenario, the phonon-displacement operators corresponding to the split phonon modes with frequencies $\omega_{ph}^+$ and $\omega_{ph}^-$ of the interacting Green's function matrix are given by
\begin{align}
Q_+ & =\frac{1}{\sqrt{2}}(Q_a-iQ_b),\\
Q_- & =\frac{1}{\sqrt{2}}(Q_a+iQ_b),
\end{align}
where $Q_a\sim(a+a^\dagger)$ and  $Q_b\sim(b+b^\dagger)$ were the displacement operators corresponding to linearly polarized phonon modes $a$ and $b$. This shows that the new modes, $Q_+$ and $Q_-$, correspond to circular superpositions of the two orthogonal components and have opposite chiralities. We would like to emphasize at this point that these zone-centered phonon modes here become chiral due to time-reversal breaking. On the other hand, inversion symmetry breaking can allow chiral phonons at other high-symmetry points in the Brilliouin zone, as studied in many two-dimensional hexagonal lattices~\cite{zhang:2015,Zhu2018}.

These chiral phonons exhibit Zeeman splitting as their energies change with an applied magnetic field. In the limit $B\rightarrow0$, this splitting becomes linear in the magnetic field, which allows us to attribute an effective magnetic moment to the chiral phonons. We denote the effective magnetic moment of chiral phonons by $\mu_{ph}$, and the splitting is accordingly given by
\begin{equation}
    \omega_{ph}^\pm=\omega_0\pm \mu_{ph}B.
\end{equation}
whereas the phonon magnetic moment can be expressed as
\begin{equation}\label{phononmagmom}
    \mu_{ph}=\frac{1}{2}\frac{\partial(\omega_{ph}^+-\omega_{ph}^-)}{\partial B}\bigg|_{B\rightarrow 0},
\end{equation}
which can be evaluated by using Eq.~\eqref{split1a}.
The magnitude of phonon Zeeman splitting obtained here relies on the coupling between orbital excitations and phonons. In order to have strong coupling between electronic and phonon degrees of freedom, the energies of these excitations should be of the same order of magnitude as the phonon energies. In the following sections, we will apply this model to rare-earth trihalide paramagnets and transition-metal oxide magnets to predict the Zeeman splitting and effective magnetic moments of chiral phonons in these materials.


\section{Chiral phonons in 4$f$ paramagnets\label{sec:III}}
The splitting of optical phonons in paramagnetic rare-earth compounds was extensively studied in the 1970s in a series of papers~\cite{schaack:1975,schaack:1976,schaack:1977,Thalmeier1977}. Amongst other compounds, it was shown that the rare-earth trihalide CeCl$_3$ exhibits a large splitting of doubly degenerate phonon modes in an external magnetic field. Recently, using the early experimental data on orbit-lattice coupling, it was predicted in Ref.~\cite{Juraschek2022_giantphonomag}, that chiral phonons in this material can produce effective magnetic fields on the order of tens of tesla when coherently excited with ultrashort laser pulses. Subsequently, CeCl$_3$ has emerged as an interesting candidate to study magneto-phononic and phono-magnetic properties of chiral phonons. We will determine the microscopic origin of the orbit-lattice coupling and apply the model derived in the previous section to predict the Zeeman splittings and effective magnetic moments in this material. We stress that this is the first quantitative prediction using only microscopic parameters and {\it ab-initio} results, without the need for phenomenological theory or experimental data.

\subsection{Structural and electronic properties of CeCl$_3$}
\begin{figure*}
    \centering
    \includegraphics[scale=0.27]{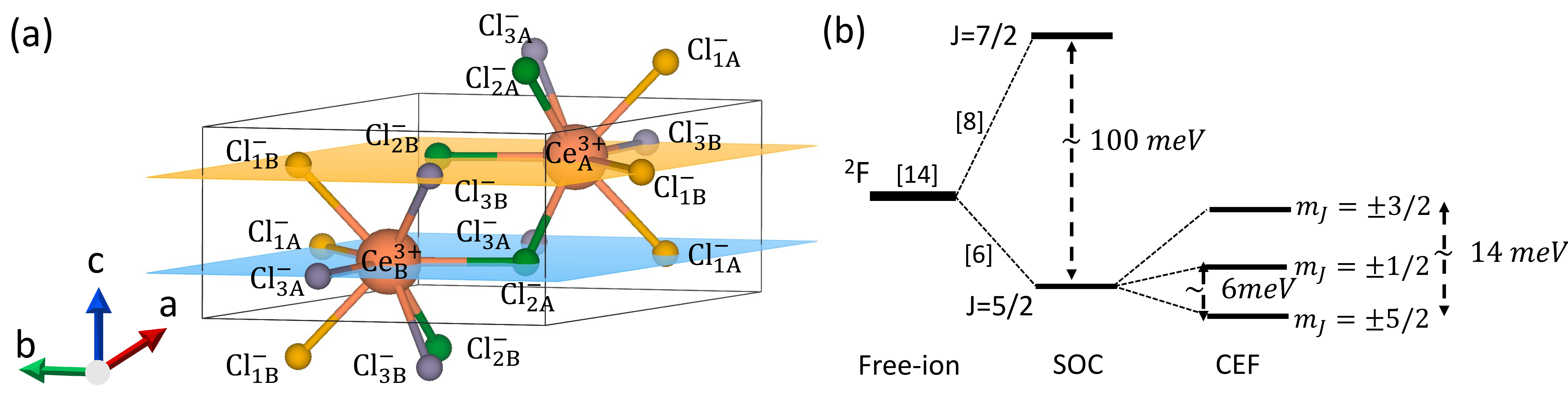}
    \caption{\textbf{Properties of CeCl$_3$}: (a) Hexagonal crystal structure of CeCl$_3$ with point group 6/$m$ and precise ionic labels. (b) Electronic energy levels of the Ce$^{3+}$ ion. The energy levels compared to the free ion are split by spin-orbit coupling and by the crystal electric field, resulting in three Kramers doublets, of which $\pm$5/2 is the ground state.}
    \label{Crystal}
\end{figure*}

The rare-earth trihalide CeCl$_3$ (Fig.~\ref{Crystal}(a)) belongs to space group no. 176 (point group $6/m$) and its primitive unit cell contains eight atoms, two Ce$^{3+}$ ions located at the 2$c$ Wyckoff positions (shown as Ce$^{3+}_A$ and Ce$^{3+}_B$) and six Cl$^{-}$ ions  at the 6$h$ Wyckoff positions (shown as Cl$^{-}_{1A}$,Cl$^{-}_{2A}$,Cl$^{-}_{3A}$,Cl$^{-}_{1B}$,Cl$^{-}_{2B}$,Cl$^{-}_{3B}$). The eight-atom unit cell leads to 21 optical phonons consisting of irreducible representations $2A_g+ 1A_u+2B_g+2B_u+1E_{1g}+3E_{2g}+2E_{1u}+1E_{2u}$~\cite{Juraschek2022_giantphonomag}.
Each Ce$^{3+}$ ion has nine nearest neighbors arranged in three different planes as shown in Fig.~\ref{Crystal}(a) for the Ce$^{3+}$ ion $A$. 

The ground-state configuration of the Ce$^{3+}$ ($4f^1$) ion is given by a nearly free-ion configuration of a $L=3,S=1/2$ state in accordance with Hund's rule. The spin-orbit coupling splits this 14 dimensional space into $J=5/2$ and $J=7/2$ total angular momentum sectors and the ground-state is given by the six-dimensional $J=5/2$ $(^2F_{5/2})$ state. Since there is only one electron in the $4f$ orbitals, the wavefunctions of different states in this multiplet can be written as
    \begin{align}
    \begin{split}
    \ket{J=5/2,m_j=\pm5/2}=&-\sqrt{\frac{1}{7}}\ket{m_l=\pm2,m_s=\pm1/2}\\&+\sqrt{\frac{6}{7}}\ket{m_l=\pm3,m_s=\mp1/2},\end{split}\label{state52}\\
    \begin{split}
    \ket{J=5/2,m_j=\pm3/2}=&-\sqrt{\frac{2}{7}}\ket{m_l=\pm1,m_s=\pm1/2}\\&+\sqrt{\frac{5}{7}}\ket{m_l=\pm2,m_s=\mp1/2},\end{split}\label{state32}\\
    \begin{split}\ket{J=5/2,m_j=\pm1/2}=&-\sqrt{\frac{3}{7}}\ket{m_l=\pm0,m_s=\pm1/2}\\&+\sqrt{\frac{4}{7}}\ket{m_l=\pm1,m_s=\mp1/2},\end{split}\label{state12}
    \end{align}
where $\ket{m_l,m_s}$ is a $4f$ orbital state with orbital quantum number $m_l$ and spin quantum number $m_s$. The CEF further splits it into three Kramers doublets $\ket{\pm5/2}$, $\ket{\pm1/2}$, and $\ket{\pm3/2}$ with energies 0~meV, 5.82~meV, 14.38~meV, respectively~\cite{schaack:1977}, as shown in Fig.~\ref{Crystal}(b).

\subsection{Microscopic model for the orbit-lattice coupling}

Previous Raman studies have shown that the doubly degenerate modes $E_{1g}$ and $E_{2g}$ split into left- and right-handed circularly polarized chiral phonon modes when a magnetic field is applied along the $c$-axis of the crystal, perpendicular to the plane of the components of the doubly degenerate phonon modes~\cite{Thalmeier1977,schaack:1977}. In CeCl$_3$, the $E_{1g}$ mode shows the largest splitting in experiment, and we will therefore first focus our analysis on this mode, which involves the displacement of Cl$^{-}$ ion along the $c$-axis. As there is only one $E_{1g}$ phonon, the displacement pattern for this mode can be obtained directly from group theory and is given by
\begin{align}
     E_{1g}^1(a)&=\frac{Q_a}{2\sqrt{6}}\left(0,0, 2 \hat{z},-\hat{z},-\hat{z},-2 \hat{z},\hat{z},\hat{z}\right),\label{E1gmodead}\\
     E_{1g}^1(b)&=\frac{Q_b}{2\sqrt{2}}\left(0,0,0,\hat{z},-\hat{z},0,-\hat{z},\hat{z}\right),\label{E1gmodebd}
\end{align}
in the basis (Ce$_A^{3+}$,Ce$_B^{3+}$,Cl$_{1A}^{-}$,Cl$_{2A}^{-}$,Cl$_{3A}^{-}$,Cl$_{1B}^{-}$,Cl$_{2B}^{-}$,Cl$_{3B}^{-}$) and $Q_{a/b}$ are the normal mode coordinates (amplitudes) of the two components $a$ and $b$ in units of \AA$\sqrt{\mathrm{amu}}$, where amu is the atomic mass unit. We show the atomic displacements in Fig.~\ref{CeCl3phonon}(a). The displacements of the Cl$^-$ ions modify the Coulomb potential around the Ce$^{3+}$ ions which perturb the electronic Hamiltonian on the magnetic ion.

We use a point-charge model to describe the crystal electric field of the system, in which the potential energy of an electron at position $\mathbf{r}$ from Ce$^{3+}$ nucleus due to the $n^{th}$ Cl$^-$ ion is given by:
\begin{equation}
    V(\mathbf{r},\mathbf{R}_n)=\frac{e^2}{4\pi\epsilon_0}\frac{1}{|\mathbf{R}_n-\mathbf{r}|}
    \label{Coulomb}
\end{equation}
where $\mathbf{R}_n=\mathbf{R}_{0,n}+\mathbf{u}_{n}$ is the displacement of the $n^{th}$ ligand ion from Ce$^{3+}$ nucleus which depends on the equilibrium displacement $\mathbf{R}_{0,n}$ and the relative lattice displacement $\mathbf{u}_{n}$ arising from the phonon. Now, the perturbation introduced by a given phonon mode can be obtained by a Taylor expansion in lattice displacement $\mathbf{u}_{n}$ and after expressing these dispalcements in terms of normal coordinates $Q_{a,b}$ and summing over all nearest neighbor ligands, the first order term is given by: 
\begin{align}
V(E_{1g}(a))&=\left[-0.06 xz+0.16 yz\right]Q_a~\frac{\mathrm{eV}}{\text{\AA}^3\sqrt{\mathrm{amu}}},\label{V1e1g}\\
    V(E_{1g}(b))&=\left[0.16 xz+0.06 yz\right]Q_b~\frac{\mathrm{eV}}{\text{\AA}^3\sqrt{\mathrm{amu}}}\label{V1e1gb}.
\end{align}
Now, using spherical coordinates to express $xz (yz)=r^2\cos\theta\sin\theta \cos\phi (\sin\phi)$ and writing the states in Eq.~\eqref{state52}-\eqref{state12} in terms of $4f$ basis states with wavefunction $\langle r|m_l,m_s\rangle=R(r)Y_3^{m_l}(\theta,\phi)$ where $R(r)$ is a radial part and $Y_3^{m_l}(\theta,\phi)$
 is the spherical harmonic (see Appendix for details), we evaluate the matrix elements of above perturbations which are given by:
\begin{figure*}
    \centering
  \includegraphics[scale=0.37]{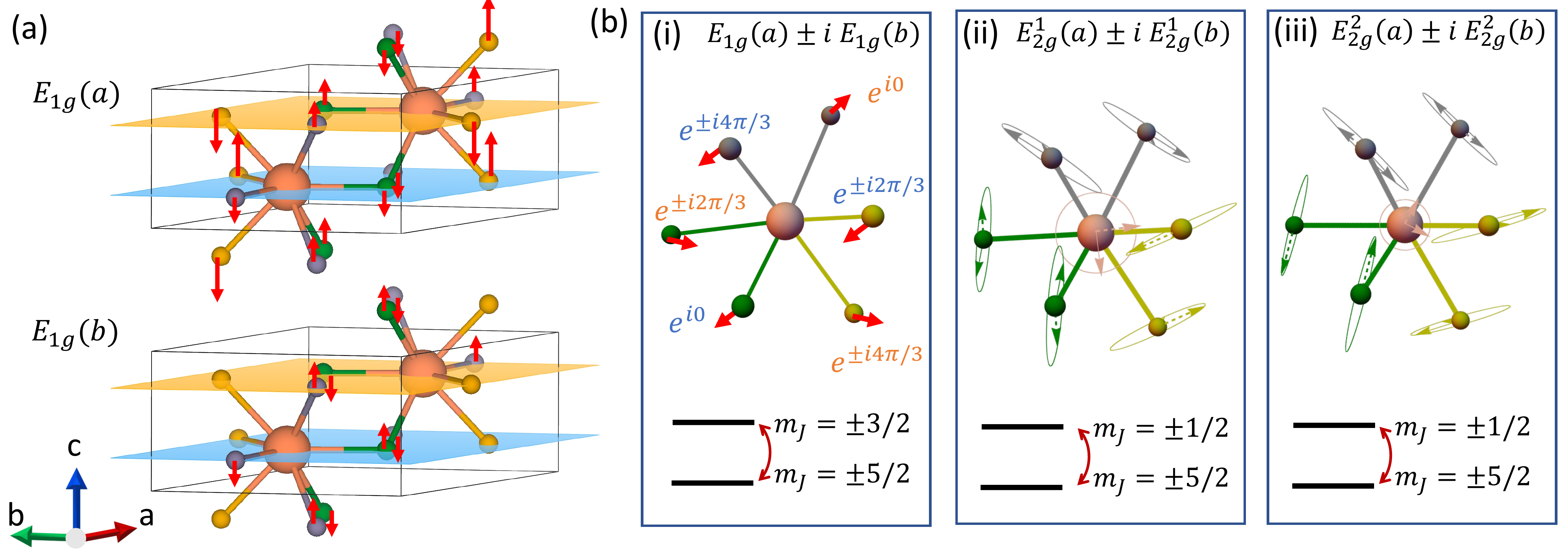}
    \caption{\textbf{Chiral phonons in CeCl$_3$}: (a) Atomic displacements for the two components, $a$ and $b$, of the doubly degenerate $E_{1g}$ mode. Notably, both components involve only motion of the ions along the $c$-direction of the crystal. (b) Displacement patterns for chiral phonons in the $ab$-plane of the crystal arise from superpositions of the two components of the doubly degenerate $E_{1g}$ and $E_{2g}$ modes with a $\pi/2$ phase difference. We show the orbital transitions between the Kramers doublets that couple to the respective chiral phonon modes. 
    (i) For $E_{1g}$, the Cl$^{-}$ ions in the upper and lower plane around the Ce$^{3+}$ ion move in opposite directions along the $c$-direction of the crystal with a relative phase of $\pm 2\pi/3$ between atoms lying in the same plane. (ii,iii) For the $E_{2g}^{1}$ and $E_{2g}^{2}$ modes, all ions move on circular orbits around their equilibrium position in the $ab$ plane of the crystal with different phase differences between different atoms.}
    \label{CeCl3phonon}
\end{figure*}
\begin{align}
    H_1(xz)&=-\frac{2}{7\sqrt{5}}\langle r^2\rangle\begin{pmatrix}
    &\left|\frac{5}{2},\pm\frac{5}{2}\right>&\left|\frac{5}{2},\pm\frac{3}{2}\right>\\
    \left|\frac{5}{2}\pm\frac{5}{2}\right>&0&\pm 1\\[1.2em]\left|\frac{5}{2},\pm\frac{3}{2}\right>&\pm 1&0
    \end{pmatrix},
    \label{xzpr}\\
     H_1(yz)&=\frac{2}{7\sqrt{5}}\langle r^2\rangle\begin{pmatrix}
    &\left|\frac{5}{2},\pm\frac{5}{2}\right>&\left|\frac{5}{2},\pm\frac{3}{2}\right>\\
    \left|\frac{5}{2},\pm\frac{5}{2}\right>&0&i\\[1.2em]\left|\frac{5}{2},\pm\frac{3}{2}\right>&-i&0
    \end{pmatrix},
    \label{yzpr}
\end{align}
where $\langle r^2\rangle=\int_0^\infty r^2|R(r)|^2r^2dr$ is the mean of the square of $4f$ electron radius. Now, we express the phonon displacement as 
\begin{align}
    Q_a & =\frac{\hbar}{\sqrt{\hbar\omega_{0}}}\left(a+a^\dagger\right)=\frac{0.06\text{\AA}\sqrt{\mathrm{eV.amu}}}{\sqrt{\hbar\omega_{{0}}}}\left(a+a^\dagger\right), \label{Qab}\\
    Q_b & =\frac{\hbar}{\sqrt{\hbar\omega_{0}}}\left(b+b^\dagger\right)=\frac{0.06\text{\AA}\sqrt{\mathrm{eV. amu}}}{\sqrt{\hbar\omega_{{0}}}}\left(b+b^\dagger\right), \label{Qba}
\end{align}
where we restored $\hbar$ and $\hbar\omega_{{0}}$ is the energy of phonon. 

The orbit-lattice coupling operators connecting different electronic states from Eq.~\eqref{coupling1} and Eq.~\eqref{coupling2} become
\begin{align}
\hat{O}_a & =ge^{i\theta}\bigg|+\frac{5}{2}\bigg\rangle\bigg\langle+\frac{3}{2}\bigg|-ge^{-i\theta}\bigg|-\frac{5}{2}\bigg\rangle\bigg\langle-\frac{3}{2}\bigg|+\mathrm{h.c.},\label{couplingE1ga}\\
\hat{O}_b & =ige^{i\theta}\bigg|+\frac{5}{2}\bigg\rangle\bigg\langle+\frac{3}{2}\bigg|+ige^{-i\theta}\bigg|-\frac{5}{2}\bigg\rangle\bigg\langle-\frac{3}{2}\bigg|+\mathrm{h.c.}.
\label{couplingE1gb}
\end{align}
Here, we combined Eqs.~(\ref{V1e1g})-(\ref{Qba}) in order to obtain $g = -\sqrt{0.16^2+0.06^2}\frac{2}{7\sqrt{5}}\langle r^2\rangle \frac{0.06}{\sqrt{\omega_0}} {\mathrm{eV}}^{3/2}/\text{\AA}^2$ 
and $\tan(\theta)=0.16/0.06$. Compared to the general expression of the orbit-lattice coupling in Eq.~\eqref{coupling1} and Eq.~\eqref{coupling2}, we find that $g_a=ig_b=ge^{i\theta}$. Following our previously derived model, this orbit-lattice coupling leads to a splitting of the $E_{1g}$ mode into two circularly polarized phonon modes with opposite chirality. The split modes have orbital angular momenta of $\pm\hbar$, arising from the superpositions $E_{1g}(a)\pm i E_{1g}(b)$ obtained from Eq.~(\ref{E1gmodead}) and Eq.~(\ref{E1gmodebd}). This can also be seen by applying a $C_3(z)$ rotation operation around each Ce$^{3+}$ site, and the mode is an eigenstate of the $C_3(z)$ operator with eigenvalue $e^{i2\pi/3}$. The displacements associated with the two chiral phonon modes are depicted in Fig.~\ref{CeCl3phonon}(b) alongside the orbital transitions with which they hybridize. The orbital angular momentum for these modes arises from the relative phase between neighboring atoms. 

\subsection{Phonon Zeeman splitting and effective phonon magnetic moment}
We now proceed to compute the energy splitting and effective phonon magnetic moments that can be associated with the chiral phonon modes. For the $E_{1g}$ phonon mode, $\omega_0=22.75$~meV and $g= 7~\text{meV/\AA}^2 \left<r^2\right>$. With a mean-square radius of $\left<r^2\right>\sim 0.1$ \AA$^2$, we get $g\sim 0.7$~meV. Now, the splitting of the phonon modes is given by Eq.~\eqref{split1a}, where the relative contributions of the two additive terms depend on the values of electron-phonon coupling, $g$, the energy difference between the electronic excitation and the phonon mode, $\omega_0-\Delta$, and the occupancy difference, $f_{21}$. CeCl$_3$ is a paramagnetic system and the population difference for two states in the ground-state Kramer doublet is therefore given by, 
\begin{equation}
    f_{21}=\tanh\left(\frac{5}{2}\frac{g^{el}_{5/2}B}{k_BT}\right),
\end{equation}
where 
$g^{el}_{5/2}\approx 4/5\mu_B$ is the electronic $g$-factor for~$J=5/2$ states. In the regime, $\mu_B B\ll k_BT$, we can approximate the population difference to linear order, $f_{21}\approx 2\mu_B B/k_BT$. 
In order to calculate the splitting, we assume an off-resonant condition as the electronic excitation energy, $\Delta\approx 16$~meV results in $|\omega_0-\Delta|\gg \tilde{g}$ which allows us to approximate Eq.~\eqref{split1a} by Eq.~(\ref{off_resonant}) which gives,
\begin{equation}
\gamma B (\Omega_{ph}^2-\omega^2_0)+\tilde{g}^2 f_{21}\omega_0\approx \tilde{g}^2 \omega_0 \left(\frac{\gamma B}{\omega_0-\Delta_0}+\frac{2\mu_B B}{k_BT}\right),
\label{contr}
\end{equation}
where $\gamma=(5/2-3/2)g^{el}_{5/2}=2/5~\mu_B$.
As a result, the relative contribution of the two terms in Eq.~\eqref{contr} depends on the temperature. In the present case for the $E_{1g}$ mode, the second term dominates in the low-temperature regime, but the two contributions become equal when $k_B T\approx 5(\omega_0-\Delta_0)=50$~meV, i.e. at 500~K. 
We therefore only consider the contribution from the second term which depends on the difference in occupation of two states of the Kramers doublet $\ket{\pm5/2}$.

\begin{figure*}
    \centering
    \includegraphics[scale=0.28]{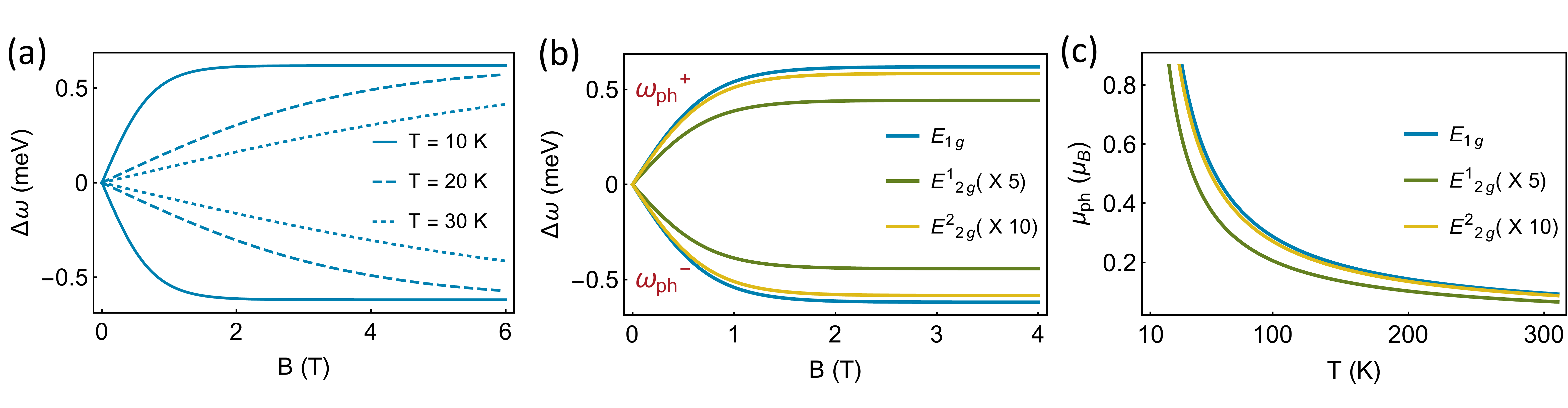}
    \caption{\textbf{Phonon Zeeman effect in CeCl$_3$}: (a) Splitting of the doubly degenerate $E_{1g}$ mode with frequency $\omega_0$ into left- and right-handed circularly polarized phonon modes with frequencies $\omega_{ph}^\pm$, as a function of the applied magnetic field, $B$. For strong magnetic fields, the magnetization and therefore the splitting saturates. The lower the temperature, the lower magnetic field strengths are required to reach saturation. (b) Phonon Zeeman splitting for three different $E_g$ phonon modes of CeCl$_3$ at 2 K. (c) Phonon magnetic moment as a function of temperature for three different $E_g$ phonon modes of CeCl$_3$.}
    \label{CeCl3splitting}
\end{figure*}

The resulting splitting is shown in Fig.~\ref{CeCl3splitting}(a). We obtain a value of 1.24 meV (10 cm$^{-1}$) for the saturated phonon splitting, which is reasonably close to the value of 18~cm$^{-1}$ observed in Ref.~\cite{schaack:1977}. The relative splitting reaches more than 5\%{} at saturation, and the magnitude of the magnetic fields required to reach saturation increases with increasing temperature according to the dependence of $\tanh(\chi B)$ on $f_{21}$. The value of the effective phonon magnetic moment $\mu_{ph}$ is inversely proportional to the temperature as shown in Fig.~\ref{CeCl3splitting}(c), and our model predicts $\mu_{ph}=2.9\,\mu_B$ at $T=10$~K, several orders of magnitude higher than those produced by purely ionic circular charge currents \cite{juraschek2:2017,Juraschek2019,Geilhufe2021,Zabalo2022}. The value of the saturation splitting and $\mu_{ph}$ at three different temperatures are presented in Table~\ref{tab:cecl3}, which range between 0.1~$\mu_B$ at room temperature and up to 9.3~$\mu_B$ at 2~K.

A similar analysis can be done for the $E_{2g}^1$ (12.1~meV) and $E_{2g}^2$ (21.5~meV) phonon modes, whose displacements are depicted in Fig.~\ref{CeCl3phonon}(c). Here, because of the existence of two phonon modes with the same symmetry, the displacements cannot be unambiguously determined from group theory, and we compute the phonon eigenvectors using density functional theory calculations published in prior work \cite{Juraschek2022_giantphonomag}. Using the point-charge model, we calculate the orbit-lattice coupling for these phonons, please see the Appendix for details. These phonons couple with orbital excitations between $\ket{m_j=\pm5/2}$ and $\ket{m_j=\pm1/2}$, as illustrated in Fig.~\ref{CeCl3phonon} (c) and it leads to phonon Zeeman splitting, as shown in Fig.~\ref{CeCl3splitting}(b). 

Our model predicts effective phonon magnetic moments of $\mu_{ph}=0.4~\mu_B$ and $\mu_{ph}=0.27~\mu_B$ at $T=10$~K and a saturation splitting of 0.18~meV (1.5 cm$^{-1}$) and 0.12~meV (1 cm$^{-1}$) for $E_{2g}^1$ and $E_{2g}^2$, respectively as shown in Table~\ref{tab:cecl3}. According to Ref.~\cite{schaack:1977}, the observed saturation splitting of the $E_{2g}^1$ mode is 0.87~meV (7 cm$^{-1}$) which is about four times the values obtained from our microscopic model. On the other hand, no splitting was observed for $E_{2g}^2$ mode in the same experiment. Here again, the observed phonon magnetic moment decreases with temperature as shown in Fig.~\ref{CeCl3splitting}(c). This disagreement could be either due to the crudeness of our point-charge model or due to the resolution of this experiment which is limited to $1$ cm$^{-1}$.

\begin{table*}[t]
\centering
\bgroup
\def\arraystretch{1.3}
\caption{
Calculated saturation splitting, phonon Zeeman splitting in an applied magnetic field of $B=1$~T, as well as effective phonon magnetic moments at liquid helium, liquid nitrogen, and room temperature, for the different doubly degenerate phonon modes in CeCl$_3$.
}
\begin{tabular}{l c c c c c}
\hline\hline
Phonon mode\;\;\;\;\; & Saturation splitting\;\;\;\;\; & Splitting at 1~T and 10 K\;\;\;\;\;& $\mu_{ph}(2~\mathrm{K})$\;\;\;\;\; & $\mu_{ph}(77~\mathrm{K})$\;\;\;\;\; & $\mu_{ph}(295~\mathrm{K}) $ \\
\hline
$E_{1g}$(22.75~meV) &  1.24~meV & 0.32~meV & 9.3~$\mu_B$ & 0.37~$\mu_B$ & 0.1~$\mu_B$\\
$E^{1}_{2g}$(12.1~meV) & 0.18~meV & 0.05~meV & 1.3~$\mu_B$ & 0.05~$\mu_B$ & 0.01~$\mu_B$\\
$E^{2}_{2g}$(21.5~meV) & 0.12~meV & 0.02~meV & 0.88~$\mu_B$ & 0.03~$\mu_B$ & 0.01~$\mu_B$\\
\hline\hline
\end{tabular}
\label{tab:cecl3}
\egroup
\end{table*}

\section{Chiral phonons in $3d$ magnets \label{sec:IV}}

In the previous section, we discussed the example of rare-earth trihalides, where the giant magnetic response of chiral phonons originates from the coupling of CEF-split electronic levels with chiral optical phonons. In this section, we show that chiral optical phonons in $3d$-electron magnets with octahedral ligand configuration can yield a similarly strong response. We begin with a general analysis of orbital configurations and then perform calculations for the concrete example of CoTiO$_3$.

\begin{figure}
    \centering
    \includegraphics[scale=0.33]{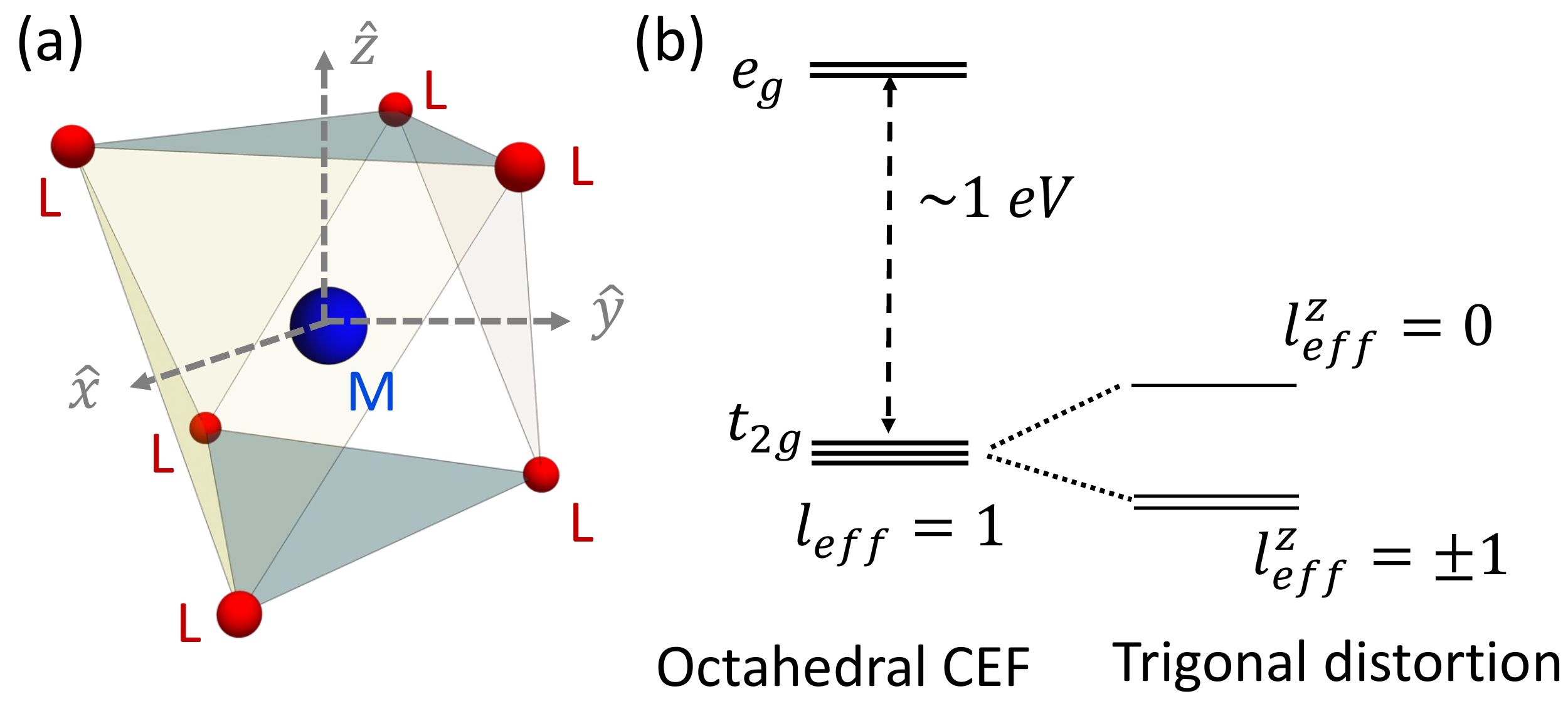}
    \includegraphics[scale=0.33]{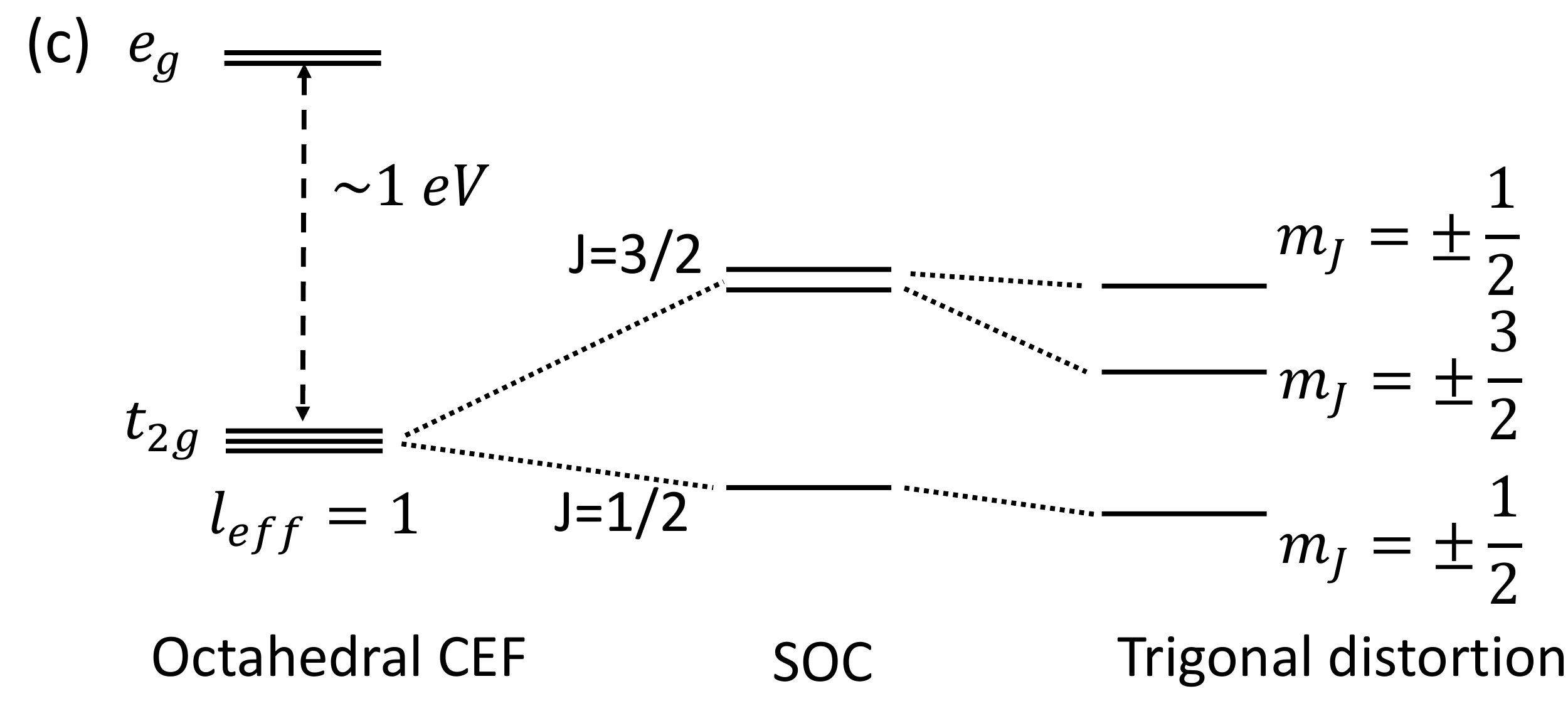}
    \caption{(a) A magnetic transition-metal ion $M$ surrounded by a trigonally distorted octahedral cage of ligand ions $L$. The site symmetry of the magnetic ion is $C_3$. (b) The splitting of $d$-orbitals due to a trigonal crystal field from the ligand arrangement on the left. (c) The splitting of $d$-orbitals when spin-orbit coupling
is stronger than the trigonal distortion.
    }
    \label{trigonal}
\end{figure}

In materials with octahedral ligand configurations around the magnetic ion, the CEF is usually strong with a splitting of $e_g$ and $t_{2g}$ orbitals of the order of a few eV for most materials. This renders the coupling between the $e_g-t_{2g}$ electronic excitations and phonons weak and thus makes a magnetic response arising directly from these transitions unfeasible. In many materials, however, the $t_{2g}$ and $e_g$ manifolds are split further due to either lattice distortions or spin-orbit coupling ~\cite{khomskii2016role,goodenough1968}. Both cases host electronic transitions with energies comparable to those of the optical phonon modes in the system that allows them to couple strongly and hybridize. Consider a magnetic transition-metal ion surrounded by a trigonally-distorted octahedron of ligand ions, as depicted in Fig.~\ref{trigonal}(a), which are common in face-sharing octahedral geometries.
In this scenario, the site symmetry for a magnetic ion is reduced to 
$C_3$ from $O_h$ and the $t_{2g}$ manifold splits according to $l_z$, where the $z$-axis is oriented along the $C_3$ rotation axis, as shown in Fig.~\ref{trigonal}(b).
The energy of the states depends on the sign of the trigonal distortion, and the $t_{2g}$ orbitals split into the following two manifolds 
\begin{align}
    \ket{l_z=\pm 1}&=-\frac{1}{\sqrt{3}}(d_{xy}\mp id_{x^2-y^2})\pm\frac{i}{\sqrt{6}}(d_{xz}\mp id_{yz}),\\
    \ket{l_z=0}&=d_{3z^2-r^2}.
\end{align}

On the other hand, if spin-orbit coupling is stronger than the splitting induced by the trigonal distortion, we need to consider eigenstates characterized by the total angular momentum, $J$, as shown in Fig.~\ref{trigonal}(c). In this limit, the trigonal distortion introduces a perturbation of the form $H_{tri}=\delta J_z^2$ to the Hamiltonian of magnetic ion which splits $J=3/2$ multiple into two manifolds with $m_j\pm 1/2$ and $m_j=\pm 3/2$, but does not affect the $J=1/2$ states.
In both cases, there are low-lying electronic transitions that involve a transfer of angular momentum of $\Delta m=\pm 1$, namely the transition from $\ket{l_z=\pm1}$ to $\ket{l_z=0}$ in the case of a trigonal distortion, and $\ket{J=1/2,m_j=\pm1/2}$ to $\ket{J=3/2,m_j=\pm3/2}$ in the case of spin-orbit coupling. These transitions are similar in nature to transitions from $\ket{m_j=\pm 5/2}$ to $\ket{m_j=\pm 3/2}$ in the case of the rare-earth trihalides in the previous section.

Next, let us consider an optical phonon mode that can be characterized by the $E_g$ irreducible representation of the $C_3$ point group. There are many basis functions (corresponding to displacement patterns) that transform according to this irreducible representation for the system shown in Fig.~\ref{trigonal}. One such possibility is the $xy$ in-plane motion of a magnetic ion $M$ located at the center of the trigonally distorted octahedra. The two components of the $E_g$  mode in this case are simply represented by the motion of the ion, $M$, in the $x$ and $y$ directions, respectively, and result in the following form of perturbation to the CEF,
\begin{align}
V(E_{g}(a))&\propto xz \,Q_a~ \frac{\mathrm{eV}}{\text{\AA}^3\sqrt{\mathrm{amu}}}, \label{dpert1}\\
    V(E_{g}(b))&\propto yz\, Q_b \frac{\mathrm{eV}}{\text{\AA}^3\sqrt{\mathrm{amu}}}, \label{dpert2}
\end{align}
where $Q_{a/b}$ 
 are the normal mode coordinates associated with $E_g(a/b)$, similar to Eqs.~\eqref{V1e1g} and \eqref{V1e1gb} in the case of the rare-earth trihalides discussed in the previous section. This perturbation ultimately results in a coupling similar to the one discussed in Eqs.~\eqref{coupling1} and \eqref{coupling2} of Sec.~\ref{sec:II} and can lead to phonon chirality and a phonon Zeeman effect.
 
 The magnitude of this effect depends on the phonon energies, electronic wavefunctions of states involved in low-energy excitations and their energies which would be material specific. However, as discussed in the previous section, the phonon magnetic moment can be significantly larger if the electronic excitation energy is closer to phonon energy. The typical energy scale associated with SOC and trigonal distortion is usually less in the range of 10-100~meV for $d$ electron systems which puts these electronic excitations in close proximity with optical phonons and hence makes the above effect feasible.

Additionally, most transition-metal systems have significant superexchange interactions with neighboring spins originating in the large spatial extent of $d$-orbitals. As a result, one should expect a rather different temperature trend for the phonon magnetic moment $\mu_{ph}$  below magnetic ordering temperatures, which can be evaluated by including the exchange mean-field contributions.

We next perform calculations for the concrete example of the $XY$-quantum magnet CoTiO$_3$ which is known to have spin-orbit excitations with energies comparable to a range of optical phonons in the system~\cite{Yuan2020,yuan2020dirac,elliot2021order,dubrovin2021lattice}. 
 
\begin{figure*}
    \centering
\includegraphics[scale=0.4]{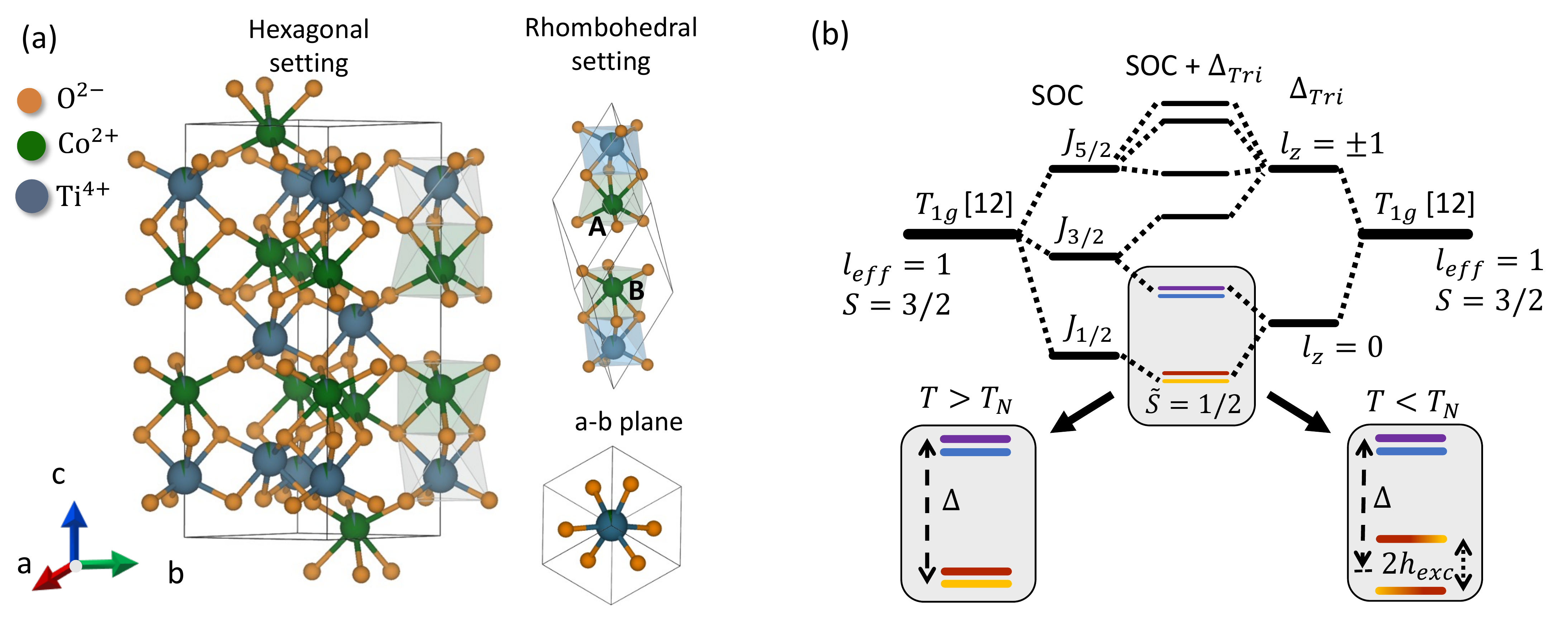}
    \caption{\textbf{Properties of CoTiO$_3$}: (a) Crystal structure of transition-metal oxide CoTiO$_3$ in the ilmenite structure with trigonal space group $R\bar{3}$. Each of the Co$^{2+}$ ions is surrounded by a trigonally distorted octahedral cage of O$^{2-}$ ions. (b) Electronic energy levels of the Co$^{2+}$ ion which are split due to a combination of the octahedral crystal field caused by the O$^{2-}$ ions, spin-orbit coupling, and trigonal distortion of the octahderal cage. Below the N\'eel temperature, the ground state manifold is further split by exchange means-fields.}
    \label{CrystalCTO}
\end{figure*}

\subsection{ Structural and electronic properties of CoTiO$_3$}

The transition-metal oxide CoTiO$_3$ crystallizes in an ilmenite structure with a trigonal space group R$\overline{3}$ (point group 3). Each of the Co$^{2+}$ ions is surrounded by a trigonally distorted octahedral cage of O$^{2-}$ ions, as shown in Fig.~\ref{CrystalCTO}(a). The rhombohedral unit cell contains two Co$^{2+}$ ions, which we denote by $A$ and $B$. The Co$^{2+}$ ions are arranged in two-dimensional slightly buckled honeycomb lattices, which are stacked in an ABC sequence along the $c$-axis, with neighboring planes displaced diagonally by one-third of the unit cell. Below the N\'eel temperature of $T_N$ = 38 K, the magnetic moments order ferromagnetically within the $ab$-planes and are coupled antiferromagnetically along the $c$-axis~\cite{Yuan2020,yuan2020dirac,elliot2021order}. The rhombohedral unit cell contains 10 ions, and group theory predicts ten Raman-active phonons, $\Gamma_{R}=5A_g\oplus5E_g$ and eight infrared-active modes $\Gamma_{IR}=4A_u\oplus4E_u$ where $A$ and $E$ are non-degenerate and doubly degenerate modes, respectively~\cite{kroumova2003bilbao,dubrovin2021lattice}. 

The magnetic properties of CoTiO$_3$ are determined by the three unpaired spins on the magnetic Co$^{2+}$ ($3d^7$). 
The spin-orbit coupling and trigonal distortion have the same energy scale in CoTiO$_3$~\cite{Yuan2020}, and the ground state of Co$^{2+}$ ($3d^7$), S=3/2 can be considered an effective $\tilde{S}=1/2$ spin state, as shown in Fig.~\ref{CrystalCTO} (b). The two low-energy manifolds  are predominantly composed of $j_\mathrm{eff}=1/2$ and $j_\mathrm{eff}=3/2$ angular momentum states, respectively, and their wavefunctions are given by: 
\begin{align}
   \ket{\psi_{1/2}} =&  \left|J=\frac{1}{2},m_j=\pm\frac{1}{2}\right> \nonumber\\
   = & \frac{1}{\sqrt{2}}\ket{m_l=\mp\tilde{1},m_s=\pm\frac{3}{2}} \nonumber\\
   & -\frac{1}{\sqrt{3}}\ket{m_l=\tilde{0},m_s=\pm\frac{1}{2}} \nonumber\\
   &+\frac{1}{\sqrt{6}}\ket{m_l=\pm\tilde{1},m_s=\mp\frac{1}{2}}
     \label{gdstate}
\end{align}
and
\begin{align}
\ket{\psi_{3/4}}= & \left|J=\frac{3}{2},m_j=\pm\frac{3}{2}\right> \nonumber\\
= & \sqrt{\frac{3}{5}}\ket{m_l=\tilde{0},m_s=\pm\frac{3}{2}} \nonumber\\
 & -\sqrt{\frac{2}{5}}\ket{m_l=\pm\tilde{1},m_s=\pm\frac{1}{2}}
\label{excstate}
\end{align}
where the states $\ket{m_l=\tilde{i},m_s}$ arise from the effective $l_\mathrm{eff}=1$ and $S=3/2$ states comprised of three holes and $m_l,m_s$ denote the magnetic quantum number along the $z$-direction of the local coordinate system of the two Co$^{2+}$ ions (check Fig.~\ref{local} in the Appendix). 
For $T>T_N$, both manifolds remain doubly degenerate and the two manifolds are separated in energy by 23.5~meV~\cite{Yuan2020} as measured in neutron-diffraction experiments. These low-energy excitations between spin-orbit split states are very close in energy with two $E_g$ optical phonons at 26~meV and 33~meV \cite{dubrovin2021lattice}. This close proximity in energy enables the hybridization between phonons and spin-orbit excitations, which in turn can produce phonon chirality and therefore a phonon Zeeman effect. 

\begin{figure*}
    \centering
\includegraphics[scale=0.37]{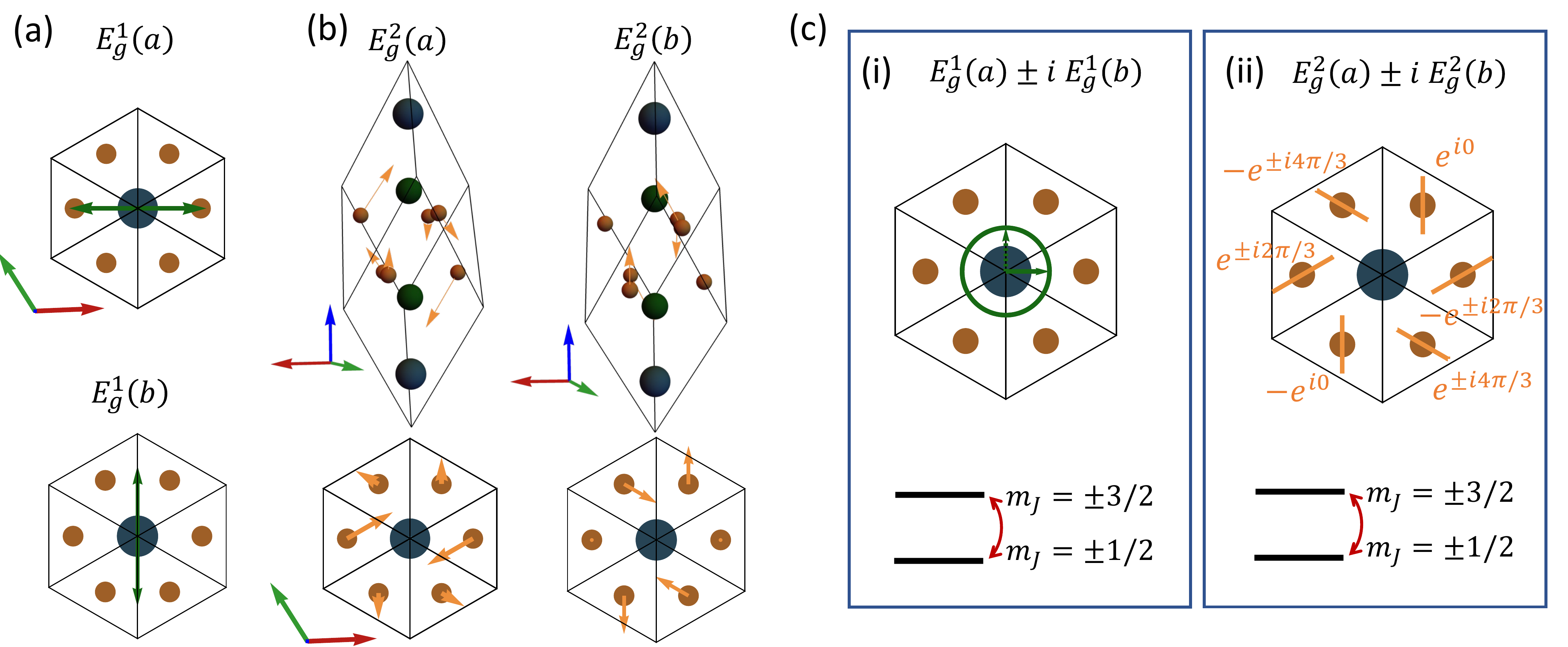}
    \caption{ \textbf{Chiral phonons in CoTiO$_3$}: (a) Atomic displacements for the two components, $a$ and $b$, of the doubly degenerate $E_{g}^1$ mode. This mode involves only the motion of Co$^{2+}$ ion in $a-b$ plane in two perpendicular directions. (b) Side and top view of atomic displacements for the two components, $a$ and $b$, of the doubly degenerate $E_{g}^2$ mode. (c) Displacement patterns for chiral phonons in the $ab$ plane of the crystal arise from superpositions of the two components of the doubly degenerate $E_{g}^1$ and $E_{g}^2$ modes with a $\pi/2$ phase difference. We show the orbital transitions between the Kramers doublets that couple to the respective chiral phonon modes.  
    (i) For the $E_{g}^{1}$ mode, Co$^{2+}$ ions move on circular orbits around their equilibrium position in the $ab$ plane of the crystal.
    (ii) For $E_{g}^2$, the O$^{2-}$ ions in the upper and lower plane around the Co$^{2+}$ ion move in opposite directions with a relative phase of $\pm 2\pi/3$ between atoms lying in the same plane.  }
    \label{phononmodeCTO}
\end{figure*}

\subsection{Microscopic model for the orbit-lattice coupling}
Phonons associated with irreducible representations other than fully symmetric  ones lower the site symmetry of the Co$^{2+}$ and hence can mix different electronic states. Here, we consider the two $E_g$ modes with energies of 26~meV and 33~meV that are close to the orbital transitions.
As in Sec.~\ref{sec:III}, we first evaluate the strength of the coupling using a point-charge model with atomic displacements of phonons obtained from group theory and first-principles calculations. 

We first find the basis functions for different phonon modes by using projection operators for the irreducible representation $E_g$. The phonon displacements of two $E_g$ modes under consideration are a superposition of these basis functions and cannot be obtained from purely group theoretical tools. 
However, first-principal calculations in previous works~\cite{dubrovin2021lattice} allowed us to approximate the lattice displacements in terms of the basis functions we obtained. The first $E_g$ mode is predominantly associated with the motion of the Co$^{2+}$ ion in the $ab$-plane, which, in the basis function of the two components of this $E_g$ mode, can be approximated as $x$ and $y$ motion of the Co$^{2+}$ ion, as shown in Fig.~\ref{phononmodeCTO}(a).

For the $E_g^2$ (33~meV) mode, we used a superposition of basis functions that matches the displacements (Fig.~\ref{phononmodeCTO}(b)) to the second $E_g$ mode in previous first-principles work~\cite{dubrovin2021lattice}. 
For the second mode, we tried several superpositions of different basis functions and have considered the one which showed displacements (Fig.~\ref{phononmodeCTO}(b)) similar to the second $E_g$ mode in previous first-principles work~\cite{dubrovin2021lattice}. 
This $E_g$ mode primarily includes the motion of ligand O$^{2-}$ ions.
For the displacements shown in  Fig.~\ref{phononmodeCTO}(b), we find that the modification of the CEF around the Co$^{2+}$ ions on the $A$/$B$ sites is given by:
\begin{align}
V^{A/B}&(E_{g}^1(a))=\left[-0.56~ \mathrm{eV}xz-0.51~\mathrm{eV} (x^2-y^2)+\mathcal{O}(r^3)\right]Q_a, \label{CoBelphmodeEg1a}\\
V^{A/B}&(E_{g}^1(b))=\pm \left[1.0~\mathrm{eV} xy-0.56~\mathrm{eV} yz+\mathcal{O}(r^3)\right]Q_b, 
\end{align}
for the $E_g^{1}$ mode, and
\begin{align}
  V^{A/B}&(E_{g}^2(a))=\left[-0.04~\mathrm{eV} xy-0.61~ eV xz\right]Q_a +\nonumber\\
  &\left[0.72~\mathrm{eV} yz-0.14~\mathrm{eV} (x^2-y^2)+\mathcal{O}(r^3)\right]Q_a \\
  V^{A/B}&(E_{g}^2(b))=\pm \left[0.28~\mathrm{eV} xy-0.72~\mathrm{eV} xz-0.61~\mathrm{eV} yz\right]Q_b \nonumber\\
  & \pm\left[-0.02~\mathrm{eV} (x^2-y^2)+\mathcal{O}(r^3)\right]Q_b,  \label{CoBelphmode5b}
\end{align}
for the $E_g^{2}$ mode
where $Q_{a/b}$ are the normal mode coordinates associated with two-components of the $E_g$ phonon expressed in units of \AA{}.$\sqrt{\text{amu}}$ in a local coordinate system around each Co$^{2+}$ ion. The $z$- and $y$-axis for $B$ sites are opposite to that of $A$ sites (see the Appendix for more details) which explains an extra negative sign in the $b$ component correction for $B$ sites. 
In the basis, $\ket{\psi_{1/2}}=\ket{J=1/2,m_j=\pm1/2}$ and $\ket{\psi_{3/4}}=\ket{J=3/2,m_j=\pm3/2}$, the resulting coupling (also shown in Fig.~\ref{phononmodeCTO}(c) takes the following form for the two components of $E_g$ mode:
\begin{align}
\hat{O}^{A/B}_{E_g^{a}}&=ge^{i\phi_{ab}}\ket{\psi_1^A}\bra{\psi_3^A}-ge^{-i\phi_{ab}}\ket{\psi_2^A}\bra{\psi_4^A}+ \mathrm{h.c.},\label{CTOelpha}\\
\hat{O}^{A/B}_{E_g^{b}}&=\pm ige^{i\phi_{ab}}\ket{\psi_1^A}\bra{\psi_3^A}\pm ige^{- i\phi_{ab}}\ket{\psi_2^A}\bra{\psi_4^A}+ \mathrm{h.c.},
\label{CTOelphb}
\end{align} 
where $g$ for two $E_g$ modes are 
\begin{equation}
    g_{E_g^{(1)}}\approx 0.3   r_0^2/\text{\AA}^2 \mathrm{meV}, ~~~\mathrm{and}~~~   g_{E_g^{(2)}}\approx 0.4 r_0^2/\text{\AA}^2 \mathrm{meV},
    \label{E1g_gestimate}
\end{equation}  
where $r_0^2=\langle r^2 \rangle\approx 1~\text{\AA}^2$ for $3d$ orbitals in Co$^{2+}$~\cite{shannon1976revised} (see Appendix~\ref{AppendixC}). The phase $\phi_{ab}$ depends on the ratio of coefficients of $xz (xy)$ and $yz (x^2-y^2)$ terms but does not affect the self-energy terms. 

To obtain the values of coupling strength $g$, we first express the phonon displacements in terms of the phononic creation and annihilation operators, $a, a^\dagger, b,b^\dagger$,
\begin{equation}
     Q_a=\frac{\hbar}{\sqrt{\hbar\omega_{0}}}\left(a+a^\dagger\right)=\frac{0.06\text{\AA}\sqrt{\mathrm{eV~amu}}}{\sqrt{\hbar\omega_{0}}}\left(a+a^\dagger\right).
\end{equation}
where  $\hbar\omega_{0}$ is the phonon energy (we restored $\hbar$ for the purpose of this equation), and then using Eqs.~(\ref{CoBelphmodeEg1a}-\ref{CoBelphmode5b}) and by expressing the states in Eq.~\eqref{gdstate} and Eq.~\eqref{excstate} in terms of three-particle $d$-orbital states, we evaluate the matrix elements for the crystal-field perturbation term between different states (see Appendix for details).  This coupling term maps to Eq.~\eqref{coupling1} and Eq.~\eqref{coupling2} in Sec.~\ref{sec:II}, and accordingly, the $E_{g}$ mode will split into two circularly polarized modes with opposite chirality when a magnetic field is applied along the $c$-axis.  The split modes will have angular momentum of $\pm\hbar$ along the $c$-axis which arises from the orbital angular momentum possessed by the superposition $E_{g}(a)\pm i E_{g}(b)$ shown in Fig.~\ref{phononmodeCTO}(c).

\subsection{Phonon Zeeman splitting and effective phonon magnetic moment}
In order to  evaluate the phonon Zeeman splitting for both of these $E_g$ modes, we can apply the model discussed in Sec.~\ref{sec:II} with slight modifications. For $T>T_N$, the system is in the paramagnetic phase and the eigenstates for the two manifolds involved in electronic excitations are identical to the basis states $\ket{\psi_{1/2}}$ and $\ket{\psi_{3/4}}$ described in Eq.~\eqref{gdstate} and Eq.~\eqref{excstate}, respectively. In this case, after accounting for the electron-phonon interaction described in Eq.~\eqref{CTOelpha} and Eq.~\eqref{CTOelphb}, the non-interacting Green's function is given by:
\begin{widetext}
\begin{equation}
    \mathbf{D}^{-1}(\omega)=\begin{pmatrix}
    \frac{\omega^2-\omega_0^2}{2\omega_0}-\tilde{g}^2\left(\frac{f_{1}^A\Delta_1^A}{\omega^2-(\Delta_1^A)^2}+\frac{f_2^A\Delta^A_2}{\omega^2-(\Delta_2^A)2}+A\Longleftrightarrow B\right)&i\tilde{g}^2\left(-\frac{f_1^A\omega}{\omega^2-(\Delta_1^A)^2}+\frac{f_2^A\omega}{\omega^2-(\Delta_2^A)^2}-A\Longleftrightarrow B\right)\\
    -i\tilde{g}^2\left(-\frac{f_1^A\omega}{\omega^2-(\Delta_1^A)^2}+\frac{f_2^A\omega}{\omega^2-(\Delta_2^A)^2}-A\Longleftrightarrow B\right)&\frac{\omega^2-\omega_0^2}{2\omega_0}-\tilde{g}^2\left(\frac{f_1^A\Delta_1^A}{\omega^2-(\Delta_1^A)^2}+\frac{f_2^A\Delta_2^A}{\omega^2-(\Delta_2^A)^2}+A\Longleftrightarrow B\right),
    \end{pmatrix}
\end{equation}
\end{widetext}
where again $\tilde{g}^2=4\pi g^2$. 

The correction to the non-interacting phonon Green's function here is similar to that in Eq.~\eqref{newg} discussed in Sec.~\ref{sec:II}. The only difference is that the off-diagonal term has contributions from two magnetic ions denoted by $A$ and $B$. The contributions from the two ions come with opposite signs, but it is also worth noticing that the electron-phonon interactions were evaluated within the local coordinate systems of each of the ions. The local $z$-coordinates for two ions point in opposite directions and thus for a magnetic field applied along the $c$ axis of the crystal, the population difference for two states in the lower Kramers doublet ($J=1/2$) are opposite as well - hence the two contributions add up. 

\begin{figure*}
    \centering
 \includegraphics[scale=0.28]{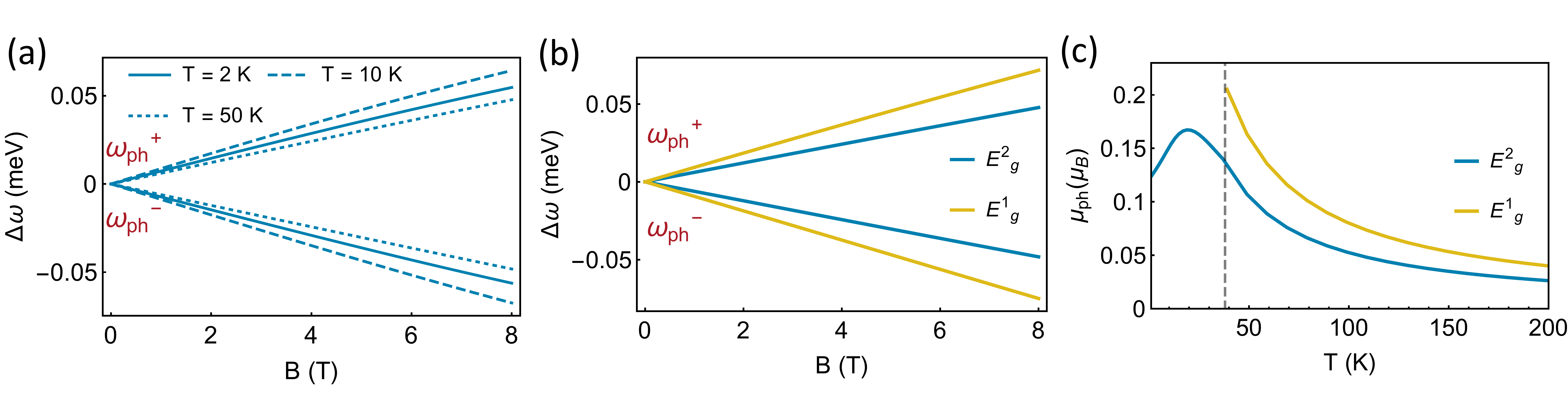}
    \caption{\textbf{Phonon Zeeman effect in CoTiO$_3$}: (a) Splitting of the doubly degenerate $E^2_{g}$ mode with frequency $\omega_0$ into left- and right-handed circularly polarized phonon modes with frequencies $\omega_{ph}^\pm$, as a function of the applied magnetic field, $B$ along $c$ axis at three different temperatures. (b) Phonon Zeeman splitting for $E_{g}^1$ and $E_{g}^2$ mode at 50 K (c) Phonon magnetic moment, $\mu_{ph}$, as a function of temperature for $E_g^1$ and $E_g^2$ mode. The phonon magnetic moment increases with decreasing temperature in the paramagnetic regime. The $E_g^2$ mode shows very distinct behavior below and above N\'eel temperature, $T_N=38 K$ shown by the dotted line. The $E_g^1$ mode is very close in energy to the spin orbit excitation and both energies are affected by the magnetic transition which would significantly influence the hybridization mechanism and is not captured in our model.}
    \label{splittingCTO}
\end{figure*}

Following the same procedure as in Sec.~\ref{sec:II} and Sec.~\ref{sec:III}, we obtain the phonon Zeeman splitting shown in Fig.~\ref{splittingCTO}(a), where we have used the values of orbit-lattice coupling from Eq.~\eqref{E1g_gestimate}, $\Delta_1=\Delta_2= 23.5$~meV, 
magnetic moment  $\mu^{gd}_{el}=1.9\,\mu_B$ for the $J=1/2$ states (ground state manifold) of Co$^{2+}$ on the basis of range mentioned in Ref.~\cite{Yuan2020}, and $\omega_{ph}=26$~meV and $\omega_{ph}=33$~meV for $E_g^{1}$ and $E_g^{1}$ modes, respectively. 
At T=50 K, this leads to significant splitting for both modes as shown in Fig.~\ref{splittingCTO}(b) yielding a phonon magnetic moments of $\mu_{ph}=0.2\,\mu_B$ and $0.1\, \mu_B$ for $E_g^{1}$ and $E_g^{2}$, respectively. In the paramagnetic regime, we expect that $\mu_{ph}$ scales as 1/T with temperature according to Eq.~\eqref{contr}, and hence drops sharply when $T$ is increased, as shown in Fig.~\ref{splittingCTO}(c) and also in table~\ref{tab:cotio3}. 

However, this analysis does not work below the Neel temperature. For $T<T_N$, magnetic order sets in, and spins develop a finite in-plane magnetic moment. The exchange field arising from spin-ordering alters the single-ion energy levels and their eigenstates. Without loss of generality, we can assume that the resulting mean-field points in the $x$-direction, which splits up the lower
ground-state Kramers doublet even in the absence of the external magnetic field, as illustrated in Fig.~\ref{CrystalCTO}(b). The upper manifold is not affected, but new eigenstates are formed for the lower manifold that are given by
\begin{equation}
\begin{pmatrix}
    \mu^{gd}_{el}B_z^\alpha&h_{{ex}}(T)\\
    h_{{ex}}(T)&-\mu^{gd}_{el}B_z^\alpha
\end{pmatrix}\ket{\psi_{\tilde{1}/\tilde{2}}^\alpha}=E_{\tilde{1}/\tilde{2}}\ket{\psi_{\tilde{1}/\tilde{2}}^\alpha}
\end{equation}
where the Hamiltonian is written in the basis $\left(\ket{\psi_{1}^\alpha},\ket{\psi_{2}^\alpha}\right)$ defined in Eq.~\eqref{gdstate}, 
$\alpha=A,B$ denotes the Co$^{2+}$ ion site, $h_{{ex}}(T)$ is the exchange mean-field and $B_z$ is the external magnetic field, $E_{\tilde{2}}=-E_{\tilde{1}}=\sqrt{(\mu^{gd}_{el}B_z)^2+h_{ex}^2(T)}$
\begin{table*}[t]
\centering
\bgroup
\def\arraystretch{1.3}
\caption{
Calculated saturation splitting, phonon Zeeman splitting in an applied magnetic field of $B=1$~T at 10~K, as well as effective phonon magnetic moments at liquid helium, two different temperatures below and above the Neel temperature, and room temperature, for  doubly degenerate $E_g^2$ phonon mode in CoTiO$_3$.}
\begin{tabular}{l c c c c c}
\hline\hline
Phonon mode\;\;\;\;\;\; & Saturation splitting\;\;\;\;\;\; & Zeeman splitting at 1~T and 10~K\;\;\;\;\;\; & $\mu_{ph}(2~\mathrm{K})$\;\;\;\;\;\; & $\mu_{ph}(40~\mathrm{K})$\;\;\;\;\;\; & $\mu_{ph}(77~\mathrm{K}) $ \\
\hline
$E^2_g$(33~meV) &$>0.1$~meV  &0.02~meV  & 0.13~$\mu_B$ & 0.14~$\mu_B$ & 0.07~$\mu_B$\\
\hline\hline
\end{tabular}
\label{tab:cotio3}
\egroup
\end{table*}
We can now apply our orbit-lattice coupling model for these new eigenstates and obtain the phonon energies by solving $\mathrm{Det}(\mathbf{D}^{-1}(\omega))=0$. The diagonal components are given by
  
\begin{align}
    &\mathbf{D}^{-1}|_{\alpha\alpha}= \nonumber\\
    & \frac{\omega^2-\omega_0^2}{2\omega_0}-2\tilde{g}^2\left(\frac{f_{\tilde{1}}E_{{\tilde{1}}3}\left(\cos\frac{\theta}{2}\right)^2}{\omega^2-E_{{\tilde{1}}3}^2}+\frac{f_{\tilde{1}}E_{{\tilde{1}}4}\left(\sin\frac{\theta}{2}\right)^2}{\omega^2-E_{{\tilde{1}}4}^2}\right) \nonumber\\
    & -2\tilde{g}^2\left(\frac{f_{\tilde{2}}E_{{\tilde{2}}3}\left(\sin\frac{\theta}{2}\right)^2}{\omega^2-E_{{\tilde{2}}3}^2}+\frac{f_1E_{24}\left(\cos\frac{\theta}{2}\right)^2}{\omega^2-E_{{\tilde{2}}4}^2}\right),
    \label{phononGexchangefielddiag}
\end{align} 
for $\alpha=a,b$ and off-diagonal components  are given by:
\begin{align}
     \mathbf{D}^{-1}|_{ab}&=-\mathbf{D}^{-1}|_{ba}= \nonumber \\
    & 2i\tilde{g}^2\left(-\frac{f_{\tilde{1}}\omega\left(\cos\frac{\theta}{2}\right)^2}{\omega^2-E_{{\tilde{1}}3}^2}+\frac{f_{\tilde{1}}\omega\left(\sin\frac{\theta}{2}\right)^2}{\omega^2-E_{{\tilde{1}}4}^2}\right) \nonumber\\
    &+2i\tilde{g}^2\left(-\frac{f_{\tilde{2}}\omega\left(\sin\frac{\theta}{2}\right)^2}{\omega^2-E_{{\tilde{2}}3}^2}+\frac{f_{\tilde{2}}\omega\left(\cos\frac{\theta}{2}\right)^2}{\omega^2-E_{{\tilde{2}}4}^2}\right),
    \label{phononGexchangefieldoffdiag}
\end{align} 
where $\cos(\theta)=\mu^{gd}_{el}B\sqrt{(\mu^{gd}_{el}B)^2+h_{ex}^2(T)}$,  $E_{{\tilde{1}}3}=E_{{\tilde{1}}4}=\Delta_0-E_{\tilde{1}}$, and  $E_{{\tilde{2}}3}=E_{{\tilde{2}}4}=\Delta_0-E_{\tilde{2}}$, and $h_{ex}(T)=h_0\sqrt{\left(1-T/T_N\right)}$  is the mean-field as a function of temperature. 

Here, the structure of eigenstates ($\ket{\psi_{\tilde{1}}},\ket{\psi_{\tilde{2}}}$ ) is such that the off-diagonal term comes out to be zero when these states are an equal superposition of $\ket{\psi_1}$ and $\ket{\psi_2}$,
i.e., for $B=0$ and hence no phonon energy splitting can occur in the absence of a magnetic field as expected. However, once we apply a magnetic field along the $c$-axis, the eigenstates in the ground-state manifold are no longer an equal superposition of up and down spin  and develop a net magnetic moment along the $c$-direction, and as a result, the off-diagonal term becomes proportional to $B$, which leads to a splitting of the previously degenerate phonon modes as shown in Fig.~\ref{splittingCTO}(a,b). Interestingly, the splitting does not saturate even at very low temperatures, which can be understood on the basis of exchange interactions.

In this case, the temperature dependence enters in two different ways: on the one hand, it determines the thermal population of the two states in the ground state manifold, and on the other hand also, the spin-polarization of each state through the temperature-dependence of the exchange mean-field. This makes the problem analytically untractable, and hence we obtain the phonon frequencies by numerically evaluating the poles of Green's functions. We show the combined temperature dependence of $\mu_{ph}$ above and below the N\'eel temperature in Fig.~\ref{splittingCTO}(c), where we have considered a maximum exchange mean-field of $h_0=3$~meV on the basis of the values presented in Ref.~\cite{Yuan2020}.

In both cases, $T>T_N$ and $T<T_N$, the phonon splitting is non-zero only if the populations and eigenstates of the ground-state manifold are such that the magnetic ion carries a net magnetic moment along the $c$-axis. In the paramagnetic case, the magnetic moment is directly related to the population difference of the two states in the lower Kramers doublet, as the two states have a magnetic moment along the $c$-axis but with opposite signs. In contrast, in the antiferromagnetically ordered state, we further need to take into account the net magnetic moment of each state in addition to the population difference between the two states. The direction of the net magnetic moment of each state is determined by the combined effect of the in-plane exchange mean field and out-of-plane applied magnetic field. As a result, the net magnetization of the sample increases at a slower rate with the applied magnetic field in the $T<T_N$ case compared to the paramagnetic region. This rate keeps on decreasing as the temperature is decreased further and leads to a maximum of the phonon magnetic moment below the N\'eel temperature, in contrast to the pure paramagnetic case in the rare-earth trihalides. Overall, we can expect that the temperature trend of the phonon $g$-factor should be similar to the temperature dependence of the magnetic susceptibility along the $c$-direction.


Our calculations for CoTiO$_3$ have shown two things: 1) applying an external magnetic field can produce chiral phonons with large effective magnetic moments on the order of $0.1\mu_B$, and 2) the phonon $g$-factor follows the same trend as the magnetic suscpetibility. This intuition can be extended to ferromagnets, where we expect that the phonon Zeeman splitting would saturate very quickly near the critical temperature and chiral phonons would remain split with fixed energy separation below $T_C$. It also indicates that in some cases, it is possible to have chirality-dependent phonon energy splitting even in the absence of an external magnetic field~\cite{boniniprl2023}.


\section{Discussion\label{sec:V}}

In summary, we have developed a microscopic model that describes the hybridization of doubly degenerate phonon modes with electronic orbital transitions between doublet states. An applied magnetic field splits the degeneracy of the doublets and therefore that of the phonons coupled to it, resulting in circularly polarized phonon modes with opposite chirality. The splitting is determined by the population asymmetry between the ground-state doublets, which makes the mechanism temperature dependent. The splitting of the phonon frequencies is linear in the limit of small magnetic fields, which is consistent with the phenomenological notion of the phonon Zeeman effect \cite{juraschek2:2017,Juraschek2019}, and which allows us to assign an effective magnetic moment to the chiral phonon modes. The specific form of the orbit-lattice coupling leading to these phenomena depends on the point-group symmetry and orbital configuration of the material. Furthermore, in order for the mechanism to be significant, the phonon modes and orbital transitions need to be on similar energy scales. We have therefore applied the model and computed phonon Zeeman splittings and effective phonon magnetic moments for the specific cases of CeCl$_3$, a $4f$-electron paramagnet, in which the orbital transition between the doublet states are determined by the crystal electric field, as well as CoTiO$_3$, a $3d$-electron antiferromagnet, in which the orbital transitions are determined by spin-orbit coupling and a trigonal distortion.

In the case of CeCl$_3$, the effective phonon magnetic moment increases monotonically with decreasing temperature over the entire investigated temperature spectrum, because the spins of the Ce$^{3+}$ ions order only at very low temperatures of $<0.1$~K \cite{Landau1973}, not considered here. We predict values of several $\mu_B$ at cryogenic temperatures that correspond to phonon frequency splittings of the order of 10~cm$^{-1}$, corroborating early experimental measurements \cite{schaack:1976,schaack:1977}. Even at room temperature, the effective magnetic moments of the phonon modes range between 0.01~$\mu_B$-0.1~$\mu_B$, orders of magnitude larger than those generated by purely ionic charge currents \cite{ceresoli:2002,juraschek2:2017,Hamada2018,Juraschek2019,Geilhufe2021,Zabalo2022}. In the case of CoTiO$_3$, we distinguish between the high-temperature paramagnetic and the low-temperature antiferromagnetic phases. The paramagnetic phase behaves similar to the case of CeCl$_3$, with a monotonically increasing effective phonon magnetic moment for decreasing temperatures down to the N\'eel temperature. Below the N\'eel temperature however, the value of the magnetic moment peaks at approximately $0.17\mu_B$, because the exchange mean-field in the ordered state exhibits an additional, competing temperature dependence that produces a global maximum of the effective magnetic moment of the phonon. Despite the decreasing trend of the magnetic moment at high temperatures, even at room temperature, we still obtain $\mu_{ph}=0.07\mu_B$, an order of magnitude larger than that predicted by ionic charge currents. Combining both temperature dependencies below and above $T_N$, the effective phonon magnetic moment follows roughly the trend of the magnetic susceptibility.

While we have looked at two particular examples in this manuscript, the theory developed in Sec.~\ref{sec:II} is general to all materials that exhibit doubly degenerate phonon modes and orbital transitions between doublet states with comparable energy scales. We therefore expect that in particular more $3d$ transition-metal oxide compounds may show the proposed phenomena, which host a variety of materials with trigonal point-group symmetries and octahedral coordinations of ligand ions. Beyond transition-metal oxides, $4d$-electron magnets could be interesting candidates, because they possess larger spin-orbit couplings and therefore higher transition energies between doublet states, which allows hybridizations with high-frequency phonon modes above 10~THz (40 meV). RuCl$_3$, for example, possesses orbital excitations on the order of 100~meV \cite{Sandiland2016} as well as doubly degenerate phonon modes with energies around 50~meV.

In magnetically ordered materials, the details of the mechanism for phonon Zeeman splitting further depend on the exchange interactions between the magnetic ions, and an upcoming challenge will be to investigate how the phenomenon unfolds when interactions such as superexchange, itinerant electrons, or ring-exchange interactions \cite{Takahashi1977,Fedorova2018} are present. Beyond CEF and spin-orbit excitations, the mechanism can be extended to include hybridizations between doubly degenerate phonon modes and other electronic or collective excitations, such as low-energy charge-transfer excitations or magnons, which show strong magnetic-field dependence \cite{Liu2021_FePS3,Vaclavkova2021_FePS3}.

Finally, we point out that we have investigated only Raman-active phonon modes with $E_{(i)g}$ symmetries in this study. The same evaluation can be done for infrared-active phonon modes, $E_{(i)u}$ symmetries in our investigated materials, which can be resonantly driven with ultrashort laser pulses in the terahertz and mid-infrared spectral range. A recent study proposed that infrared-active phonon modes driven in paramagnetic CeCl$_3$ can potentially produce giant effective magnetic fields through the effective magnetic moment of the phonons \cite{Juraschek2022_giantphonomag}, and this phenomenon was subsequently experimentally demonstrated in a paramagnetic oxide-ferrimagnetic garnet heterostructure \cite{Davies2022}. This mechanism should be readily applicable to the $E^1_u$ and $E^2_u$ modes in CoTiO$_3$ and offers the potential for an unprecedented spin-switching protocol that could establish a new paradigm in ultrafast spintronics and data processing.


\acknowledgements
We thank Sebastian Stepanow (ETH Zurich), Xiaoqin (Elaine) Li, David Lujan, and Jeongheon Choe for useful discussions. This research was primarily supported by the National Science Foundation through the Center for Dynamics and Control of Materials: an NSF MRSEC under Cooperative Agreement No.\ DMR-1720595.  G.A.F.\ acknowledges additional support from NSF DMR-2114825 and from the Alexander von Humboldt Foundation.


\onecolumngrid

\appendix


\section{Derivation of the phonon Zeeman splitting and effective magnetic moments of phonons}
In this section, we provide more details on the derivations of the equations in the main text. We begin by considering a degenerate phonon mode with two components that is described by the Hamiltonian
    \begin{equation}
        H_{ph}=\omega_0(a^\dagger a+ b^\dagger b).
    \end{equation}
We only consider zone-centered phonon modes and can accordingly drop the momentum dependence in the phonon operators and energies. In order to account for the effect of electron-phonon interactions on the phonon spectrum, we use a Green's function formalism. For the non-interacting system, the Green's function matrix is given by
    
    \begin{equation}
        \mathbf{D}_0(\omega)=\begin{pmatrix}
        D_0^{aa}(\omega)&0\\0&D_0^{bb}(\omega)
        \end{pmatrix},
    \end{equation}
    where the components are given by
    \begin{equation}
        D_0^{aa}(\omega)=D_0^{bb}(\omega)=\frac{2\omega_0}{\omega^2-\omega_0^2}.
    \end{equation}
    The phonon frequency, $\omega_0$, can be trivially retrieved by solving $\mathrm{Det}(\mathbf{D}_0^{-1}(\omega))=0$.
    We obtain these Green's functions from the Fourier transform of time-dependent phonon propagators,
\begin{align}
    D^{aa}_0( t-t')&=-i\theta(t-t')\bra{0}|A(t)A(t')\ket{0}-i\theta(t'-t)\bra{0}A(t')A(t)\ket{0},\\
    D^{bb}_0( t-t')&=-i\theta(t-t')\bra{0}|B(t)B(t')\ket{0}-i\theta(t'-t)\bra{0}B(t')B(t)\ket{0},
\end{align}
where $A(t)=a(t)+a^\dagger(t)$ and $B(t)=b(t)+b^\dagger(t)$ with $a(t)=ae^{-i\omega_0t}$ and $b_k(t)=be^{-i\omega_0t}$.

We next consider the electronic Hamiltonian, which can be expressed in second quantization as
    \begin{equation}
        H_{el}=\sum_{i=1}^4\varepsilon_ic_i^\dagger c_i,
    \end{equation}
    where $c_i^\dagger$  and $c_i$ are the creation and annihilation operators for electrons in state $i$ on the magnetic ion. Their Green's functions read
    \begin{equation}
        G_0^{ii}(t-t')=-i\left<0|\mathcal{T}\left[c_i(t)c_i^\dagger(t')\right]|0\right>=-i\left[\theta(t-t')\left<0|c_i(t)c_i^\dagger(t')|0\right>-\theta(t'-t)\left<0|c_i^\dagger(t')c_i(t)|0\right>\right],
    \end{equation}
    where the time-dependent operators are given by
    \begin{equation}
        c_i(t)=e^{i\varepsilon_ic_i^\dagger c_it}c_ie^{-i\varepsilon_ic_i^\dagger c_it}=e^{-i\varepsilon_it}c_i.
    \end{equation}
    We can therefore write the Green's function as
    \begin{equation}
        G_0^{ii}(t-t')=-i\left[\theta(t-t')(1-f_i)-\theta(t'-t)f_i\right]e^{-i\varepsilon_i(t-t')},
    \end{equation}
    where $f_i=\left<0|c_i^\dagger c_i|0\right>$ is the occupation number for state $i$, given by the Fermi-Dirac distribution. A Fourier transform of this expression yields
    \begin{equation}
        G_0^{ii}(\omega)=\frac{1-f_i}{\omega-\varepsilon_i+i\eta}+\frac{f_i}{\omega-\varepsilon_i-i\eta}.
    \end{equation}
The electron-phonon interaction of Eqs.~\eqref{elph1} and \eqref{coupling2} from the main text can therefore be expressed in second quantization as
\begin{equation}
    H_{el-ph} =V^a+V^b,
\end{equation}
where
\begin{align}
        V^a&=\sum_{i,j}g(a^\dagger+a)\Gamma^a_{ij}=g(a^\dagger+a)(c_3^\dagger c_1+c_1^\dagger c_3)-g(a^\dagger+a)(c_4^\dagger c_2+c_2^\dagger c_3), \label{vertices1}\\
        V^b&=\sum_{i,j}g(a^\dagger+a)\Gamma^a_{ij}=ig(b^\dagger+b)(c_3^\dagger c_1-c_1^\dagger c_3)+ig(b^\dagger+b)(c_4^\dagger c_2-c_4^\dagger c_2).\label{vertices2}        
\end{align}
Including the electron-phonon interaction, the interacting phonon propagator, $\mathbf{D}$, and electronic propagator, $\mathbf{G}$, are given by
    \begin{equation}
    \mathbf{D}(q,\omega)=D_0(q,\omega)+D_0(q,\omega)\mathbf{\Pi}(q,\omega)\mathbf{D}(q,\omega),
\end{equation}
where $\mathbf{\Pi}(q,\omega)=D_0^{-1}-\mathbf{D}^{-1}$ is the phonon self energy (the main quantity of interest for us), as well as
\begin{equation}
    \mathbf{G}(q,\omega)=G_0(q,\omega)+G_0(q,\omega)\mathbf{\Sigma}(q,\omega)\mathbf{G}(q,\omega).
\end{equation}

In a perturbative treatment, $\mathbf{\Pi}$ and $\mathbf{\Sigma}$ can be calculated from non-interacting Green's functions. The full expression is given by

\begin{equation}
    \begin{pmatrix}
    \mathbf{D}^{aa}&\mathbf{D}^{ab}\\
    \mathbf{D}^{ba}&\mathbf{D}^{bb}
    \end{pmatrix}=\begin{pmatrix}
    D_0^{aa}&D_0^{ab}\\
    D_0^{ba}&D_0^{bb}
    \end{pmatrix}+\begin{pmatrix}
    D_0^{aa}&D_0^{ab}\\
    D_0^{ba}&D_0^{bb}
    \end{pmatrix}\mathbf{\Pi}\begin{pmatrix}
    \mathbf{D}^{aa}&\mathbf{D}^{ab}\\
    \mathbf{D}^{ba}&\mathbf{D}^{bb}
    \end{pmatrix},
    \label{dyson}
\end{equation}
where the self-energy term is given by
\begin{equation}
    \mathbf{\Pi}=\begin{pmatrix}
    \mathbf{\Pi}_{13}^{aa}+\mathbf{\Pi}_{31}^{aa}+ \mathbf{\Pi}_{24}^{aa}+\mathbf{\Pi}_{42}^{aa}&\mathbf{\Pi}_{13}^{ab}+\mathbf{\Pi}_{31}^{ab}+\mathbf{\Pi}_{24}^{ab}+\mathbf{\Pi}_{42}^{ab}\\
    \mathbf{\Pi}_{13}^{ba}+\mathbf{\Pi}_{31}^{ba}+\mathbf{\Pi}_{24}^{ba}+\mathbf{\Pi}_{42}^{ba}&\mathbf{\Pi}_{13}^{bb}+\mathbf{\Pi}_{31}^{bb}+\mathbf{\Pi}_{24}^{bb}+\mathbf{\Pi}_{42}^{bb}
    \end{pmatrix}.
\end{equation}
We now approximate this term with its non-interacting value,
\begin{equation}
    \mathbf{\Pi}_{ij}^{\alpha\chi}(\omega)\approx\Pi_{ij}^{\alpha\chi}(\omega)=i\int d\omega' G_0^{ii}(\omega+\omega')G_0^{jj}(\omega')\Gamma_{ij}^\alpha\Gamma_{ji}^\chi.
\end{equation}
Together with the frequency-dependent electronic Green's function,
\begin{equation}
    G_0^{ii}(\omega)=\frac{1-f_i}{\omega-\varepsilon_i+i\eta}+\frac{f_i}{\omega-\varepsilon_i-i\eta},
\end{equation}
and using Eqs.~\eqref{vertices1} and \eqref{vertices2} to write down $\Gamma_{13}^a=\Gamma_{31}^a=-\Gamma_{24}^a=-\Gamma^a_{42}=g$ and $\Gamma_{13}^b=-\Gamma_{31}^b=\Gamma_{24}^b=-\Gamma^b_{42}=ig$.

\begin{align}
   & \int d\omega' G_0^{ii}(\omega+\omega')G_0^{jj}(\omega')=\int d\omega'\left( \frac{1-f_i}{\omega+\omega'-\varepsilon_i+i\eta}+\frac{f_i}{\omega+\omega'-\varepsilon_i-i\eta}\right)\left( \frac{1-f_j}{\omega'-\varepsilon_j+i\eta}+\frac{f_j}{\omega'-\varepsilon_j-i\eta}\right)\\&=
    \int d\omega'\left[ \left(\frac{(1-f_i)(1-f_j)}{(\omega+\omega'-\varepsilon_i+i\eta)(\omega'-\varepsilon_j+i\eta)}\right)+\left(\frac{(1-f_i)f_j}{(\omega+\omega'-\varepsilon_i+i\eta)(\omega'-\varepsilon_j-i\eta)}\right)\right]\\
    &+ \int d\omega'\left[\left(\frac{f_i(1-f_j)}{(\omega+\omega'-\varepsilon_i-i\eta)(\omega'-\varepsilon_j+i\eta)}\right)+\left(\frac{f_if_j}{(\omega+\omega'-\varepsilon_i-i\eta)(\omega'-\varepsilon_j-i\eta)}\right)\right]
\end{align}
We utilize the relation $\lim_{\eta\rightarrow 0}\frac{1}{x+i\eta}=P(\frac{1}{x})+i\pi\delta(x)$ and obtain 
\begin{align}
    \int d\omega'\left[ \left(\frac{(1-f_i)(1-f_j)}{(\omega+\omega'-\varepsilon_i+i\eta)(\omega'-\varepsilon_j+i\eta)}\right)\right]&=\frac{(1-f_i)(1-f_j)}{\omega-\varepsilon_{ij}}\int d\omega'\left(\frac{1}{\omega'-\varepsilon_j+i\eta}-\frac{1}{\omega+\omega'-\varepsilon_i+i\eta}\right) \nonumber\\
    & =0,\\
    \int d\omega'\left[ \left(\frac{(1-f_i)f_j}{(\omega+\omega'-\varepsilon_i+i\eta)(\omega'-\varepsilon_j-i\eta)}\right)\right]&=\frac{(1-f_i)f_j}{\omega-\varepsilon_{ij}+2i\eta}\int d\omega'\left(\frac{1}{\omega'-\varepsilon_j-i\eta}-\frac{1}{\omega+\omega'-\varepsilon_i+i\eta}\right) \nonumber\\
    & = -i2\pi\frac{(1-f_i)f_j}{\omega-\varepsilon_{ij}+2i\eta},\\
    \int d\omega'\left[ \left(\frac{f_i(1-f_j)}{(\omega+\omega'-\varepsilon_i+i\eta)(\omega'-\varepsilon_j+i\eta)}\right)\right]&=\frac{f_i(1-f_j)}{\omega-\varepsilon_{ij}-2i\eta}\int d\omega'\left(\frac{1}{\omega'-\varepsilon_j+i\eta}-\frac{1}{\omega+\omega'-\varepsilon_i-i\eta}\right), \nonumber\\
    & =i2\pi\frac{f_i(1-f_j)}{\omega-\varepsilon_{ij}-2i\eta}\\
    \int d\omega'\left[ \left(\frac{f_if_j}{(\omega+\omega'-\varepsilon_i-i\eta)(\omega'-\varepsilon_j-i\eta)}\right)\right]&=\frac{f_if_j}{\omega-\varepsilon_{ij}}\int d\omega'\left(\frac{1}{\omega'-\varepsilon_j-i\eta}-\frac{1}{\omega+\omega'-\varepsilon_i-i\eta}\right)\nonumber\\
    & =0,
\end{align}
as well as
\begin{equation}
    \int d\omega' G_0^{ii}(\omega+\omega')G_0^{jj}(\omega')=-i2\pi\frac{(1-f_i)f_j}{\omega-\varepsilon_{ij}+2i\eta}+i2\pi\frac{f_i(1-f_j)}{\omega-\varepsilon_{ij}-2i\eta}.
\end{equation}
In the limit $\eta\rightarrow0$, this expression becomes
\begin{equation}
    \int d\omega' G_0^{ii}(\omega+\omega')G_0^{jj}(\omega')=i2\pi\frac{f_i-f_j}{\omega-\varepsilon_{ij}}.
\end{equation}
This allows us to obtain closed forms for the components of the self energy,
\begin{align}
    \Pi_{13}^{aa}+ \Pi_{31}^{aa}&= \Pi_{13}^{bb}+ \Pi_{31}^{bb}=2\pi g^2 (f_1-f_3)\left(\frac{1}{\omega-\varepsilon_{13}}-\frac{1}{\omega-\varepsilon_{31}}\right)=2\pi \frac{2g^2 (f_1-f_3)\varepsilon_{13}}{\omega^2-\varepsilon_{31}^2},\\
    \Pi_{24}^{aa}+ \Pi_{42}^{aa}&=\Pi_{24}^{bb}+ \Pi_{42}^{bb}=2\pi g^2 (f_2-f_4)\left(\frac{1}{\omega-\varepsilon_{24}}-\frac{1}{\omega-\varepsilon_{42}}\right)=2\pi \frac{2g^2 (f_2-f_4)\varepsilon_{24}}{\omega^2-\varepsilon_{42}^2},\\
     \Pi_{13}^{ab}+ \Pi_{31}^{ab}&=-(\Pi_{13}^{ba}+ \Pi_{31}^{ba})=-2\pi ig^2 (f_1-f_3)\left(\frac{1}{\omega-\varepsilon_{13}}+\frac{1}{\omega-\varepsilon_{31}}\right)=2\pi \frac{-2ig^2 (f_1-f_3)\omega}{\omega^2-\varepsilon_{31}^2},\\
     \Pi_{24}^{ab}+ \Pi_{42}^{ab}&=-(\Pi_{24}^{ba}+ \Pi_{42}^{ba})=2\pi ig^2 (f_2-f_4)\left(\frac{1}{\omega-\varepsilon_{13}}+\frac{1}{\omega-\varepsilon_{31}}\right)=2\pi \frac{2ig^2 (f_2-f_4)\omega}{\omega^2-\varepsilon_{42}^2}.
\end{align}
Inserting these expressions into Eq.~\eqref{dyson}, we get 
\begin{equation}
    \mathbf{D}^{-1}=\begin{pmatrix}
    \frac{\omega^2-\omega_0^2}{2\omega_0}-\tilde{g}^2\left(\frac{f_1\Delta_1}{\omega^2-\Delta_1^2}+\frac{f_2\Delta_2}{\omega^2-\Delta_2^2}\right)&i\tilde{g}^2\left(-\frac{f_1\omega}{\omega^2-\Delta_1^2}+\frac{f_2\omega}{\omega^2-\Delta_2^2}\right)\\
    -i\tilde{g}^2\left(-\frac{f_1\omega}{\omega^2-\Delta_1^2}+\frac{f_2\omega}{\omega^2-\Delta_2^2}\right)&\frac{\omega^2-\omega_0^2}{2\omega_0}-\tilde{g}^2\left(\frac{f_1\Delta_1}{\omega^2-\Delta_1^2}+\frac{f_2\Delta_2}{\omega^2-\Delta_2^2}\right)
    \end{pmatrix},
    \label{Dvals}
\end{equation}
where $\tilde{g}^2=4\pi g^2$, $\Delta_1=\varepsilon_{31}$, $\Delta_2=\varepsilon_{42}$ and we assume the excited state to be unoccupied, $f_3=f_4=0$. The modified energies can then be obtained by solving $\mathrm{Det}(\mathbf{D}^{-1})=0$. The result of this equation depends on the application of a magnetic field, and we consider two cases.\\

\textbf{Case 1:} $B=0$. Here, $f_1=f_2=f_0/2$  and $\Delta_1=\Delta_2=\Delta$. In this scenario, the off-diagonal term in $\mathbf{D}^{-1}$ is zero and the evaluation of $\mathrm{Det}(\mathbf{D}^{-1}(\omega))=0$ reduces to:
\begin{align}
    (\omega^2-\omega_0^2)(\omega^2-\Delta^2)-2\tilde{g}^2f_0\omega_0\Delta & =0,\\
    (\omega^2-\omega_0^2)(\omega^2-\Delta^2)-2\tilde{g}^2f_0\omega_0\Delta & =0,
\end{align}
which have identical solutions, indicating the doubly degenerate nature of phonon and electronic excitations. The solutions corresponding to the phonon and electronic excitation branches are respectively given by:
\begin{align}
    \Omega_{ph}\equiv\omega_{ph}(B=0)=\left(\frac{\omega_0^2+\Delta^2}{2}+\sqrt{\left(\frac{\omega_0^2-\Delta^2}{2}\right)^2+2\tilde{g}^2f_0\omega_0\Delta}\right)^{1/2},\\
    \Omega_{{el}}\equiv\omega_{el}(B=0)=\left(\frac{\omega_0^2+\Delta^2}{2}-\sqrt{\left(\frac{\omega_0^2-\Delta^2}{2}\right)^2+2\tilde{g}^2f_0\omega_0\Delta}\right)^{1/2}.
\end{align}
This coupling modifies the phonon and electronic excitation energies but doesn't lift the degeneracy of two excitations. However, if the bare frequencies for two excitations are close, then even a weak electron-phonon coupling term can introduce significant mixing and the excitations are no longer purely phononic or electronic in nature. We have focused only on the off-resonant case where these aspects can be safely ignored.

\textbf{Case 2:} $B\neq 0$. We next apply an external magnetic field, $\mathbf{B}=B\,\hat{z}$, which lifts the degeneracies of the Kramers doublets, $\varepsilon_{12}\neq 0$ and $\varepsilon_{34}\neq 0$. This subsequently modifies the electronic transition energies as
\begin{equation}
    \Delta_1=\Delta-\gamma B,~~\Delta_2=\Delta+\gamma B,
\end{equation}
where $\gamma=\mu^{el}_{ex}-\mu^{el}_{gs}$ depends on the magnetic moment of the ground- and excited-state doublets. Lifting the degeneracy of the ground-state doublet leads to  asymmetric populations of the ground-state energy levels, $f_{12}\neq 0$. 
 Accordingly, the secular equation, $\mathrm{Det}(\mathbf{D}^{-1}(\omega))=0$ for $\mathbf{D}^{-1}$ given by Eq.~\eqref{Dvals}  becomes:
 \begin{align}
    (\omega^2-\omega_0^2)(\omega^2-\Delta^2)-2\tilde{g}^2f_0\omega_0\Delta+2 \omega \left(B\gamma (\omega^2-\omega^2_0)+\tilde{g}^2\omega_0f_{21}\right)+\gamma B\left(\gamma B (\omega^2-\omega^2_0)+2\tilde{g}^2\omega_0f_{21}\right)=0\label{sc1}\\
    (\omega^2-\omega_0^2)(\omega^2-\Delta^2)-2\tilde{g}^2f_0\omega_0\Delta-2 \omega \left(B\gamma (\omega^2-\omega^2_0)+\tilde{g}^2\omega_0f_{21}\right)+\gamma B\left(\gamma B (\omega^2-\omega^2_0)+2\tilde{g}^2\omega_0f_{21}\right)=0
    \label{seculareq}
\end{align}
These two equations are not equivalent and there is a term linear in $\omega$ that indicates a frequency splitting of phonon and electronic excitations. Given that the electron-phonon coupling is weak and the electronic excitations are off-resonant from phonons, we can assume that phonon energies are modified only slightly and have the following form

 \begin{align}
     w_{ph}^{\pm}=\Omega_{ph}\left(1\mp \eta \right),\\
     w_{el}^{\pm}=\Omega_{el}\left(1\mp \epsilon \right).
\end{align}
For the case of a paramagnetic system, the population difference is given by 

 \begin{align}
     f_{21}   = -\tanh\left(\frac{\mu^{el}_{gs} B}{k_BT}\right).
 \end{align}
 where $\mu^{el}_{gs}$ is the magnetic moment of ground state manifold.
We can plug the above equations into Eq.~\eqref{sc1} and Eq.~\eqref{sc2} in order to obtain 
\begin{equation}
    \Omega_{ph}\eta = \frac{\gamma B(\Omega_+^2-\omega^2_0)+\tilde{g}^2\omega_0 f_{21}}{\Omega_{ph}^2-\Omega_{{el}}^2+\gamma^2B^2} = \frac{\gamma B(\Omega_+^2-\omega^2_0)+\tilde{g}^2\omega_0\tanh\left(\frac{\mu^{el}_\mathrm{gs}B}{k_BT}\right)}{\sqrt{\left(\omega_0^2-\Delta^2\right)^2+8\tilde{g}^2f_0\omega_0\Delta}+\gamma^2B^2}.
\end{equation}
For the off-resonant case, we can assume $|\Delta-\omega_0|\gg \gamma B$ and therefore neglect the linear $B$ term in the numerator and the quadratic one in the denominator. The off-resonant case is a reasonable assumption, as $\gamma B\sim 0.5$~meV in strong magnetic fields of $B=10$~T, whereas often $|\Delta-\omega_0|>10$~meV. As a result, the splitting of the phonon frequencies can be written as
\begin{equation}
   \frac{\omega_{ph}^+-\omega_{ph}^-}{\omega_{ph}(B=0)}\approx\frac{2\tilde{g}^2}{\sqrt{\left(\omega_0^2-\Delta^2\right)^2+8\tilde{g}^2f_0\omega_0\Delta}}\tanh\left(\frac{\mu^{el}_{gs}B}{k_BT}\right),
   \label{split1}
\end{equation}

which retrieves the early result by Thalmeier and Fulde \cite{Thalmeier1977}.


\section{Orbit-lattice coupling of the $E_{2g}$ modes in CeCl$_3$}
As discussed in main text, the phonon lowers the symmetry around the magnetic ion and for the lattice distortion induced by $E_{1g}$ phonon, the first order term for the change in Coulomb potential is given by: 
\begin{align}
V(E_{1g}(a))&=\left[-0.06 xz+0.16 yz\right]Q_a~\frac{\mathrm{eV}}{\text{\AA}^3\sqrt{\mathrm{amu}}},\label{V1e1g}\\
    V(E_{1g}(b))&=\left[0.16 xz+0.06 yz\right]Q_b~\frac{\mathrm{eV}}{\text{\AA}^3\sqrt{\mathrm{amu}}}\label{V1e1gb}.
\end{align} 
Now, we can express $xz=r^2\sin\theta\cos\theta \cos\phi$ and $yz=r^2\sin\theta\cos\theta \cos\phi$ in spherical coordinates. The electronic states on Ce$^{3+}$ ion can be written in terms of $\ket{L=3,m=m_l}$ which have wavefunction $\langle r|L=3,m=m_l \rangle =R (r) Y_3^{m_l}(\theta,\phi)$. This allows us to calculate the matrix elements between different $4f$ states and the only non-zero terms are given by:
\begin{align}
\langle m=\pm 3|xz|m=\pm 2\rangle=\mp \langle r^2\rangle\frac{1}{3\sqrt{6}}\\
\langle m=\pm 2|xz|m=\pm 1\rangle=\mp \langle r^2\rangle\frac{1}{3\sqrt{10}}\\
\langle m=\pm 1|xz|m=\pm 0\rangle=\mp \langle r^2\rangle\frac{1}{3\sqrt{75}}
\end{align}

\begin{align}
\langle \pm m=3|yz|m= \pm 2\rangle= \langle r^2\rangle\frac{i}{3\sqrt{6}}\\
\langle \pm m=2|yz|m=\pm 1\rangle=\langle r^2\rangle\frac{i}{3\sqrt{10}}\\
\langle m= \pm 1|yz|m=0\rangle= \langle r^2\rangle\frac{i}{3\sqrt{75}}
\end{align}

Using these values for states given in Eq.~\eqref{state52}-\eqref{state12}, we get:
\begin{align}
    H_1(xz)&=-\frac{2}{7\sqrt{5}}\langle r^2\rangle\begin{pmatrix}
    &\left|\frac{5}{2},\pm\frac{5}{2}\right>&\left|\frac{5}{2},\pm\frac{3}{2}\right>\\
    \left|\frac{5}{2}\pm\frac{5}{2}\right>&0&\pm 1\\[1.2em]\left|\frac{5}{2},\pm\frac{3}{2}\right>&\pm 1&0
    \end{pmatrix},
    \\
     H_1(yz)&=\frac{2}{7\sqrt{5}}\langle r^2\rangle\begin{pmatrix}
    &\left|\frac{5}{2},\pm\frac{5}{2}\right>&\left|\frac{5}{2},\pm\frac{3}{2}\right>\\
    \left|\frac{5}{2},\pm\frac{5}{2}\right>&0&i\\[1.2em]\left|\frac{5}{2},\pm\frac{3}{2}\right>&-i&0
    \end{pmatrix}.
\end{align}

Next, we use the same microscopic model to calculate the Zeeman effect for other phonons as well. Here, we consider $E_{2g}^1$ (12~meV) and $E_{2g}^2$ (21.5~meV) phonons of CeCl$_3$.  The phonon eigenvectors are obtained  from density functional theory \cite{Juraschek2022_giantphonomag}.  As mentioned in the main text (Eq.~\ref{Coulomb}), these lattice displacements perturb the CEF around the Ce$^{3+}$ ions, and the resulting  modifications can be described as

\begin{align}
V(E_{2g}^1(a))&=\left[-0.05 xy-0.007 (x^2-y^2)\right]  Q_a~\mathrm{eV}/(\text{\AA}^2\sqrt{\mathrm{amu}}),\\
    V(E_{2g}^1(b))&=\left[0.014 xy-0.025 (x^2-y^2)\right] Q_b~\mathrm{eV}/(\text{\AA}^2\sqrt{\mathrm{amu}}),
\end{align}
\begin{align}
V(E_{2g}^2(a))&=\left[0.08 xy+0.01 (x^2-y^2)\right]  Q_a~\mathrm{eV}/(\text{\AA}^2\sqrt{\mathrm{amu}}),\\
    V(E_{2g}^2(b))&=\left[-0.02 xy+0.04 (x^2-y^2)\right] Q_b~\mathrm{eV}/(\text{\AA}^2\sqrt{\mathrm{amu}}),
\end{align}
where $Q_{a,b}$ is the amplitude of the phonon mode. Using spherical harmonics, we express this perturbation in the basis of electronic states,
using:
 \begin{align}
\langle m=\pm 3|xy|m=\pm 1\rangle= \pm\langle r^2\rangle\frac{i}{3\sqrt{15}}\\
\langle m=\pm 2|xy|m=\pm 0\rangle= \pm\langle r^2\rangle\frac{\sqrt{2}i}{3\sqrt{15}}\\
\langle m= 1|xy|m=-1\rangle= \pm \langle r^2\rangle\frac{2i}{15}
\end{align}
\begin{align}
\langle m=\pm 3|x^2-y^2|m=\pm 1\rangle= \langle r^2\rangle\frac{2}{3\sqrt{15}}\\
\langle m=\pm 2|x^2-y^2|m=\pm 0\rangle= \langle r^2\rangle\frac{\sqrt{8}}{3\sqrt{15}}\\
\langle m= 1|x^2-y^2|m=-1\rangle= \langle r^2\rangle\frac{4}{15}
\end{align}
 which gives:
\begin{align}
    H_1(x^2-y^2)&=-\frac{2\sqrt{2}}{7\sqrt{5}}\langle r^2\rangle\begin{pmatrix}
    &\left|\frac{5}{2},\frac{5}{2}\right>&\left|\frac{5}{2},\frac{1}{2}\right>\\
    \left|\frac{5}{2},\frac{5}{2}\right>&0&1\\[1.2em]\left|\frac{5}{2},\frac{3}{2}\right>&1&0
    \end{pmatrix}, \\
    H_1(x^2-y^2)&=-\frac{2\sqrt{2}}{7\sqrt{5}}\langle r^2\rangle\begin{pmatrix}
    &\left|\frac{5}{2},-\frac{5}{2}\right>&\left|\frac{5}{2},-\frac{1}{2}\right>\\
    \left|\frac{5}{2},-\frac{5}{2}\right>&0&1\\[1.2em]\left|\frac{5}{2},-\frac{1}{2}\right>&1&0
    \end{pmatrix},
    \label{xzpr}
\end{align}
\begin{align}
    H_1(xy)&=\frac{\sqrt{2}}{7\sqrt{5}}\langle r^2\rangle\begin{pmatrix}
    &\left|\frac{5}{2},\frac{5}{2}\right>&\left|\frac{5}{2},\frac{1}{2}\right>\\
    \left|\frac{5}{2},\frac{5}{2}\right>&0&i\\[1.2em]\left|\frac{5}{2},\frac{1}{2}\right>&-i&0
    \end{pmatrix}, \\
    H_1(xy)&=\frac{\sqrt{2}}{7\sqrt{5}}\langle r^2\rangle\begin{pmatrix}
    &\left|\frac{5}{2},-\frac{5}{2}\right>&\left|\frac{5}{2},-\frac{1}{2}\right>\\
    \left|\frac{5}{2},-\frac{5}{2}\right>&0&-i\\[1.2em]\left|\frac{5}{2},-\frac{1}{2}\right>&i&0
    \end{pmatrix},
    \label{yzpr}
\end{align}
where $\langle r^2\rangle$ is mean-square radius for $4f$ orbitals. The form of the Hamiltonian is similar to what we obtained for the $E_{1g}$ mode in the main text, except for the fact that the $E_{2g}$ modes couple the electronic orbitals $\ket{\pm5/2}$ with $\ket{\pm1/2}$. This coupling results in a significant phonon Zeeman effect and leads to chiral phonons  as shown in Fig.~~\ref{CeCl3phonon} (c).


\section{Orbital configuration and orbital-lattice coupling in CoTiO$_3$}
\label{AppendixC}
The Co$^{2+}$ is a $3d^3$ system with three unpaired spins and for a free ion, Hund's coupling dictates that the ground state manifold has $L=3, S=3/2$ and thus 28 degenerate states. These twenty-eight states can be obtained from linear combinations of the following seven states ($m_S=+3/2 $):
\begin{align}
    &\ket{L=3, m_L=3}=\dddi{2}{1}{0}\ket{0}\label{d1}\\
    &\ket{L=3, m_L=2}=\dddi{2}{1}{-1}\ket{0}\\
    &\ket{L=3, m_L=1}=\frac{1}{\sqrt{5}}\left(\sqrt{3}\dddi{2}{0}{-1}+\sqrt{2}\dddi{2}{1}{-2}\right)\ket{0}\\
    &\ket{L=3, m_L=0}=\frac{1}{\sqrt{5}}\left(\dddi{1}{0}{-1}+2\dddi{2}{0}{-2}\right)\ket{0}\\
    &\ket{L=3, m_L=-1}=-\frac{1}{\sqrt{5}}\left(\sqrt{3}\dddi{-2}{0}{1}+\sqrt{2}\dddi{-2}{-1}{2}\right)\ket{0}\\
    &\ket{L=3, m_L=-2}=-\dddi{-2}{-1}{1}\ket{0}\\
    &\ket{L=3, m_L=-3}=-\dddi{-2}{-1}{0}\ket{0}
    \label{d7}
\end{align}
where $\di{j}$ creates a $d$-orbital state:
\begin{equation}
    \di{j}\ket{0}\equiv \left|{L=2,m_l=j,S=\frac{1}{2},m_s=+\frac{1}{2}}\right>
\end{equation}

These seven states are split by the octahedral CEF of the ligand ions. This CEF effect can be incorporated by introducing the following term to the Hamiltonian:
\begin{equation}
    H_{Oh}=\Delta_0 \left(\mathcal{P}_{T_{2g}}-\frac{3}{2}\mathcal{P}_{E_g}\right)
\end{equation}
where $\mathcal{P}_{T_{2g}/E_g}$ is the projection operator on the $T_{2g}$ and $E_g$ $d$ orbitals. In terms of creation and annihilation operators for $d$ orbitals, this term can be expressed as follows, 
\begin{equation}
    H_{Oh}=\Delta_0\sum_{\sigma=\uparrow,\downarrow}\left(\dci{xy}\dai{xy}+\dci{yz}\dai{yz}+\dci{xz}\dai{xz}-\frac{3}{2}\dci{x^2-y^2}\dai{x^2-y^2}-\frac{3}{2}\dci{z^2}\dai{z^2}\right),
\end{equation}
which splits the seven-dimensional Hilbert space into four different manifolds. The ground-state sector is spanned by the following three states:
\begin{align}
    &\ket{\tilde{1}}=\sqrt{\frac{3}{8}}\ket{3,1}+\sqrt{\frac{5}{8}}\ket{3,-3}\\
   & \ket{\tilde{0}}=-\ket{3,0}\\
    &\ket{-\tilde{1}}=\sqrt{\frac{3}{8}}\ket{3,-1}+\sqrt{\frac{5}{8}}\ket{3,3},
\end{align}
which we denote by $T_{1g}$ in Fig.~\ref{CrystalCTO} (b),
and the other three-sectors (irrelevant for our calculations) are given by:
\begin{align}
    &\sqrt{\frac{5}{8}}\ket{3,1}-\sqrt{\frac{3}{8}}\ket{3,-3}\\
    &\sqrt{\frac{5}{8}}\ket{3,-1}-\sqrt{\frac{3}{8}}\ket{3,3},
\end{align}
\begin{align}
    \sqrt{\frac{1}{2}}\ket{3,2}+\sqrt{\frac{1}{2}}\ket{3,-2},
\end{align}
and
\begin{align}
    \sqrt{\frac{1}{2}}\ket{3,2}-\sqrt{\frac{1}{2}}\ket{3,-2}.
\end{align}
As the crystal field splitting arising from the octahedral field in this case is of the order of $1$~eV, we are going to focus only on the ground-state sector which can be represented as an effective angular momentum, $l_\mathrm{eff}=1$, sector with $\vec{l}=-\frac{2}{3}\vec{L}$. 

\begin{figure}
    \centering
    \includegraphics[scale=0.3]{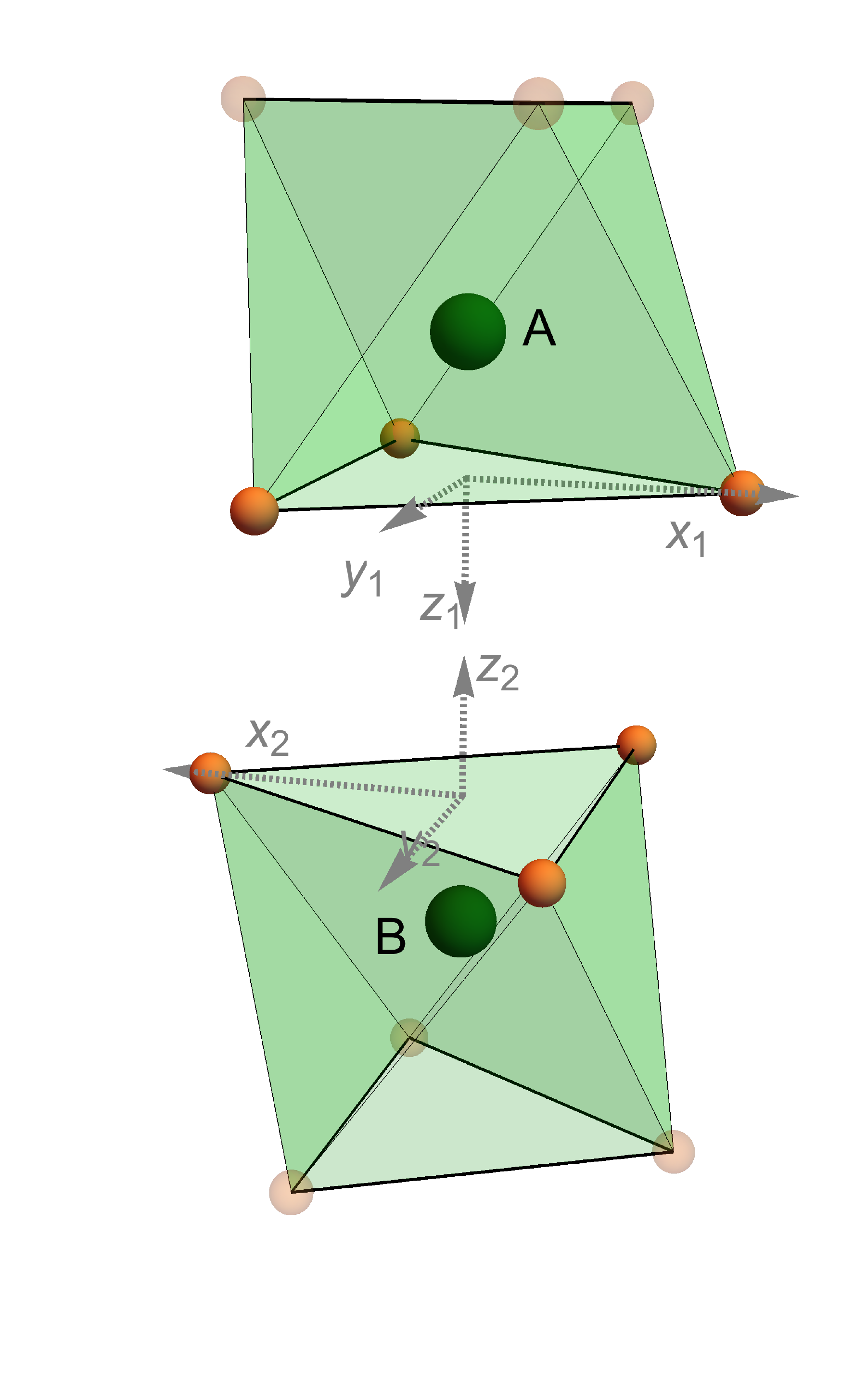}
    \caption{Local coordinate system around each of the two Co$^{2+}$ sites, $A$ and $B$, of the rhombohedral unit cell.}
    \label{local}
\end{figure}
Next, we take into account the effect of spin-orbit coupling,
\begin{equation}
    H_{SO}=\frac{3}{2}\lambda \vec{l}\cdot\vec{S},
\end{equation}
which splits this $l_\mathrm{eff}=1$ manifold further into
\begin{itemize}
    \item $j_{\mathrm{eff}}=5/2$ manifold: 
\begin{align}
    &\left|\frac{5}{2},\frac{5}{2}\right>=\ket{m_l=\tilde{1},m_s=\frac{3}{2}}\\
     &\left|\frac{5}{2},\frac{3}{2}\right>=\sqrt{\frac{2}{5}}\ket{m_l=\tilde{0},m_s=\frac{3}{2}}+\sqrt{\frac{3}{5}}\ket{m_l=\tilde{1},m_s=\frac{1}{2}}\\
     &\left|\frac{5}{2},\frac{1}{2}\right>=\frac{1}{\sqrt{10}}\left(\ket{m_l=-\tilde{1},m_s=\frac{3}{2}}+\sqrt{6}\ket{m_l=\tilde{0},m_s=\frac{1}{2}}+\sqrt{3}\ket{m_l=\tilde{1},m_s=-\frac{1}{2}}\right)\\
     &\left|\frac{5}{2},-\frac{1}{2}\right>=\frac{1}{\sqrt{10}}\left(\ket{m_l=\tilde{1},m_s=-\frac{3}{2}}+\sqrt{6}\ket{m_l=\tilde{0},m_s=-\frac{1}{2}}+\sqrt{3}\ket{m_l=-\tilde{1},m_s=\frac{1}{2}}\right)\\
     &\left|\frac{5}{2},-\frac{3}{2}\right>=\sqrt{\frac{2}{5}}\ket{m_l=\tilde{0},m_s=-\frac{3}{2}}+\sqrt{\frac{3}{5}}\ket{m_l=-\tilde{1},m_s=-\frac{1}{2}}\\
   & \left|\frac{5}{2},-\frac{5}{2}\right>=\ket{m_l=-\tilde{1},m_s=-\frac{3}{2}}
\end{align}
\item $j_\mathrm{eff}=3/2$ manifold:
\begin{align}
     &\left|\frac{3}{2},\frac{3}{2}\right>=\sqrt{\frac{3}{5}}\ket{m_l=\tilde{0},m_s=\frac{3}{2}}-\sqrt{\frac{2}{5}}\ket{m_l=\tilde{1},m_s=\frac{1}{2}}\\
     &\left|\frac{3}{2},\frac{1}{2}\right>=\frac{1}{\sqrt{15}}\left(\sqrt{6}\ket{m_l=-\tilde{1},m_s=\frac{3}{2}}+\ket{m_l=\tilde{0},m_s=\frac{1}{2}}-\sqrt{8}\ket{m_l=\tilde{1},m_s=-\frac{1}{2}}\right)\\
     &\left|\frac{3}{2},-\frac{1}{2}\right>=\frac{1}{\sqrt{15}}\left(\sqrt{6}\ket{m_l=\tilde{1},m_s=-\frac{3}{2}}+\ket{m_l=\tilde{0},m_s=-\frac{1}{2}}-\sqrt{8}\ket{m_l=-\tilde{1},m_s=\frac{1}{2}}\right)\\
     &\left|\frac{3}{2},-\frac{3}{2}\right>=\sqrt{\frac{3}{5}}\ket{m_l=\tilde{0},m_s=-\frac{3}{2}}-\sqrt{\frac{2}{5}}\ket{m_l=-\tilde{1},m_s=-\frac{1}{2}}
\end{align}
\item $j_\mathrm{eff}=1/2$ manifold:
\begin{align}
     &\left|\frac{1}{2},\frac{1}{2}\right>=\frac{1}{\sqrt{6}}\left(\sqrt{3}\ket{m_l=-\tilde{1},m_s=\frac{3}{2}}-\sqrt{2}\ket{m_l=\tilde{0},m_s=\frac{1}{2}}+\ket{m_l=\tilde{1},m_s=-\frac{1}{2}}\right)\\
     &\left|\frac{1}{2},-\frac{1}{2}\right>=\frac{1}{\sqrt{6}}\left(\sqrt{3}\ket{m_l=\tilde{1},m_s=-\frac{3}{2}}-\sqrt{2}\ket{m_l=\tilde{0},m_s=-\frac{1}{2}}+\ket{m_l=-\tilde{1},m_s=\frac{1}{2}}\right)
\end{align}
\end{itemize}
\textbf{Trigonal distortion} :
In CoTiO$_3$, the octahedral cage is trigonally distorted which reduces the Co$^{2+}$ site symmetry from $O_h$  to $C_3$. 
This distortion is significant and it introduces a perturbation of the form $\delta {L_z^2}_{tr}$ along the $z$ direction of the trigonal coordinate system, which splits the $j_\mathrm{eff}=3/2$ and $j_\mathrm{eff}=5/2$ manifolds further, according to the expectation value of $j_z$, where $z$ axis is parallel (antiparallel) to $c$ axis for site $A$ ($B$) of the Co$^{2+}$ ions, as shown in Fig.~\ref{local}. The two lower manifolds can still be characterized by $m_j=\pm1/2$ and $m_j=\pm3/2$ and are predominantly composed of $j_\mathrm{eff}=1/2$ and $j_\mathrm{eff}=3/2$, respectively. 

\textbf{Orbital lattice coupling}: In order to calculate the orbital-lattice coupling for Eq.~\eqref{CTOelpha}-\eqref{CTOelphb}, we first express the electronic states $\ket{1/2,\pm1/2}$ and $\ket{3/2,\pm3/2}$ in terms of constituent $d$ orbital states in Eq.~\eqref{d1}-\eqref{d7} which in turn are expressed in the form of Spherical harmonics and this gives
\begin{equation}
    H_1(xz)=r_0^2\frac{1}{70\sqrt{60}}\begin{pmatrix}
    &\left|\frac{1}{2},\pm\frac{1}{2}\right>&\left|\frac{3}{2},\pm\frac{3}{2}\right>\\
    \left|\frac{1}{2},\pm\frac{1}{2}\right>&0&\pm1\\[1.2em]\left|\frac{3}{2},\pm\frac{3}{2}\right>&\pm1&0
    \end{pmatrix}
\end{equation}
\begin{equation}
    H_1(yz)=r_0^2\frac{1}{70\sqrt{60}}\begin{pmatrix}
    &\left|\frac{1}{2},\pm\frac{1}{2}\right>&\left|\frac{3}{2},\pm\frac{3}{2}\right>\\
    \left|\frac{1}{2},\pm\frac{1}{2}\right>&0&i\\[1.2em]\left|\frac{3}{2},\pm\frac{3}{2}\right>&-i&0
    \end{pmatrix}
\end{equation}
for different perturbations in the Coulomb potential.


\bibliography{dominik.bib}

\begin{thebibliography}{63}%
\makeatletter
\providecommand \@ifxundefined [1]{%
 \@ifx{#1\undefined}
}%
\providecommand \@ifnum [1]{%
 \ifnum #1\expandafter \@firstoftwo
 \else \expandafter \@secondoftwo
 \fi
}%
\providecommand \@ifx [1]{%
 \ifx #1\expandafter \@firstoftwo
 \else \expandafter \@secondoftwo
 \fi
}%
\providecommand \natexlab [1]{#1}%
\providecommand \enquote  [1]{``#1''}%
\providecommand \bibnamefont  [1]{#1}%
\providecommand \bibfnamefont [1]{#1}%
\providecommand \citenamefont [1]{#1}%
\providecommand \href@noop [0]{\@secondoftwo}%
\providecommand \href [0]{\begingroup \@sanitize@url \@href}%
\providecommand \@href[1]{\@@startlink{#1}\@@href}%
\providecommand \@@href[1]{\endgroup#1\@@endlink}%
\providecommand \@sanitize@url [0]{\catcode `\\12\catcode `\$12\catcode
  `\&12\catcode `\#12\catcode `\^12\catcode `\_12\catcode `\%12\relax}%
\providecommand \@@startlink[1]{}%
\providecommand \@@endlink[0]{}%
\providecommand \url  [0]{\begingroup\@sanitize@url \@url }%
\providecommand \@url [1]{\endgroup\@href {#1}{\urlprefix }}%
\providecommand \urlprefix  [0]{URL }%
\providecommand \Eprint [0]{\href }%
\providecommand \doibase [0]{https://doi.org/}%
\providecommand \selectlanguage [0]{\@gobble}%
\providecommand \bibinfo  [0]{\@secondoftwo}%
\providecommand \bibfield  [0]{\@secondoftwo}%
\providecommand \translation [1]{[#1]}%
\providecommand \BibitemOpen [0]{}%
\providecommand \bibitemStop [0]{}%
\providecommand \bibitemNoStop [0]{.\EOS\space}%
\providecommand \EOS [0]{\spacefactor3000\relax}%
\providecommand \BibitemShut  [1]{\csname bibitem#1\endcsname}%
\let\auto@bib@innerbib\@empty
\bibitem [{\citenamefont {Zhang}\ and\ \citenamefont
  {Averitt}(2014)}]{Zhang2014}%
  \BibitemOpen
  \bibfield  {author} {\bibinfo {author} {\bibfnamefont {J.}~\bibnamefont
  {Zhang}}\ and\ \bibinfo {author} {\bibfnamefont {R.}~\bibnamefont
  {Averitt}},\ }\bibfield  {title} {\bibinfo {title} {{Dynamics and Control in
  Complex Transition Metal Oxides}},\ }\href
  {https://doi.org/10.1146/annurev-matsci-070813-113258} {\bibfield  {journal}
  {\bibinfo  {journal} {Annu. Rev. Mater. Res.}\ }\textbf {\bibinfo {volume}
  {44}},\ \bibinfo {pages} {19} (\bibinfo {year} {2014})}\BibitemShut {NoStop}%
\bibitem [{\citenamefont {Zhu}\ \emph {et~al.}(2018)\citenamefont {Zhu},
  \citenamefont {Yi}, \citenamefont {Li}, \citenamefont {Xiao}, \citenamefont
  {Zhang}, \citenamefont {Yang}, \citenamefont {Kaindl}, \citenamefont {Li},
  \citenamefont {Wang},\ and\ \citenamefont {Zhang}}]{Zhu2018}%
  \BibitemOpen
  \bibfield  {author} {\bibinfo {author} {\bibfnamefont {H.}~\bibnamefont
  {Zhu}}, \bibinfo {author} {\bibfnamefont {J.}~\bibnamefont {Yi}}, \bibinfo
  {author} {\bibfnamefont {M.-Y.}\ \bibnamefont {Li}}, \bibinfo {author}
  {\bibfnamefont {J.}~\bibnamefont {Xiao}}, \bibinfo {author} {\bibfnamefont
  {L.}~\bibnamefont {Zhang}}, \bibinfo {author} {\bibfnamefont {C.-W.}\
  \bibnamefont {Yang}}, \bibinfo {author} {\bibfnamefont {R.~A.}\ \bibnamefont
  {Kaindl}}, \bibinfo {author} {\bibfnamefont {L.-J.}\ \bibnamefont {Li}},
  \bibinfo {author} {\bibfnamefont {Y.}~\bibnamefont {Wang}},\ and\ \bibinfo
  {author} {\bibfnamefont {X.}~\bibnamefont {Zhang}},\ }\bibfield  {title}
  {\bibinfo {title} {{Observation of chiral phonons}},\ }\href
  {https://doi.org/10.1126/science.aar2711} {\bibfield  {journal} {\bibinfo
  {journal} {Science}\ }\textbf {\bibinfo {volume} {582}},\ \bibinfo {pages}
  {579} (\bibinfo {year} {2018})}\BibitemShut {NoStop}%
\bibitem [{\citenamefont {Streib}(2021)}]{Streib2021}%
  \BibitemOpen
  \bibfield  {author} {\bibinfo {author} {\bibfnamefont {S.}~\bibnamefont
  {Streib}},\ }\bibfield  {title} {\bibinfo {title} {{Difference between
  angular momentum and pseudoangular momentum}},\ }\href
  {https://doi.org/10.1103/PhysRevB.103.L100409} {\bibfield  {journal}
  {\bibinfo  {journal} {Phys. Rev. B}\ }\textbf {\bibinfo {volume} {103}},\
  \bibinfo {pages} {L100409} (\bibinfo {year} {2021})}\BibitemShut {NoStop}%
\bibitem [{\citenamefont {Ishito}\ \emph {et~al.}(2023)\citenamefont {Ishito},
  \citenamefont {Mao}, \citenamefont {Kousaka}, \citenamefont {Togawa},
  \citenamefont {Iwasaki}, \citenamefont {Zhang}, \citenamefont {Murakami},
  \citenamefont {Kishine},\ and\ \citenamefont {Satoh}}]{ishito2023truly}%
  \BibitemOpen
  \bibfield  {author} {\bibinfo {author} {\bibfnamefont {K.}~\bibnamefont
  {Ishito}}, \bibinfo {author} {\bibfnamefont {H.}~\bibnamefont {Mao}},
  \bibinfo {author} {\bibfnamefont {Y.}~\bibnamefont {Kousaka}}, \bibinfo
  {author} {\bibfnamefont {Y.}~\bibnamefont {Togawa}}, \bibinfo {author}
  {\bibfnamefont {S.}~\bibnamefont {Iwasaki}}, \bibinfo {author} {\bibfnamefont
  {T.}~\bibnamefont {Zhang}}, \bibinfo {author} {\bibfnamefont
  {S.}~\bibnamefont {Murakami}}, \bibinfo {author} {\bibfnamefont {J.-i.}\
  \bibnamefont {Kishine}},\ and\ \bibinfo {author} {\bibfnamefont
  {T.}~\bibnamefont {Satoh}},\ }\bibfield  {title} {\bibinfo {title} {Truly
  chiral phonons in $\alpha$-hgs},\ }\href
  {https://www.nature.com/articles/s41567-022-01790-x} {\bibfield  {journal}
  {\bibinfo  {journal} {Nature Physics}\ }\textbf {\bibinfo {volume} {19}},\
  \bibinfo {pages} {35} (\bibinfo {year} {2023})}\BibitemShut {NoStop}%
\bibitem [{\citenamefont {Ueda}\ \emph {et~al.}(2023)\citenamefont {Ueda},
  \citenamefont {Garc{\'\i}a-Fern{\'a}ndez}, \citenamefont {Agrestini},
  \citenamefont {Romao}, \citenamefont {van~den Brink}, \citenamefont
  {Spaldin}, \citenamefont {Zhou},\ and\ \citenamefont
  {Staub}}]{ueda2023chiral}%
  \BibitemOpen
  \bibfield  {author} {\bibinfo {author} {\bibfnamefont {H.}~\bibnamefont
  {Ueda}}, \bibinfo {author} {\bibfnamefont {M.}~\bibnamefont
  {Garc{\'\i}a-Fern{\'a}ndez}}, \bibinfo {author} {\bibfnamefont
  {S.}~\bibnamefont {Agrestini}}, \bibinfo {author} {\bibfnamefont {C.~P.}\
  \bibnamefont {Romao}}, \bibinfo {author} {\bibfnamefont {J.}~\bibnamefont
  {van~den Brink}}, \bibinfo {author} {\bibfnamefont {N.~A.}\ \bibnamefont
  {Spaldin}}, \bibinfo {author} {\bibfnamefont {K.-J.}\ \bibnamefont {Zhou}},\
  and\ \bibinfo {author} {\bibfnamefont {U.}~\bibnamefont {Staub}},\ }\bibfield
   {title} {\bibinfo {title} {Chiral phonons in quartz probed by x-rays},\
  }\href {https://www.nature.com/articles/s41586-023-06016-5} {\bibfield
  {journal} {\bibinfo  {journal} {Nature}\ ,\ \bibinfo {pages} {1}} (\bibinfo
  {year} {2023})}\BibitemShut {NoStop}%
\bibitem [{\citenamefont {Rebane}(1983)}]{Rebane1983}%
  \BibitemOpen
  \bibfield  {author} {\bibinfo {author} {\bibfnamefont {Y.~T.}\ \bibnamefont
  {Rebane}},\ }\bibfield  {title} {\bibinfo {title} {{Faraday effect produced
  in the residual-ray region by the magnetic moment of an optical phonon in an
  ionic crystal}},\ }\href
  {http://www.jetp.ras.ru/cgi-bin/e/index/e/57/6/p1356?a=list} {\bibfield
  {journal} {\bibinfo  {journal} {Sov. Phys. JETP}\ }\textbf {\bibinfo {volume}
  {57}},\ \bibinfo {pages} {1356} (\bibinfo {year} {1983})}\BibitemShut
  {NoStop}%
\bibitem [{\citenamefont {Juraschek}\ \emph {et~al.}(2017)\citenamefont
  {Juraschek}, \citenamefont {Fechner}, \citenamefont {Balatsky},\ and\
  \citenamefont {Spaldin}}]{juraschek2:2017}%
  \BibitemOpen
  \bibfield  {author} {\bibinfo {author} {\bibfnamefont {D.~M.}\ \bibnamefont
  {Juraschek}}, \bibinfo {author} {\bibfnamefont {M.}~\bibnamefont {Fechner}},
  \bibinfo {author} {\bibfnamefont {A.~V.}\ \bibnamefont {Balatsky}},\ and\
  \bibinfo {author} {\bibfnamefont {N.~A.}\ \bibnamefont {Spaldin}},\
  }\bibfield  {title} {\bibinfo {title} {{Dynamical multiferroicity}},\ }\href
  {https://doi.org/10.1103/PhysRevMaterials.1.014401} {\bibfield  {journal}
  {\bibinfo  {journal} {Phys. Rev. Mater.}\ }\textbf {\bibinfo {volume} {1}},\
  \bibinfo {pages} {014401} (\bibinfo {year} {2017})}\BibitemShut {NoStop}%
\bibitem [{\citenamefont {Juraschek}\ and\ \citenamefont
  {Spaldin}(2019)}]{Juraschek2019}%
  \BibitemOpen
  \bibfield  {author} {\bibinfo {author} {\bibfnamefont {D.~M.}\ \bibnamefont
  {Juraschek}}\ and\ \bibinfo {author} {\bibfnamefont {N.~A.}\ \bibnamefont
  {Spaldin}},\ }\bibfield  {title} {\bibinfo {title} {{Orbital magnetic moments
  of phonons}},\ }\href {https://doi.org/10.1103/PhysRevMaterials.3.064405}
  {\bibfield  {journal} {\bibinfo  {journal} {Phys. Rev. Mater.}\ }\textbf
  {\bibinfo {volume} {3}},\ \bibinfo {pages} {064405} (\bibinfo {year}
  {2019})}\BibitemShut {NoStop}%
\bibitem [{\citenamefont {Xiong}\ \emph {et~al.}(2022)\citenamefont {Xiong},
  \citenamefont {Chen}, \citenamefont {Ma},\ and\ \citenamefont
  {Zhang}}]{Xiong2022}%
  \BibitemOpen
  \bibfield  {author} {\bibinfo {author} {\bibfnamefont {G.}~\bibnamefont
  {Xiong}}, \bibinfo {author} {\bibfnamefont {H.}~\bibnamefont {Chen}},
  \bibinfo {author} {\bibfnamefont {D.}~\bibnamefont {Ma}},\ and\ \bibinfo
  {author} {\bibfnamefont {L.}~\bibnamefont {Zhang}},\ }\bibfield  {title}
  {\bibinfo {title} {{Effective magnetic fields induced by chiral phonons}},\
  }\href {https://doi.org/10.1103/PhysRevB.106.144302} {\bibfield  {journal}
  {\bibinfo  {journal} {Phys. Rev. B}\ }\textbf {\bibinfo {volume} {106}},\
  \bibinfo {pages} {144302} (\bibinfo {year} {2022})}\BibitemShut {NoStop}%
\bibitem [{\citenamefont {Anastassakis}\ and\ \citenamefont
  {Burstein}(1971)}]{Anastassakis1971}%
  \BibitemOpen
  \bibfield  {author} {\bibinfo {author} {\bibfnamefont {E.}~\bibnamefont
  {Anastassakis}}\ and\ \bibinfo {author} {\bibfnamefont {E.}~\bibnamefont
  {Burstein}},\ }\bibfield  {title} {\bibinfo {title} {{Morphic effects
  II-effects of external forces on the frequencies of the $\mathbf{q} \approx
  0$ optical phonons}},\ }\href {https://doi.org/10.1016/0022-3697(71)90005-9}
  {\bibfield  {journal} {\bibinfo  {journal} {J.~Phys. Chem. Solids}\ }\textbf
  {\bibinfo {volume} {32}},\ \bibinfo {pages} {563} (\bibinfo {year}
  {1971})}\BibitemShut {NoStop}%
\bibitem [{\citenamefont {Holz}(1972)}]{Holz1972}%
  \BibitemOpen
  \bibfield  {author} {\bibinfo {author} {\bibfnamefont {A.}~\bibnamefont
  {Holz}},\ }\bibfield  {title} {\bibinfo {title} {{Phonons in a Strong Static
  Magnetic Field}},\ }\href {https://doi.org/10.1007/BF02735509} {\bibfield
  {journal} {\bibinfo  {journal} {IL Nuovo Cimento}\ }\textbf {\bibinfo
  {volume} {9}},\ \bibinfo {pages} {83} (\bibinfo {year} {1972})}\BibitemShut
  {NoStop}%
\bibitem [{\citenamefont {Strohm}\ \emph {et~al.}(2005)\citenamefont {Strohm},
  \citenamefont {Rikken},\ and\ \citenamefont {Wyder}}]{strohm:2005}%
  \BibitemOpen
  \bibfield  {author} {\bibinfo {author} {\bibfnamefont {C.}~\bibnamefont
  {Strohm}}, \bibinfo {author} {\bibfnamefont {G.~L. J.~A.}\ \bibnamefont
  {Rikken}},\ and\ \bibinfo {author} {\bibfnamefont {P.}~\bibnamefont
  {Wyder}},\ }\bibfield  {title} {\bibinfo {title} {{Phenomenological Evidence
  for the Phonon Hall Effect}},\ }\href
  {https://doi.org/10.1103/PhysRevLett.95.155901} {\bibfield  {journal}
  {\bibinfo  {journal} {Phys. Rev. Lett.}\ }\textbf {\bibinfo {volume} {95}},\
  \bibinfo {pages} {155901} (\bibinfo {year} {2005})}\BibitemShut {NoStop}%
\bibitem [{\citenamefont {Sheng}\ \emph {et~al.}(2006)\citenamefont {Sheng},
  \citenamefont {Sheng},\ and\ \citenamefont {Ting}}]{sheng:2006}%
  \BibitemOpen
  \bibfield  {author} {\bibinfo {author} {\bibfnamefont {L.}~\bibnamefont
  {Sheng}}, \bibinfo {author} {\bibfnamefont {D.~N.}\ \bibnamefont {Sheng}},\
  and\ \bibinfo {author} {\bibfnamefont {C.~S.}\ \bibnamefont {Ting}},\
  }\bibfield  {title} {\bibinfo {title} {{Theory of the Phonon Hall Effect in
  Paramagnetic Dielectrics}},\ }\href
  {https://doi.org/10.1103/PhysRevLett.96.155901} {\bibfield  {journal}
  {\bibinfo  {journal} {Phys. Rev. Lett.}\ }\textbf {\bibinfo {volume} {96}},\
  \bibinfo {pages} {155901} (\bibinfo {year} {2006})}\BibitemShut {NoStop}%
\bibitem [{\citenamefont {Zhang}\ \emph {et~al.}(2010)\citenamefont {Zhang},
  \citenamefont {Ren}, \citenamefont {Wang},\ and\ \citenamefont
  {Li}}]{Zhang2010}%
  \BibitemOpen
  \bibfield  {author} {\bibinfo {author} {\bibfnamefont {L.}~\bibnamefont
  {Zhang}}, \bibinfo {author} {\bibfnamefont {J.}~\bibnamefont {Ren}}, \bibinfo
  {author} {\bibfnamefont {J.~S.}\ \bibnamefont {Wang}},\ and\ \bibinfo
  {author} {\bibfnamefont {B.}~\bibnamefont {Li}},\ }\bibfield  {title}
  {\bibinfo {title} {{Topological Nature of the Phonon Hall Effect}},\ }\href
  {https://doi.org/10.1103/PhysRevLett.105.225901} {\bibfield  {journal}
  {\bibinfo  {journal} {Phys. Rev. Lett.}\ }\textbf {\bibinfo {volume} {105}},\
  \bibinfo {pages} {225901} (\bibinfo {year} {2010})}\BibitemShut {NoStop}%
\bibitem [{\citenamefont {Grissonnanche}\ \emph {et~al.}(2019)\citenamefont
  {Grissonnanche}, \citenamefont {Legros}, \citenamefont {Badoux},
  \citenamefont {Lefran{\c{c}}ois}, \citenamefont {Zatko}, \citenamefont
  {Lizaire}, \citenamefont {Lalibert{\'{e}}}, \citenamefont {Gourgout},
  \citenamefont {Zhou}, \citenamefont {Pyon}, \citenamefont {Takayama},
  \citenamefont {Takagi}, \citenamefont {Ono}, \citenamefont {Doiron-Leyraud},\
  and\ \citenamefont {Taillefer}}]{Grissonnanche2019}%
  \BibitemOpen
  \bibfield  {author} {\bibinfo {author} {\bibfnamefont {G.}~\bibnamefont
  {Grissonnanche}}, \bibinfo {author} {\bibfnamefont {A.}~\bibnamefont
  {Legros}}, \bibinfo {author} {\bibfnamefont {S.}~\bibnamefont {Badoux}},
  \bibinfo {author} {\bibfnamefont {E.}~\bibnamefont {Lefran{\c{c}}ois}},
  \bibinfo {author} {\bibfnamefont {V.}~\bibnamefont {Zatko}}, \bibinfo
  {author} {\bibfnamefont {M.}~\bibnamefont {Lizaire}}, \bibinfo {author}
  {\bibfnamefont {F.}~\bibnamefont {Lalibert{\'{e}}}}, \bibinfo {author}
  {\bibfnamefont {A.}~\bibnamefont {Gourgout}}, \bibinfo {author}
  {\bibfnamefont {J.~S.}\ \bibnamefont {Zhou}}, \bibinfo {author}
  {\bibfnamefont {S.}~\bibnamefont {Pyon}}, \bibinfo {author} {\bibfnamefont
  {T.}~\bibnamefont {Takayama}}, \bibinfo {author} {\bibfnamefont
  {H.}~\bibnamefont {Takagi}}, \bibinfo {author} {\bibfnamefont
  {S.}~\bibnamefont {Ono}}, \bibinfo {author} {\bibfnamefont {N.}~\bibnamefont
  {Doiron-Leyraud}},\ and\ \bibinfo {author} {\bibfnamefont {L.}~\bibnamefont
  {Taillefer}},\ }\bibfield  {title} {\bibinfo {title} {{Giant thermal Hall
  conductivity in the pseudogap phase of cuprate superconductors}},\ }\href
  {https://doi.org/10.1038/s41586-019-1375-0} {\bibfield  {journal} {\bibinfo
  {journal} {Nature}\ }\textbf {\bibinfo {volume} {571}},\ \bibinfo {pages}
  {376} (\bibinfo {year} {2019})}\BibitemShut {NoStop}%
\bibitem [{\citenamefont {Grissonnanche}\ \emph {et~al.}(2020)\citenamefont
  {Grissonnanche}, \citenamefont {Th{\'{e}}riault}, \citenamefont {Gourgout},
  \citenamefont {Boulanger}, \citenamefont {Lefran{\c{c}}ois}, \citenamefont
  {Ataei}, \citenamefont {Lalibert{\'{e}}}, \citenamefont {Dion}, \citenamefont
  {Zhou}, \citenamefont {Pyon}, \citenamefont {Takayama}, \citenamefont
  {Takagi}, \citenamefont {Doiron-Leyraud},\ and\ \citenamefont
  {Taillefer}}]{Grissonnanche2020}%
  \BibitemOpen
  \bibfield  {author} {\bibinfo {author} {\bibfnamefont {G.}~\bibnamefont
  {Grissonnanche}}, \bibinfo {author} {\bibfnamefont {S.}~\bibnamefont
  {Th{\'{e}}riault}}, \bibinfo {author} {\bibfnamefont {A.}~\bibnamefont
  {Gourgout}}, \bibinfo {author} {\bibfnamefont {M.~E.}\ \bibnamefont
  {Boulanger}}, \bibinfo {author} {\bibfnamefont {E.}~\bibnamefont
  {Lefran{\c{c}}ois}}, \bibinfo {author} {\bibfnamefont {A.}~\bibnamefont
  {Ataei}}, \bibinfo {author} {\bibfnamefont {F.}~\bibnamefont
  {Lalibert{\'{e}}}}, \bibinfo {author} {\bibfnamefont {M.}~\bibnamefont
  {Dion}}, \bibinfo {author} {\bibfnamefont {J.~S.}\ \bibnamefont {Zhou}},
  \bibinfo {author} {\bibfnamefont {S.}~\bibnamefont {Pyon}}, \bibinfo {author}
  {\bibfnamefont {T.}~\bibnamefont {Takayama}}, \bibinfo {author}
  {\bibfnamefont {H.}~\bibnamefont {Takagi}}, \bibinfo {author} {\bibfnamefont
  {N.}~\bibnamefont {Doiron-Leyraud}},\ and\ \bibinfo {author} {\bibfnamefont
  {L.}~\bibnamefont {Taillefer}},\ }\bibfield  {title} {\bibinfo {title}
  {{Chiral phonons in the pseudogap phase of cuprates}},\ }\href
  {https://doi.org/10.1038/s41567-020-0965-y} {\bibfield  {journal} {\bibinfo
  {journal} {Nat. Phys.}\ }\textbf {\bibinfo {volume} {16}},\ \bibinfo {pages}
  {1108} (\bibinfo {year} {2020})}\BibitemShut {NoStop}%
\bibitem [{\citenamefont {Park}\ and\ \citenamefont
  {Yang}(2020)}]{Park2020_phononhall}%
  \BibitemOpen
  \bibfield  {author} {\bibinfo {author} {\bibfnamefont {S.}~\bibnamefont
  {Park}}\ and\ \bibinfo {author} {\bibfnamefont {B.~J.}\ \bibnamefont
  {Yang}},\ }\bibfield  {title} {\bibinfo {title} {{Phonon Angular Momentum
  Hall Effect}},\ }\href {https://doi.org/10.1021/acs.nanolett.0c03220}
  {\bibfield  {journal} {\bibinfo  {journal} {Nano Lett.}\ }\textbf {\bibinfo
  {volume} {20}},\ \bibinfo {pages} {7694} (\bibinfo {year}
  {2020})}\BibitemShut {NoStop}%
\bibitem [{\citenamefont {Saito}\ \emph {et~al.}(2019)\citenamefont {Saito},
  \citenamefont {Misaki}, \citenamefont {Ishizuka},\ and\ \citenamefont
  {Nagaosa}}]{Saito2019}%
  \BibitemOpen
  \bibfield  {author} {\bibinfo {author} {\bibfnamefont {T.}~\bibnamefont
  {Saito}}, \bibinfo {author} {\bibfnamefont {K.}~\bibnamefont {Misaki}},
  \bibinfo {author} {\bibfnamefont {H.}~\bibnamefont {Ishizuka}},\ and\
  \bibinfo {author} {\bibfnamefont {N.}~\bibnamefont {Nagaosa}},\ }\bibfield
  {title} {\bibinfo {title} {{Berry Phase of Phonons and Thermal Hall Effect in
  Nonmagnetic Insulators}},\ }\href
  {https://doi.org/10.1103/PhysRevLett.123.255901} {\bibfield  {journal}
  {\bibinfo  {journal} {Physical Review Letters}\ }\textbf {\bibinfo {volume}
  {123}},\ \bibinfo {pages} {255901} (\bibinfo {year} {2019})}\BibitemShut
  {NoStop}%
\bibitem [{\citenamefont {Flebus}\ and\ \citenamefont
  {MacDonald}(2021)}]{Flebus2021}%
  \BibitemOpen
  \bibfield  {author} {\bibinfo {author} {\bibfnamefont {B.}~\bibnamefont
  {Flebus}}\ and\ \bibinfo {author} {\bibfnamefont {A.~H.}\ \bibnamefont
  {MacDonald}},\ }\bibfield  {title} {\bibinfo {title} {{Charged Defects and
  Phonon Hall Effects in Ionic Crystals}},\ }\href
  {http://arxiv.org/abs/2106.13889} {\bibfield  {journal} {\bibinfo  {journal}
  {arXiv:2106.13889}\ } (\bibinfo {year} {2021})}\BibitemShut {NoStop}%
\bibitem [{\citenamefont {Flebus}\ and\ \citenamefont
  {MacDonald}(2022)}]{Flebus2022}%
  \BibitemOpen
  \bibfield  {author} {\bibinfo {author} {\bibfnamefont {B.}~\bibnamefont
  {Flebus}}\ and\ \bibinfo {author} {\bibfnamefont {A.~H.}\ \bibnamefont
  {MacDonald}},\ }\bibfield  {title} {\bibinfo {title} {{The phonon Hall
  viscosity of ionic crystals}},\ }\href {http://arxiv.org/abs/2205.13666}
  {\bibfield  {journal} {\bibinfo  {journal} {arXiv:2205.13666}\ } (\bibinfo
  {year} {2022})}\BibitemShut {NoStop}%
\bibitem [{\citenamefont {Nova}\ \emph {et~al.}(2017)\citenamefont {Nova},
  \citenamefont {Cartella}, \citenamefont {Cantaluppi}, \citenamefont
  {F{\"{o}}rst}, \citenamefont {Bossini}, \citenamefont {Mikhaylovskiy},
  \citenamefont {Kimel}, \citenamefont {Merlin},\ and\ \citenamefont
  {Cavalleri}}]{nova:2017}%
  \BibitemOpen
  \bibfield  {author} {\bibinfo {author} {\bibfnamefont {T.~F.}\ \bibnamefont
  {Nova}}, \bibinfo {author} {\bibfnamefont {A.}~\bibnamefont {Cartella}},
  \bibinfo {author} {\bibfnamefont {A.}~\bibnamefont {Cantaluppi}}, \bibinfo
  {author} {\bibfnamefont {M.}~\bibnamefont {F{\"{o}}rst}}, \bibinfo {author}
  {\bibfnamefont {D.}~\bibnamefont {Bossini}}, \bibinfo {author} {\bibfnamefont
  {R.~V.}\ \bibnamefont {Mikhaylovskiy}}, \bibinfo {author} {\bibfnamefont
  {A.~V.}\ \bibnamefont {Kimel}}, \bibinfo {author} {\bibfnamefont
  {R.}~\bibnamefont {Merlin}},\ and\ \bibinfo {author} {\bibfnamefont
  {A.}~\bibnamefont {Cavalleri}},\ }\bibfield  {title} {\bibinfo {title} {{An
  effective magnetic field from optically driven phonons}},\ }\href
  {https://doi.org/10.1038/nphys3925} {\bibfield  {journal} {\bibinfo
  {journal} {Nat. Phys.}\ }\textbf {\bibinfo {volume} {13}},\ \bibinfo {pages}
  {132} (\bibinfo {year} {2017})}\BibitemShut {NoStop}%
\bibitem [{\citenamefont {Juraschek}\ and\ \citenamefont
  {Narang}(2020)}]{Juraschek2020}%
  \BibitemOpen
  \bibfield  {author} {\bibinfo {author} {\bibfnamefont {D.~M.}\ \bibnamefont
  {Juraschek}}\ and\ \bibinfo {author} {\bibfnamefont {P.}~\bibnamefont
  {Narang}},\ }\bibfield  {title} {\bibinfo {title} {{Shaken not strained}},\
  }\href {https://doi.org/10.1038/s41567-020-0937-2} {\bibfield  {journal}
  {\bibinfo  {journal} {Nat. Phys.}\ }\textbf {\bibinfo {volume} {16}},\
  \bibinfo {pages} {900} (\bibinfo {year} {2020})}\BibitemShut {NoStop}%
\bibitem [{\citenamefont {Juraschek}\ \emph {et~al.}(2021)\citenamefont
  {Juraschek}, \citenamefont {Wang},\ and\ \citenamefont
  {Narang}}]{Juraschek2021}%
  \BibitemOpen
  \bibfield  {author} {\bibinfo {author} {\bibfnamefont {D.~M.}\ \bibnamefont
  {Juraschek}}, \bibinfo {author} {\bibfnamefont {D.~S.}\ \bibnamefont
  {Wang}},\ and\ \bibinfo {author} {\bibfnamefont {P.}~\bibnamefont {Narang}},\
  }\bibfield  {title} {\bibinfo {title} {{Sum-frequency excitation of coherent
  magnons}},\ }\href {https://doi.org/10.1103/PhysRevB.103.094407} {\bibfield
  {journal} {\bibinfo  {journal} {Phys. Rev. B}\ }\textbf {\bibinfo {volume}
  {103}},\ \bibinfo {pages} {094407} (\bibinfo {year} {2021})}\BibitemShut
  {NoStop}%
\bibitem [{\citenamefont {Juraschek}\ \emph {et~al.}(2022)\citenamefont
  {Juraschek}, \citenamefont {Neuman},\ and\ \citenamefont
  {Narang}}]{Juraschek2022_giantphonomag}%
  \BibitemOpen
  \bibfield  {author} {\bibinfo {author} {\bibfnamefont {D.~M.}\ \bibnamefont
  {Juraschek}}, \bibinfo {author} {\bibfnamefont {T.}~\bibnamefont {Neuman}},\
  and\ \bibinfo {author} {\bibfnamefont {P.}~\bibnamefont {Narang}},\
  }\bibfield  {title} {\bibinfo {title} {{Giant effective magnetic fields from
  optically driven chiral phonons in 4$f$ paramagnets}},\ }\href
  {https://doi.org/10.1103/PhysRevResearch.4.013129} {\bibfield  {journal}
  {\bibinfo  {journal} {Phys. Rev. Research}\ }\textbf {\bibinfo {volume}
  {4}},\ \bibinfo {pages} {013129} (\bibinfo {year} {2022})}\BibitemShut
  {NoStop}%
\bibitem [{\citenamefont {Geilhufe}\ \emph {et~al.}(2021)\citenamefont
  {Geilhufe}, \citenamefont {Juri{\v{c}}i{\'{c}}}, \citenamefont {Bonetti},
  \citenamefont {Zhu},\ and\ \citenamefont {Balatsky}}]{Geilhufe2021}%
  \BibitemOpen
  \bibfield  {author} {\bibinfo {author} {\bibfnamefont {R.~M.}\ \bibnamefont
  {Geilhufe}}, \bibinfo {author} {\bibfnamefont {V.}~\bibnamefont
  {Juri{\v{c}}i{\'{c}}}}, \bibinfo {author} {\bibfnamefont {S.}~\bibnamefont
  {Bonetti}}, \bibinfo {author} {\bibfnamefont {J.-X.}\ \bibnamefont {Zhu}},\
  and\ \bibinfo {author} {\bibfnamefont {A.~V.}\ \bibnamefont {Balatsky}},\
  }\bibfield  {title} {\bibinfo {title} {{Dynamically induced magnetism in
  KTaO$_3$}},\ }\href {https://doi.org/10.1103/PhysRevResearch.3.L022011}
  {\bibfield  {journal} {\bibinfo  {journal} {Phys. Rev. Research}\ }\textbf
  {\bibinfo {volume} {3}},\ \bibinfo {pages} {L022011} (\bibinfo {year}
  {2021})}\BibitemShut {NoStop}%
\bibitem [{\citenamefont {Geilhufe}(2022)}]{Geilhufe2022}%
  \BibitemOpen
  \bibfield  {author} {\bibinfo {author} {\bibfnamefont {R.~M.}\ \bibnamefont
  {Geilhufe}},\ }\bibfield  {title} {\bibinfo {title} {{Dynamic electron-phonon
  and spin-phonon interactions due to inertia}},\ }\href
  {https://doi.org/10.1103/PhysRevResearch.4.L012004} {\bibfield  {journal}
  {\bibinfo  {journal} {Phys. Rev. Research}\ }\textbf {\bibinfo {volume}
  {4}},\ \bibinfo {pages} {L012004} (\bibinfo {year} {2022})}\BibitemShut
  {NoStop}%
\bibitem [{\citenamefont {Geilhufe}\ and\ \citenamefont
  {Hergert}(2022)}]{Geilhufe2022electronmagmom}%
  \BibitemOpen
  \bibfield  {author} {\bibinfo {author} {\bibfnamefont {R.~M.}\ \bibnamefont
  {Geilhufe}}\ and\ \bibinfo {author} {\bibfnamefont {W.}~\bibnamefont
  {Hergert}},\ }\bibfield  {title} {\bibinfo {title} {{Electron magnetic moment
  of chiral phonons}},\ }\href {https://arxiv.org/abs/2208.05746} {\bibfield
  {journal} {\bibinfo  {journal} {arXiv:2208.05746}\ } (\bibinfo {year}
  {2022})}\BibitemShut {NoStop}%
\bibitem [{\citenamefont {Basini}\ \emph {et~al.}(2022)\citenamefont {Basini},
  \citenamefont {Pancaldi}, \citenamefont {Wehinger}, \citenamefont {Udina},
  \citenamefont {Tadano}, \citenamefont {Hoffmann}, \citenamefont {Balatsky},\
  and\ \citenamefont {Bonetti}}]{Basini2022}%
  \BibitemOpen
  \bibfield  {author} {\bibinfo {author} {\bibfnamefont {M.}~\bibnamefont
  {Basini}}, \bibinfo {author} {\bibfnamefont {M.}~\bibnamefont {Pancaldi}},
  \bibinfo {author} {\bibfnamefont {B.}~\bibnamefont {Wehinger}}, \bibinfo
  {author} {\bibfnamefont {M.}~\bibnamefont {Udina}}, \bibinfo {author}
  {\bibfnamefont {T.}~\bibnamefont {Tadano}}, \bibinfo {author} {\bibfnamefont
  {M.~C.}\ \bibnamefont {Hoffmann}}, \bibinfo {author} {\bibfnamefont {A.~V.}\
  \bibnamefont {Balatsky}},\ and\ \bibinfo {author} {\bibfnamefont
  {S.}~\bibnamefont {Bonetti}},\ }\bibfield  {title} {\bibinfo {title}
  {{Terahertz electric-field driven dynamical multiferroicity in SrTiO$_3$}},\
  }\href {https://arxiv.org/abs/2210.01690} {\bibfield  {journal} {\bibinfo
  {journal} {arXiv:2210.01690}\ } (\bibinfo {year} {2022})}\BibitemShut
  {NoStop}%
\bibitem [{\citenamefont {Davies}\ \emph {et~al.}(2022)\citenamefont {Davies},
  \citenamefont {Fennema}, \citenamefont {Tsukamoto},\ and\ \citenamefont
  {Kirilyuk}}]{Davies2022}%
  \BibitemOpen
  \bibfield  {author} {\bibinfo {author} {\bibfnamefont {C.~S.}\ \bibnamefont
  {Davies}}, \bibinfo {author} {\bibfnamefont {N.}~\bibnamefont {Fennema}},
  \bibinfo {author} {\bibfnamefont {A.}~\bibnamefont {Tsukamoto}},\ and\
  \bibinfo {author} {\bibfnamefont {A.}~\bibnamefont {Kirilyuk}},\ }\bibfield
  {title} {\bibinfo {title} {{Ultrafast Helicity-Dependent Magnetic Switching
  by Optical Phonons Driven at Resonance}},\ }\href
  {https://doi.org/10.1109/IRMMW-THz50927.2022.9895594} {\bibfield  {journal}
  {\bibinfo  {journal} {International Conference on Infrared, Millimeter, and
  Terahertz Waves, IRMMW-THz}\ ,\ \bibinfo {pages} {43035}} (\bibinfo {year}
  {2022})}\BibitemShut {NoStop}%
\bibitem [{\citenamefont {Luo}\ \emph {et~al.}(2023)\citenamefont {Luo},
  \citenamefont {Lin}, \citenamefont {Zhang}, \citenamefont {Chen},
  \citenamefont {Blackert}, \citenamefont {Xu}, \citenamefont {Yakobson},\ and\
  \citenamefont {Zhu}}]{luo2023large}%
  \BibitemOpen
  \bibfield  {author} {\bibinfo {author} {\bibfnamefont {J.}~\bibnamefont
  {Luo}}, \bibinfo {author} {\bibfnamefont {T.}~\bibnamefont {Lin}}, \bibinfo
  {author} {\bibfnamefont {J.}~\bibnamefont {Zhang}}, \bibinfo {author}
  {\bibfnamefont {X.}~\bibnamefont {Chen}}, \bibinfo {author} {\bibfnamefont
  {E.~R.}\ \bibnamefont {Blackert}}, \bibinfo {author} {\bibfnamefont
  {R.}~\bibnamefont {Xu}}, \bibinfo {author} {\bibfnamefont {B.~I.}\
  \bibnamefont {Yakobson}},\ and\ \bibinfo {author} {\bibfnamefont
  {H.}~\bibnamefont {Zhu}},\ }\bibfield  {title} {\bibinfo {title} {Large
  effective magnetic fields from chiral phonons in rare-earth halides},\ }\href
  {https://arxiv.org/pdf/2306.03852.pdf} {\bibfield  {journal} {\bibinfo
  {journal} {arXiv:2306.03852}\ } (\bibinfo {year} {2023})}\BibitemShut
  {NoStop}%
\bibitem [{\citenamefont {Ceresoli}(2002)}]{ceresoli:2002}%
  \BibitemOpen
  \bibfield  {author} {\bibinfo {author} {\bibfnamefont {D.}~\bibnamefont
  {Ceresoli}},\ }\href
  {https://iris.sissa.it/handle/20.500.11767/4017#.W826MRx9jAI} {\bibinfo
  {title} {{Berry Phase Calculations of the Rotational and Pseudorotational
  g-factor in Molecules and Solids}}},\ \bibinfo {howpublished} {PhD thesis}
  (\bibinfo {year} {2002})\BibitemShut {NoStop}%
\bibitem [{\citenamefont {Hamada}\ \emph {et~al.}(2018)\citenamefont {Hamada},
  \citenamefont {Minamitani}, \citenamefont {Hirayama},\ and\ \citenamefont
  {Murakami}}]{Hamada2018}%
  \BibitemOpen
  \bibfield  {author} {\bibinfo {author} {\bibfnamefont {M.}~\bibnamefont
  {Hamada}}, \bibinfo {author} {\bibfnamefont {E.}~\bibnamefont {Minamitani}},
  \bibinfo {author} {\bibfnamefont {M.}~\bibnamefont {Hirayama}},\ and\
  \bibinfo {author} {\bibfnamefont {S.}~\bibnamefont {Murakami}},\ }\bibfield
  {title} {\bibinfo {title} {{Phonon Angular Momentum Induced by the
  Temperature Gradient}},\ }\href
  {https://doi.org/10.1103/PhysRevLett.121.175301} {\bibfield  {journal}
  {\bibinfo  {journal} {Phys. Rev. Lett.}\ }\textbf {\bibinfo {volume} {121}},\
  \bibinfo {pages} {175301} (\bibinfo {year} {2018})}\BibitemShut {NoStop}%
\bibitem [{\citenamefont {Zabalo}\ \emph {et~al.}(2022)\citenamefont {Zabalo},
  \citenamefont {Dreyer},\ and\ \citenamefont {Stengel}}]{Zabalo2022}%
  \BibitemOpen
  \bibfield  {author} {\bibinfo {author} {\bibfnamefont {A.}~\bibnamefont
  {Zabalo}}, \bibinfo {author} {\bibfnamefont {C.~E.}\ \bibnamefont {Dreyer}},\
  and\ \bibinfo {author} {\bibfnamefont {M.}~\bibnamefont {Stengel}},\
  }\bibfield  {title} {\bibinfo {title} {{Rotational g factors and Lorentz
  forces of molecules and solids from density functional perturbation
  theory}},\ }\href {https://doi.org/10.1103/physrevb.105.094305} {\bibfield
  {journal} {\bibinfo  {journal} {Phys. Rev. B}\ }\textbf {\bibinfo {volume}
  {105}},\ \bibinfo {pages} {094305} (\bibinfo {year} {2022})}\BibitemShut
  {NoStop}%
\bibitem [{\citenamefont {Schaack}(1976)}]{schaack:1976}%
  \BibitemOpen
  \bibfield  {author} {\bibinfo {author} {\bibfnamefont {G.}~\bibnamefont
  {Schaack}},\ }\bibfield  {title} {\bibinfo {title} {{Observation of
  circularly polarized phonon states in an external magnetic field}},\ }\href
  {https://doi.org/10.1088/0022-3719/9/11/009} {\bibfield  {journal} {\bibinfo
  {journal} {J.~Phys. C: Solid State Phys.}\ }\textbf {\bibinfo {volume} {9}},\
  \bibinfo {pages} {297} (\bibinfo {year} {1976})}\BibitemShut {NoStop}%
\bibitem [{\citenamefont {Schaack}(1977)}]{schaack:1977}%
  \BibitemOpen
  \bibfield  {author} {\bibinfo {author} {\bibfnamefont {G.}~\bibnamefont
  {Schaack}},\ }\bibfield  {title} {\bibinfo {title} {{Magnetic Field Dependent
  Splitting of Doubly Degenerate Phonon States in Anhydrous
  Cerium-Trichloride}},\ }\href {https://doi.org/10.1007/BF01313371} {\bibfield
   {journal} {\bibinfo  {journal} {Z. Physik B}\ }\textbf {\bibinfo {volume}
  {26}},\ \bibinfo {pages} {49} (\bibinfo {year} {1977})}\BibitemShut {NoStop}%
\bibitem [{\citenamefont {Baydin}\ \emph {et~al.}(2021)\citenamefont {Baydin},
  \citenamefont {Hernandez}, \citenamefont {Rodriguez-Vega}, \citenamefont
  {Okazaki}, \citenamefont {Tay}, \citenamefont {Noe}, \citenamefont
  {Katayama}, \citenamefont {Takeda}, \citenamefont {Nojiri}, \citenamefont
  {Rappl}, \citenamefont {Abramof}, \citenamefont {Fiete},\ and\ \citenamefont
  {Kono}}]{Baydin2022}%
  \BibitemOpen
  \bibfield  {author} {\bibinfo {author} {\bibfnamefont {A.}~\bibnamefont
  {Baydin}}, \bibinfo {author} {\bibfnamefont {F.~G.~G.}\ \bibnamefont
  {Hernandez}}, \bibinfo {author} {\bibfnamefont {M.}~\bibnamefont
  {Rodriguez-Vega}}, \bibinfo {author} {\bibfnamefont {A.~K.}\ \bibnamefont
  {Okazaki}}, \bibinfo {author} {\bibfnamefont {F.}~\bibnamefont {Tay}},
  \bibinfo {author} {\bibfnamefont {G.~T.}\ \bibnamefont {Noe}}, \bibinfo
  {author} {\bibfnamefont {I.}~\bibnamefont {Katayama}}, \bibinfo {author}
  {\bibfnamefont {J.}~\bibnamefont {Takeda}}, \bibinfo {author} {\bibfnamefont
  {H.}~\bibnamefont {Nojiri}}, \bibinfo {author} {\bibfnamefont {P.~H.~O.}\
  \bibnamefont {Rappl}}, \bibinfo {author} {\bibfnamefont {E.}~\bibnamefont
  {Abramof}}, \bibinfo {author} {\bibfnamefont {G.~A.}\ \bibnamefont {Fiete}},\
  and\ \bibinfo {author} {\bibfnamefont {J.}~\bibnamefont {Kono}},\ }\bibfield
  {title} {\bibinfo {title} {{Magnetic Control of Soft Chiral Phonons in
  PbTe}},\ }\href {https://doi.org/10.1103/PhysRevLett.128.075901} {\bibfield
  {journal} {\bibinfo  {journal} {Physical Review Letters}\ }\textbf {\bibinfo
  {volume} {128}},\ \bibinfo {pages} {075901} (\bibinfo {year}
  {2021})}\BibitemShut {NoStop}%
\bibitem [{\citenamefont {Cheng}\ \emph {et~al.}(2020)\citenamefont {Cheng},
  \citenamefont {Schumann}, \citenamefont {Wang}, \citenamefont {Zhang},
  \citenamefont {Barbalas}, \citenamefont {Stemmer},\ and\ \citenamefont
  {Armitage}}]{Cheng2020}%
  \BibitemOpen
  \bibfield  {author} {\bibinfo {author} {\bibfnamefont {B.}~\bibnamefont
  {Cheng}}, \bibinfo {author} {\bibfnamefont {T.}~\bibnamefont {Schumann}},
  \bibinfo {author} {\bibfnamefont {Y.}~\bibnamefont {Wang}}, \bibinfo {author}
  {\bibfnamefont {X.}~\bibnamefont {Zhang}}, \bibinfo {author} {\bibfnamefont
  {D.}~\bibnamefont {Barbalas}}, \bibinfo {author} {\bibfnamefont
  {S.}~\bibnamefont {Stemmer}},\ and\ \bibinfo {author} {\bibfnamefont {N.~P.}\
  \bibnamefont {Armitage}},\ }\bibfield  {title} {\bibinfo {title} {{A Large
  Effective Phonon Magnetic Moment in a Dirac Semimetal}},\ }\href
  {https://doi.org/10.1021/acs.nanolett.0c01983} {\bibfield  {journal}
  {\bibinfo  {journal} {Nano Lett.}\ }\textbf {\bibinfo {volume} {20}},\
  \bibinfo {pages} {5991} (\bibinfo {year} {2020})}\BibitemShut {NoStop}%
\bibitem [{\citenamefont {Hernandez}\ \emph {et~al.}(2022)\citenamefont
  {Hernandez}, \citenamefont {Baydin}, \citenamefont {Chaudhary}, \citenamefont
  {Tay}, \citenamefont {Katayama}, \citenamefont {Takeda}, \citenamefont
  {Nojiri}, \citenamefont {Okazaki}, \citenamefont {Rappl}, \citenamefont
  {Abramof} \emph {et~al.}}]{hernandez2022chiral}%
  \BibitemOpen
  \bibfield  {author} {\bibinfo {author} {\bibfnamefont {F.~G.~G.}\
  \bibnamefont {Hernandez}}, \bibinfo {author} {\bibfnamefont {A.}~\bibnamefont
  {Baydin}}, \bibinfo {author} {\bibfnamefont {S.}~\bibnamefont {Chaudhary}},
  \bibinfo {author} {\bibfnamefont {F.}~\bibnamefont {Tay}}, \bibinfo {author}
  {\bibfnamefont {I.}~\bibnamefont {Katayama}}, \bibinfo {author}
  {\bibfnamefont {J.}~\bibnamefont {Takeda}}, \bibinfo {author} {\bibfnamefont
  {H.}~\bibnamefont {Nojiri}}, \bibinfo {author} {\bibfnamefont {A.~K.}\
  \bibnamefont {Okazaki}}, \bibinfo {author} {\bibfnamefont {P.~H.~O.}\
  \bibnamefont {Rappl}}, \bibinfo {author} {\bibfnamefont {E.}~\bibnamefont
  {Abramof}}, \emph {et~al.},\ }\bibfield  {title} {\bibinfo {title} {Chiral
  phonons with giant magnetic moments in a topological crystalline insulator},\
  }\href {https://arxiv.org/abs/2208.12235} {\bibfield  {journal} {\bibinfo
  {journal} {arXiv:2208.12235}\ } (\bibinfo {year} {2022})}\BibitemShut
  {NoStop}%
\bibitem [{\citenamefont {Sengupta}\ \emph {et~al.}(2020)\citenamefont
  {Sengupta}, \citenamefont {Lhachemi},\ and\ \citenamefont
  {Garate}}]{Sengupta2020}%
  \BibitemOpen
  \bibfield  {author} {\bibinfo {author} {\bibfnamefont {S.}~\bibnamefont
  {Sengupta}}, \bibinfo {author} {\bibfnamefont {M.~N.~Y.}\ \bibnamefont
  {Lhachemi}},\ and\ \bibinfo {author} {\bibfnamefont {I.}~\bibnamefont
  {Garate}},\ }\bibfield  {title} {\bibinfo {title} {{Phonon Magnetochiral
  Effect of Band-Geometric Origin in Weyl Semimetals}},\ }\href
  {https://doi.org/10.1103/PhysRevLett.125.146402} {\bibfield  {journal}
  {\bibinfo  {journal} {Phys. Rev. Lett.}\ }\textbf {\bibinfo {volume} {125}},\
  \bibinfo {pages} {146402} (\bibinfo {year} {2020})}\BibitemShut {NoStop}%
\bibitem [{\citenamefont {Xiao}\ \emph {et~al.}(2021)\citenamefont {Xiao},
  \citenamefont {Ren},\ and\ \citenamefont {Xiong}}]{Xiao2021}%
  \BibitemOpen
  \bibfield  {author} {\bibinfo {author} {\bibfnamefont {C.}~\bibnamefont
  {Xiao}}, \bibinfo {author} {\bibfnamefont {Y.}~\bibnamefont {Ren}},\ and\
  \bibinfo {author} {\bibfnamefont {B.}~\bibnamefont {Xiong}},\ }\bibfield
  {title} {\bibinfo {title} {{Adiabatically induced orbital magnetization}},\
  }\href {https://doi.org/10.1103/PhysRevB.103.115432} {\bibfield  {journal}
  {\bibinfo  {journal} {Phys. Rev. B}\ }\textbf {\bibinfo {volume} {103}},\
  \bibinfo {pages} {115432} (\bibinfo {year} {2021})}\BibitemShut {NoStop}%
\bibitem [{\citenamefont {Ren}\ \emph {et~al.}(2021)\citenamefont {Ren},
  \citenamefont {Xiao}, \citenamefont {Saparov},\ and\ \citenamefont
  {Niu}}]{Ren2021}%
  \BibitemOpen
  \bibfield  {author} {\bibinfo {author} {\bibfnamefont {Y.}~\bibnamefont
  {Ren}}, \bibinfo {author} {\bibfnamefont {C.}~\bibnamefont {Xiao}}, \bibinfo
  {author} {\bibfnamefont {D.}~\bibnamefont {Saparov}},\ and\ \bibinfo {author}
  {\bibfnamefont {Q.}~\bibnamefont {Niu}},\ }\bibfield  {title} {\bibinfo
  {title} {{Phonon Magnetic Moment from Electronic Topological
  Magnetization}},\ }\href {https://doi.org/10.1103/PhysRevLett.127.186403}
  {\bibfield  {journal} {\bibinfo  {journal} {Phys. Rev. Lett.}\ }\textbf
  {\bibinfo {volume} {127}},\ \bibinfo {pages} {186403} (\bibinfo {year}
  {2021})}\BibitemShut {NoStop}%
\bibitem [{\citenamefont {Zhang}\ \emph {et~al.}(2023)\citenamefont {Zhang},
  \citenamefont {Ren}, \citenamefont {Wang}, \citenamefont {Cao},\ and\
  \citenamefont {Xiao}}]{Xiao-Wei2023}%
  \BibitemOpen
  \bibfield  {author} {\bibinfo {author} {\bibfnamefont {X.-W.}\ \bibnamefont
  {Zhang}}, \bibinfo {author} {\bibfnamefont {Y.}~\bibnamefont {Ren}}, \bibinfo
  {author} {\bibfnamefont {C.}~\bibnamefont {Wang}}, \bibinfo {author}
  {\bibfnamefont {T.}~\bibnamefont {Cao}},\ and\ \bibinfo {author}
  {\bibfnamefont {D.}~\bibnamefont {Xiao}},\ }\bibfield  {title} {\bibinfo
  {title} {Gate-tunable phonon magnetic moment in bilayer graphene},\ }\href
  {https://doi.org/10.1103/PhysRevLett.130.226302} {\bibfield  {journal}
  {\bibinfo  {journal} {Phys. Rev. Lett.}\ }\textbf {\bibinfo {volume} {130}},\
  \bibinfo {pages} {226302} (\bibinfo {year} {2023})}\BibitemShut {NoStop}%
\bibitem [{\citenamefont {Thalmeier}\ and\ \citenamefont
  {Fulde}(1977)}]{Thalmeier1977}%
  \BibitemOpen
  \bibfield  {author} {\bibinfo {author} {\bibfnamefont {P.}~\bibnamefont
  {Thalmeier}}\ and\ \bibinfo {author} {\bibfnamefont {P.}~\bibnamefont
  {Fulde}},\ }\bibfield  {title} {\bibinfo {title} {{Optical Phonons of
  Rare-Earth Halides in a Magnetic Field}},\ }\href
  {https://doi.org/10.1007/BF01570742} {\bibfield  {journal} {\bibinfo
  {journal} {Z. Phys. B}\ }\textbf {\bibinfo {volume} {26}},\ \bibinfo {pages}
  {323} (\bibinfo {year} {1977})}\BibitemShut {NoStop}%
\bibitem [{\citenamefont {Thalmeier}\ and\ \citenamefont
  {Fulde}(1978)}]{thalmeier:1978}%
  \BibitemOpen
  \bibfield  {author} {\bibinfo {author} {\bibfnamefont {P.}~\bibnamefont
  {Thalmeier}}\ and\ \bibinfo {author} {\bibfnamefont {P.}~\bibnamefont
  {Fulde}},\ }\bibfield  {title} {\bibinfo {title} {{Faraday and Cotton-Mouton
  Effects for Acoustic Phonons in Paramagnetic Rare Earth Compounds}},\ }\href
  {https://doi.org/10.1007/BF01324026} {\bibfield  {journal} {\bibinfo
  {journal} {Z. Phys. B}\ }\textbf {\bibinfo {volume} {29}},\ \bibinfo {pages}
  {299} (\bibinfo {year} {1978})}\BibitemShut {NoStop}%
\bibitem [{\citenamefont {Orbach}(1961)}]{Orbach1961}%
  \BibitemOpen
  \bibfield  {author} {\bibinfo {author} {\bibfnamefont {R.}~\bibnamefont
  {Orbach}},\ }\bibfield  {title} {\bibinfo {title} {{Spin-lattice relaxation
  in rare-earth salts}},\ }\href {https://doi.org/10.1098/rspa.1961.0211}
  {\bibfield  {journal} {\bibinfo  {journal} {Proc. Roy. Soc. A}\ }\textbf
  {\bibinfo {volume} {264}},\ \bibinfo {pages} {458} (\bibinfo {year}
  {1961})}\BibitemShut {NoStop}%
\bibitem [{\citenamefont {Stevens}(1952)}]{stevens1952matrix}%
  \BibitemOpen
  \bibfield  {author} {\bibinfo {author} {\bibfnamefont {K.}~\bibnamefont
  {Stevens}},\ }\bibfield  {title} {\bibinfo {title} {Matrix elements and
  operator equivalents connected with the magnetic properties of rare earth
  ions},\ }\href {https://nilab.physics.ucla.edu/sites/default/files/cef-2.pdf}
  {\bibfield  {journal} {\bibinfo  {journal} {Proceedings of the Physical
  Society. Section A}\ }\textbf {\bibinfo {volume} {65}},\ \bibinfo {pages}
  {209} (\bibinfo {year} {1952})}\BibitemShut {NoStop}%
\bibitem [{\citenamefont {Zhang}\ and\ \citenamefont {Niu}(2015)}]{zhang:2015}%
  \BibitemOpen
  \bibfield  {author} {\bibinfo {author} {\bibfnamefont {L.}~\bibnamefont
  {Zhang}}\ and\ \bibinfo {author} {\bibfnamefont {Q.}~\bibnamefont {Niu}},\
  }\bibfield  {title} {\bibinfo {title} {{Chiral Phonons at High-Symmetry
  Points in Monolayer Hexagonal Lattices}},\ }\href
  {https://doi.org/10.1103/PhysRevLett.115.115502} {\bibfield  {journal}
  {\bibinfo  {journal} {Phys. Rev. Lett.}\ }\textbf {\bibinfo {volume} {115}},\
  \bibinfo {pages} {115502} (\bibinfo {year} {2015})}\BibitemShut {NoStop}%
\bibitem [{\citenamefont {Schaack}(1975)}]{schaack:1975}%
  \BibitemOpen
  \bibfield  {author} {\bibinfo {author} {\bibfnamefont {G.}~\bibnamefont
  {Schaack}},\ }\bibfield  {title} {\bibinfo {title} {{Magnetic-field dependent
  phonon states in paramagnetic CeF$_3$}},\ }\href
  {https://doi.org/10.1016/0038-1098(75)90488-3} {\bibfield  {journal}
  {\bibinfo  {journal} {Solid State Commun.}\ }\textbf {\bibinfo {volume}
  {17}},\ \bibinfo {pages} {505} (\bibinfo {year} {1975})}\BibitemShut
  {NoStop}%
\bibitem [{\citenamefont {Khomskii}\ \emph {et~al.}(2016)\citenamefont
  {Khomskii}, \citenamefont {Kugel}, \citenamefont {Sboychakov},\ and\
  \citenamefont {Streltsov}}]{khomskii2016role}%
  \BibitemOpen
  \bibfield  {author} {\bibinfo {author} {\bibfnamefont {D.}~\bibnamefont
  {Khomskii}}, \bibinfo {author} {\bibfnamefont {K.}~\bibnamefont {Kugel}},
  \bibinfo {author} {\bibfnamefont {A.}~\bibnamefont {Sboychakov}},\ and\
  \bibinfo {author} {\bibfnamefont {S.}~\bibnamefont {Streltsov}},\ }\bibfield
  {title} {\bibinfo {title} {Role of local geometry in the spin and orbital
  structure of transition metal compounds},\ }\href
  {https://arxiv.org/pdf/1510.05371.pdf} {\bibfield  {journal} {\bibinfo
  {journal} {Journal of experimental and theoretical physics}\ }\textbf
  {\bibinfo {volume} {122}},\ \bibinfo {pages} {484} (\bibinfo {year}
  {2016})}\BibitemShut {NoStop}%
\bibitem [{\citenamefont {Goodenough}(1968)}]{goodenough1968}%
  \BibitemOpen
  \bibfield  {author} {\bibinfo {author} {\bibfnamefont {J.~B.}\ \bibnamefont
  {Goodenough}},\ }\bibfield  {title} {\bibinfo {title} {Spin-orbit-coupling
  effects in transition-metal compounds},\ }\href
  {https://doi.org/10.1103/PhysRev.171.466} {\bibfield  {journal} {\bibinfo
  {journal} {Phys. Rev.}\ }\textbf {\bibinfo {volume} {171}},\ \bibinfo {pages}
  {466} (\bibinfo {year} {1968})}\BibitemShut {NoStop}%
\bibitem [{\citenamefont {Yuan}\ \emph
  {et~al.}(2020{\natexlab{a}})\citenamefont {Yuan}, \citenamefont {Stone},
  \citenamefont {Shu}, \citenamefont {Chou}, \citenamefont {Rao}, \citenamefont
  {Clancy},\ and\ \citenamefont {Kim}}]{Yuan2020}%
  \BibitemOpen
  \bibfield  {author} {\bibinfo {author} {\bibfnamefont {B.}~\bibnamefont
  {Yuan}}, \bibinfo {author} {\bibfnamefont {M.~B.}\ \bibnamefont {Stone}},
  \bibinfo {author} {\bibfnamefont {G.-J.}\ \bibnamefont {Shu}}, \bibinfo
  {author} {\bibfnamefont {F.~C.}\ \bibnamefont {Chou}}, \bibinfo {author}
  {\bibfnamefont {X.}~\bibnamefont {Rao}}, \bibinfo {author} {\bibfnamefont
  {J.~P.}\ \bibnamefont {Clancy}},\ and\ \bibinfo {author} {\bibfnamefont
  {Y.-J.}\ \bibnamefont {Kim}},\ }\bibfield  {title} {\bibinfo {title}
  {Spin-orbit exciton in a honeycomb lattice magnet ${\mathrm{cotio}}_{3}$:
  Revealing a link between magnetism in $d$- and $f$-electron systems},\ }\href
  {https://doi.org/10.1103/PhysRevB.102.134404} {\bibfield  {journal} {\bibinfo
   {journal} {Phys. Rev. B}\ }\textbf {\bibinfo {volume} {102}},\ \bibinfo
  {pages} {134404} (\bibinfo {year} {2020}{\natexlab{a}})}\BibitemShut
  {NoStop}%
\bibitem [{\citenamefont {Yuan}\ \emph
  {et~al.}(2020{\natexlab{b}})\citenamefont {Yuan}, \citenamefont {Khait},
  \citenamefont {Shu}, \citenamefont {Chou}, \citenamefont {Stone},
  \citenamefont {Clancy}, \citenamefont {Paramekanti},\ and\ \citenamefont
  {Kim}}]{yuan2020dirac}%
  \BibitemOpen
  \bibfield  {author} {\bibinfo {author} {\bibfnamefont {B.}~\bibnamefont
  {Yuan}}, \bibinfo {author} {\bibfnamefont {I.}~\bibnamefont {Khait}},
  \bibinfo {author} {\bibfnamefont {G.-J.}\ \bibnamefont {Shu}}, \bibinfo
  {author} {\bibfnamefont {F.~C.}\ \bibnamefont {Chou}}, \bibinfo {author}
  {\bibfnamefont {M.~B.}\ \bibnamefont {Stone}}, \bibinfo {author}
  {\bibfnamefont {J.~P.}\ \bibnamefont {Clancy}}, \bibinfo {author}
  {\bibfnamefont {A.}~\bibnamefont {Paramekanti}},\ and\ \bibinfo {author}
  {\bibfnamefont {Y.-J.}\ \bibnamefont {Kim}},\ }\bibfield  {title} {\bibinfo
  {title} {Dirac magnons in a honeycomb lattice quantum xy magnet
  ${\mathrm{cotio}}_{3}$},\ }\href {https://doi.org/10.1103/PhysRevX.10.011062}
  {\bibfield  {journal} {\bibinfo  {journal} {Phys. Rev. X}\ }\textbf {\bibinfo
  {volume} {10}},\ \bibinfo {pages} {011062} (\bibinfo {year}
  {2020}{\natexlab{b}})}\BibitemShut {NoStop}%
\bibitem [{\citenamefont {Elliot}\ \emph {et~al.}(2021)\citenamefont {Elliot},
  \citenamefont {McClarty}, \citenamefont {Prabhakaran}, \citenamefont
  {Johnson}, \citenamefont {Walker}, \citenamefont {Manuel},\ and\
  \citenamefont {Coldea}}]{elliot2021order}%
  \BibitemOpen
  \bibfield  {author} {\bibinfo {author} {\bibfnamefont {M.}~\bibnamefont
  {Elliot}}, \bibinfo {author} {\bibfnamefont {P.~A.}\ \bibnamefont
  {McClarty}}, \bibinfo {author} {\bibfnamefont {D.}~\bibnamefont
  {Prabhakaran}}, \bibinfo {author} {\bibfnamefont {R.}~\bibnamefont
  {Johnson}}, \bibinfo {author} {\bibfnamefont {H.}~\bibnamefont {Walker}},
  \bibinfo {author} {\bibfnamefont {P.}~\bibnamefont {Manuel}},\ and\ \bibinfo
  {author} {\bibfnamefont {R.}~\bibnamefont {Coldea}},\ }\bibfield  {title}
  {\bibinfo {title} {Order-by-disorder from bond-dependent exchange and
  intensity signature of nodal quasiparticles in a honeycomb cobaltate},\
  }\href {https://www.nature.com/articles/s41467-021-23851-0} {\bibfield
  {journal} {\bibinfo  {journal} {Nature Communications}\ }\textbf {\bibinfo
  {volume} {12}},\ \bibinfo {pages} {1} (\bibinfo {year} {2021})}\BibitemShut
  {NoStop}%
\bibitem [{\citenamefont {Dubrovin}\ \emph {et~al.}(2021)\citenamefont
  {Dubrovin}, \citenamefont {Siverin}, \citenamefont {Prosnikov}, \citenamefont
  {Chernyshev}, \citenamefont {Novikova}, \citenamefont {Christianen},
  \citenamefont {Balbashov},\ and\ \citenamefont
  {Pisarev}}]{dubrovin2021lattice}%
  \BibitemOpen
  \bibfield  {author} {\bibinfo {author} {\bibfnamefont {R.}~\bibnamefont
  {Dubrovin}}, \bibinfo {author} {\bibfnamefont {N.}~\bibnamefont {Siverin}},
  \bibinfo {author} {\bibfnamefont {M.}~\bibnamefont {Prosnikov}}, \bibinfo
  {author} {\bibfnamefont {V.}~\bibnamefont {Chernyshev}}, \bibinfo {author}
  {\bibfnamefont {N.}~\bibnamefont {Novikova}}, \bibinfo {author}
  {\bibfnamefont {P.}~\bibnamefont {Christianen}}, \bibinfo {author}
  {\bibfnamefont {A.}~\bibnamefont {Balbashov}},\ and\ \bibinfo {author}
  {\bibfnamefont {R.}~\bibnamefont {Pisarev}},\ }\bibfield  {title} {\bibinfo
  {title} {Lattice dynamics and spontaneous magnetodielectric effect in
  ilmenite cotio3},\ }\href@noop {} {\bibfield  {journal} {\bibinfo  {journal}
  {Journal of Alloys and Compounds}\ }\textbf {\bibinfo {volume} {858}},\
  \bibinfo {pages} {157633} (\bibinfo {year} {2021})}\BibitemShut {NoStop}%
\bibitem [{\citenamefont {Kroumova}\ \emph {et~al.}(2003)\citenamefont
  {Kroumova}, \citenamefont {Aroyo}, \citenamefont {Perez-Mato}, \citenamefont
  {Kirov}, \citenamefont {Capillas}, \citenamefont {Ivantchev},\ and\
  \citenamefont {Wondratschek}}]{kroumova2003bilbao}%
  \BibitemOpen
  \bibfield  {author} {\bibinfo {author} {\bibfnamefont {E.}~\bibnamefont
  {Kroumova}}, \bibinfo {author} {\bibfnamefont {M.}~\bibnamefont {Aroyo}},
  \bibinfo {author} {\bibfnamefont {J.}~\bibnamefont {Perez-Mato}}, \bibinfo
  {author} {\bibfnamefont {A.}~\bibnamefont {Kirov}}, \bibinfo {author}
  {\bibfnamefont {C.}~\bibnamefont {Capillas}}, \bibinfo {author}
  {\bibfnamefont {S.}~\bibnamefont {Ivantchev}},\ and\ \bibinfo {author}
  {\bibfnamefont {H.}~\bibnamefont {Wondratschek}},\ }\bibfield  {title}
  {\bibinfo {title} {Bilbao crystallographic server: useful databases and tools
  for phase-transition studies},\ }\href@noop {} {\bibfield  {journal}
  {\bibinfo  {journal} {Phase Transitions: A Multinational Journal}\ }\textbf
  {\bibinfo {volume} {76}},\ \bibinfo {pages} {155} (\bibinfo {year}
  {2003})}\BibitemShut {NoStop}%
\bibitem [{\citenamefont {Shannon}(1976)}]{shannon1976revised}%
  \BibitemOpen
  \bibfield  {author} {\bibinfo {author} {\bibfnamefont {R.~D.}\ \bibnamefont
  {Shannon}},\ }\bibfield  {title} {\bibinfo {title} {Revised effective ionic
  radii and systematic studies of interatomic distances in halides and
  chalcogenides},\ }\href
  {https://scripts.iucr.org/cgi-bin/paper?S0567739476001551} {\bibfield
  {journal} {\bibinfo  {journal} {Acta crystallographica section A: crystal
  physics, diffraction, theoretical and general crystallography}\ }\textbf
  {\bibinfo {volume} {32}},\ \bibinfo {pages} {751} (\bibinfo {year}
  {1976})}\BibitemShut {NoStop}%
\bibitem [{\citenamefont {Bonini}\ \emph {et~al.}(2023)\citenamefont {Bonini},
  \citenamefont {Ren}, \citenamefont {Vanderbilt}, \citenamefont {Stengel},
  \citenamefont {Dreyer},\ and\ \citenamefont {Coh}}]{boniniprl2023}%
  \BibitemOpen
  \bibfield  {author} {\bibinfo {author} {\bibfnamefont {J.}~\bibnamefont
  {Bonini}}, \bibinfo {author} {\bibfnamefont {S.}~\bibnamefont {Ren}},
  \bibinfo {author} {\bibfnamefont {D.}~\bibnamefont {Vanderbilt}}, \bibinfo
  {author} {\bibfnamefont {M.}~\bibnamefont {Stengel}}, \bibinfo {author}
  {\bibfnamefont {C.~E.}\ \bibnamefont {Dreyer}},\ and\ \bibinfo {author}
  {\bibfnamefont {S.}~\bibnamefont {Coh}},\ }\bibfield  {title} {\bibinfo
  {title} {Frequency splitting of chiral phonons from broken time-reversal
  symmetry in cri$_{3}$},\ }\href
  {https://doi.org/10.1103/PhysRevLett.130.086701} {\bibfield  {journal}
  {\bibinfo  {journal} {Phys. Rev. Lett.}\ }\textbf {\bibinfo {volume} {130}},\
  \bibinfo {pages} {086701} (\bibinfo {year} {2023})}\BibitemShut {NoStop}%
\bibitem [{\citenamefont {Landau}\ \emph {et~al.}(1973)\citenamefont {Landau},
  \citenamefont {Doran},\ and\ \citenamefont {Keen}}]{Landau1973}%
  \BibitemOpen
  \bibfield  {author} {\bibinfo {author} {\bibfnamefont {D.~P.}\ \bibnamefont
  {Landau}}, \bibinfo {author} {\bibfnamefont {J.~C.}\ \bibnamefont {Doran}},\
  and\ \bibinfo {author} {\bibfnamefont {B.~E.}\ \bibnamefont {Keen}},\
  }\bibfield  {title} {\bibinfo {title} {{Thermal and Magnetic Properties of
  CeCl3}},\ }\href {https://doi.org/10.1103/PhysRevB.7.4961} {\bibfield
  {journal} {\bibinfo  {journal} {Phys. Rev. B}\ }\textbf {\bibinfo {volume}
  {7}},\ \bibinfo {pages} {4961} (\bibinfo {year} {1973})}\BibitemShut
  {NoStop}%
\bibitem [{\citenamefont {Sandilands}\ \emph {et~al.}(2016)\citenamefont
  {Sandilands}, \citenamefont {Tian}, \citenamefont {Reijnders}, \citenamefont
  {Kim}, \citenamefont {Plumb}, \citenamefont {Kim}, \citenamefont {Kee},\ and\
  \citenamefont {Burch}}]{Sandiland2016}%
  \BibitemOpen
  \bibfield  {author} {\bibinfo {author} {\bibfnamefont {L.~J.}\ \bibnamefont
  {Sandilands}}, \bibinfo {author} {\bibfnamefont {Y.}~\bibnamefont {Tian}},
  \bibinfo {author} {\bibfnamefont {A.~A.}\ \bibnamefont {Reijnders}}, \bibinfo
  {author} {\bibfnamefont {H.-S.}\ \bibnamefont {Kim}}, \bibinfo {author}
  {\bibfnamefont {K.~W.}\ \bibnamefont {Plumb}}, \bibinfo {author}
  {\bibfnamefont {Y.-J.}\ \bibnamefont {Kim}}, \bibinfo {author} {\bibfnamefont
  {H.-Y.}\ \bibnamefont {Kee}},\ and\ \bibinfo {author} {\bibfnamefont {K.~S.}\
  \bibnamefont {Burch}},\ }\bibfield  {title} {\bibinfo {title} {Spin-orbit
  excitations and electronic structure of the putative kitaev magnet
  $\ensuremath{\alpha}\ensuremath{-}{\mathrm{rucl}}_{3}$},\ }\href
  {https://doi.org/10.1103/PhysRevB.93.075144} {\bibfield  {journal} {\bibinfo
  {journal} {Phys. Rev. B}\ }\textbf {\bibinfo {volume} {93}},\ \bibinfo
  {pages} {075144} (\bibinfo {year} {2016})}\BibitemShut {NoStop}%
\bibitem [{\citenamefont {Takahashi}(1977)}]{Takahashi1977}%
  \BibitemOpen
  \bibfield  {author} {\bibinfo {author} {\bibfnamefont {M.}~\bibnamefont
  {Takahashi}},\ }\bibfield  {title} {\bibinfo {title} {Half-filled hubbard
  model at low temperature},\ }\href
  {https://doi.org/10.1088/0022-3719/10/8/031} {\bibfield  {journal} {\bibinfo
  {journal} {J.~Phys. C: Solid State Phys.}\ }\textbf {\bibinfo {volume}
  {10}},\ \bibinfo {pages} {1289} (\bibinfo {year} {1977})}\BibitemShut
  {NoStop}%
\bibitem [{\citenamefont {Fedorova}\ \emph {et~al.}(2018)\citenamefont
  {Fedorova}, \citenamefont {Bortis}, \citenamefont {Findler},\ and\
  \citenamefont {Spaldin}}]{Fedorova2018}%
  \BibitemOpen
  \bibfield  {author} {\bibinfo {author} {\bibfnamefont {N.~S.}\ \bibnamefont
  {Fedorova}}, \bibinfo {author} {\bibfnamefont {A.}~\bibnamefont {Bortis}},
  \bibinfo {author} {\bibfnamefont {C.}~\bibnamefont {Findler}},\ and\ \bibinfo
  {author} {\bibfnamefont {N.~A.}\ \bibnamefont {Spaldin}},\ }\bibfield
  {title} {\bibinfo {title} {{Four-spin ring interaction as a source of
  unconventional magnetic orders in orthorhombic perovskite manganites}},\
  }\href {https://doi.org/10.1103/PhysRevB.98.235113} {\bibfield  {journal}
  {\bibinfo  {journal} {Phys. Rev. B}\ }\textbf {\bibinfo {volume} {98}},\
  \bibinfo {pages} {235113} (\bibinfo {year} {2018})}\BibitemShut {NoStop}%
\bibitem [{\citenamefont {Liu}\ \emph {et~al.}(2021)\citenamefont {Liu},
  \citenamefont {Granados~del \'Aguila}, \citenamefont {Bhowmick},
  \citenamefont {Gan}, \citenamefont {Thu Ha~Do}, \citenamefont {Prosnikov},
  \citenamefont {Sedmidubsk\'y}, \citenamefont {Sofer}, \citenamefont
  {Christianen}, \citenamefont {Sengupta},\ and\ \citenamefont
  {Xiong}}]{Liu2021_FePS3}%
  \BibitemOpen
  \bibfield  {author} {\bibinfo {author} {\bibfnamefont {S.}~\bibnamefont
  {Liu}}, \bibinfo {author} {\bibfnamefont {A.}~\bibnamefont {Granados~del
  \'Aguila}}, \bibinfo {author} {\bibfnamefont {D.}~\bibnamefont {Bhowmick}},
  \bibinfo {author} {\bibfnamefont {C.~K.}\ \bibnamefont {Gan}}, \bibinfo
  {author} {\bibfnamefont {T.}~\bibnamefont {Thu Ha~Do}}, \bibinfo {author}
  {\bibfnamefont {M.~A.}\ \bibnamefont {Prosnikov}}, \bibinfo {author}
  {\bibfnamefont {D.}~\bibnamefont {Sedmidubsk\'y}}, \bibinfo {author}
  {\bibfnamefont {Z.}~\bibnamefont {Sofer}}, \bibinfo {author} {\bibfnamefont
  {P.~C.~M.}\ \bibnamefont {Christianen}}, \bibinfo {author} {\bibfnamefont
  {P.}~\bibnamefont {Sengupta}},\ and\ \bibinfo {author} {\bibfnamefont
  {Q.}~\bibnamefont {Xiong}},\ }\bibfield  {title} {\bibinfo {title} {Direct
  observation of magnon-phonon strong coupling in two-dimensional
  antiferromagnet at high magnetic fields},\ }\href
  {https://doi.org/10.1103/PhysRevLett.127.097401} {\bibfield  {journal}
  {\bibinfo  {journal} {Phys. Rev. Lett.}\ }\textbf {\bibinfo {volume} {127}},\
  \bibinfo {pages} {097401} (\bibinfo {year} {2021})}\BibitemShut {NoStop}%
\bibitem [{\citenamefont {Vaclavkova}\ \emph {et~al.}(2021)\citenamefont
  {Vaclavkova}, \citenamefont {Palit}, \citenamefont {Wyzula}, \citenamefont
  {Ghosh}, \citenamefont {Delhomme}, \citenamefont {Maity}, \citenamefont
  {Kapuscinski}, \citenamefont {Ghosh}, \citenamefont {Veis}, \citenamefont
  {Grzeszczyk}, \citenamefont {Faugeras}, \citenamefont {Orlita}, \citenamefont
  {Datta},\ and\ \citenamefont {Potemski}}]{Vaclavkova2021_FePS3}%
  \BibitemOpen
  \bibfield  {author} {\bibinfo {author} {\bibfnamefont {D.}~\bibnamefont
  {Vaclavkova}}, \bibinfo {author} {\bibfnamefont {M.}~\bibnamefont {Palit}},
  \bibinfo {author} {\bibfnamefont {J.}~\bibnamefont {Wyzula}}, \bibinfo
  {author} {\bibfnamefont {S.}~\bibnamefont {Ghosh}}, \bibinfo {author}
  {\bibfnamefont {A.}~\bibnamefont {Delhomme}}, \bibinfo {author}
  {\bibfnamefont {S.}~\bibnamefont {Maity}}, \bibinfo {author} {\bibfnamefont
  {P.}~\bibnamefont {Kapuscinski}}, \bibinfo {author} {\bibfnamefont
  {A.}~\bibnamefont {Ghosh}}, \bibinfo {author} {\bibfnamefont
  {M.}~\bibnamefont {Veis}}, \bibinfo {author} {\bibfnamefont {M.}~\bibnamefont
  {Grzeszczyk}}, \bibinfo {author} {\bibfnamefont {C.}~\bibnamefont
  {Faugeras}}, \bibinfo {author} {\bibfnamefont {M.}~\bibnamefont {Orlita}},
  \bibinfo {author} {\bibfnamefont {S.}~\bibnamefont {Datta}},\ and\ \bibinfo
  {author} {\bibfnamefont {M.}~\bibnamefont {Potemski}},\ }\bibfield  {title}
  {\bibinfo {title} {Magnon polarons in the van der waals antiferromagnet
  feps$_3$},\ }\href {https://doi.org/10.1103/PhysRevB.104.134437} {\bibfield
  {journal} {\bibinfo  {journal} {Phys. Rev. B}\ }\textbf {\bibinfo {volume}
  {104}},\ \bibinfo {pages} {134437} (\bibinfo {year} {2021})}\BibitemShut
  {NoStop}%
\end{thebibliography}%

\end{document}